\documentclass{emulateapj}
\usepackage{lscape}

\newcommand{\nms}{\mathsurround=0pt}
\newcommand{\oversim}[2]{\lower 2pt\vbox{\baselineskip 0pt \lineskip 1pt \ialign{$\nms#1\hfil##\hfil$\crcr#2\crcr\sim\crcr}}}
\newcommand{\gtsim}{\mathrel{\mathpalette\oversim{>}}}
\newcommand{\ltsim}{\mathrel{\mathpalette\oversim{<}}}
\newcommand{\persec}{s$^{-1}\,$}

\newcommand{\HI}{H{\sc i }}

\newcommand{\HIIline}{H{\sc ii}}
\newcommand{\OIline}{O{\sc i}}
\newcommand{\OIIIline}{O{\sc iii}}

\newcommand{\OIII}{O{\sc iii }}
\newcommand{\NH}{$N_{\mathrm{H}}$ }
\newcommand{\NHI}{$N_{\mathrm{HI}}$ }
\newcommand{\NHmath}{$N_{\mathrm{H}}$}
\newcommand{\NHImath}{$N_{\mathrm{HI}}$}
\newcommand{\percmq}{cm$^{-2}\,$}
\newcommand{\percmc}{cm$^{-3}\,$}

\newcommand{\wattperhz}{W Hz$^{-1}$}
\newcommand{\tnm}{\tablenotemark}

\slugcomment{Accepted by ApJ on 2010 April 10}

\shorttitle{X-RAY GPS/CSO GALAXIES: MODELING THE SEDs}
\shortauthors{OSTORERO ET AL.}

\begin{document}

\title{X-Ray Emitting GHz-Peaked-Spectrum Galaxies: \\
       Testing a Dynamical-Radiative Model with Broad-Band Spectra}

\author{ L. Ostorero\altaffilmark{1,2}, R. Moderski\altaffilmark{3,4}, \L. Stawarz\altaffilmark{4,5},
         A. Diaferio\altaffilmark{1,2}, I. Kowalska\altaffilmark{6}, \\ C. C. Cheung\altaffilmark{7,8},
         J. Kataoka\altaffilmark{9}, M. C. Begelman\altaffilmark{10}, S. J. Wagner\altaffilmark{11} }

\altaffiltext{1}{Dipartimento di Fisica Generale ``Amedeo Avogadro'', 
                 Universit\`a degli Studi di Torino, Via P. Giuria 1, 
                 10125 Torino, Italy}
\altaffiltext{2}{Istituto Nazionale di Fisica Nucleare (INFN), Via P. 
                 Giuria 1, 10125 Torino, Italy}
\altaffiltext{3}{Nicolaus Copernicus Astronomical Center, Bartycka 18, 00-716
                  Warsaw, Poland}
\altaffiltext{4}{Kavli Institute for Particle Astrophysics and Cosmology,
                 Stanford University, Stanford CA 94305}
\altaffiltext{5}{Astronomical Observatory, Jagiellonian University, ul. Orla
                  171, 30-244  Krak\'ow, Poland}
\altaffiltext{6}{Astronomical Observatory, University of Warsaw, Al. Ujazdowskie 4, 
                 00-478 Warsaw, Poland}
\altaffiltext{7}{NASA Goddard Space Flight Center, Astrophysics Science Division, 
                  Greenbelt, MD 20771, USA}
\altaffiltext{8}{Currently: National Research Council Research Associate, 
                 Space Science Division, Naval Research Laboratory, Washington, 
                 DC 20375, USA}
\altaffiltext{9}{Research Institute for Science and Engineering, Waseda University, 
                 3-4-1 Okubo, Shinjuku, Tokyo, Japan, 169-8555}
\altaffiltext{10}{Joint Institute for Laboratory Astrophysics, University of Colorado, 
                 Boulder, CO 80309-0440, USA}
\altaffiltext{11}{Landessternwarte Heidelberg-K\"onigstuhl, K\"onigstuhl 12,
                 69117 Heidelberg, Germany}

\begin{abstract}
In a dynamical-radiative model we recently developed to describe 
the physics of compact, GHz-Peaked-Spectrum (GPS) sources, 
the relativistic jets propagate across the inner, kpc-sized region of 
the host galaxy, while the electron population of the expanding lobes evolves and 
emits synchrotron and inverse-Compton (IC) radiation.
Interstellar-medium gas clouds engulfed by the expanding lobes, 
and photoionized by the active nucleus, are responsible for the 
radio spectral turnover through free-free absorption (FFA) of the 
synchrotron photons.
The model provides a description of the evolution of the spectral energy 
distribution (SED) of GPS sources with their expansion, predicting significant 
and complex high-energy emission, from the X-ray to the $\gamma$-ray 
frequency domain. 
Here, we test this model with the broad-band SEDs 
of a sample of eleven X-ray emitting GPS galaxies with 
Compact-Symmetric-Object (CSO) morphology, and show that:
(i)  the shape of the radio continuum at frequencies lower than the 
     spectral turnover is indeed well accounted for by the FFA mechanism; 
(ii) the observed X-ray spectra can be interpreted as  
     non-thermal radiation produced via IC scattering of 
     the local radiation fields off the lobe particles, 
     providing a viable alternative to the thermal, accretion-disk dominated scenario.
We also show that the relation between the hydrogen column densities 
derived from the X-ray (\NHmath) and radio (\NHImath) data of the sources
is suggestive of a positive correlation, which, if confirmed by future observations, 
would provide further support to our scenario of high-energy 
emitting lobes.
\end{abstract}


\section{INTRODUCTION}
\label{sec_introduction}

The power released by active galactic nuclei (AGNs) is currently interpreted 
in terms of conversion of gravitational energy to radiative energy 
by accretion processes feeding the central supermassive black hole (BH) with 
environmental gas. 
The triggering, maintenance, and fading 
of the AGN activity, as well as the 
link of these processes with the physical conditions of the environment, from 
sub- to super-galactic scales, are still widely debated issues, 
and keep on stimulating a large variety of scientific investigations. 
In this context, radio galaxies are ideal laboratories,
because they offer an edge-on view of both their 
nuclei and their relativistic jets, 
launched from the galactic center and reaching up to Mpc distances. 
The variety of powers, sizes, and morphologies 
displayed by radio galaxies not only provides pieces of evidence of different 
environmental physical conditions, but also samples subsequent stages of the 
source evolution.
In particular, key sources for the investigation of the very first phases 
of the evolution of radio galaxies are the Gigahertz-Peaked-Spectrum (GPS) sources 
associated with galaxies and characterized by a Compact Symmetric Object (CSO) 
radio morphology.

GPS sources \citep[see][for a review]{odea1998}
are a class of powerful radio sources 
($P_{1.4\, {\rm GHz}}\gtsim 10^{25}$ \wattperhz) displaying convex radio spectra 
that turn over at frequencies of about 0.5--1 GHz; 
they  make up a conspicuous fraction ($\sim$10\%) of the radio sources found in 
high-frequency radio surveys, and are optically identified with either galaxies 
or quasars; their radio sizes are smaller than about 1 kpc, and 
their radio morphologies reveal either core-jet structures or 
mini-lobes embedding terminal hotspots and possibly straddling a central core.
CSOs \citep{wilkinson1994,readhead1996} have instead emerged in VLBI surveys as a class 
of radio sources with compact ($\ltsim$500 pc) and symmetric radio structures, 
making them resemble ``classical double'' radio galaxies in miniature.
Whereas there is no debate on the true compactness of CSOs, 
the compactness of a fraction of the core-jet GPS sources 
might be the result of foreshortening of extended radio sources roughly 
aligned with the line of sight;
especially (although not exclusively) in these cases, the GPS itself might be 
a transient spectral state: flux-density variability and polarization studies 
at different radio frequencies are thus instrumental to the selection of bona-fide 
samples of GPS sources \citep{tinti2005,torniainen2005,torniainen2007,orienti2008a}.
The overlap between the GPS-source and the CSO classes is however significant:
GPS sources associated with galaxies are most likely to display a CSO morphology, and 
most CSOs exhibit a GPS. This overlap is explained in terms of 
synchrotron radio spectra dominated by the emission of the mini-lobes, 
and suffering from absorption effects causing the turnover about 1 GHz
\citep{snellen2000}. 
GPS/CSOs {associated with} galaxies
are thus high-confidence candidates for truly compact sources.

Although compact GPS sources were proposed to be young objects soon after their discovery 
\citep[e.g.,][]{shklovsky1965},
a widely discussed alternative to the youth scenario  
was the frustration scenario, ascribing the small source size to confinement by
a particularly dense interstellar medium (ISM) preventing the jet 
expansion \citep{vanbreugel1984}.
Because the required ISM confining densities are not confirmed by recent studies 
\citep[e.g.,][]{morganti2008}, the confinement scenario is considered 
less likely, although it might still apply to selected objects 
\citep[e.g.,][]{garciaburillo2007}. 
Much observational evidence has instead accumulated in favour of the youth 
scenario, the most compelling measurement being the detection, in several GPS/CSO galaxies,
of hotspot advance velocities about 0.1--0.2$c$ 
\citep{owsianikconway1998,owsianik1998,tschager2000,taylor2000,gugliucci2005}, 
which indicate kinematical source ages not higher than a few $10^3$ years, 
in good agreement with spectral ageing estimates \citep{murgia1999}.
Further evidence of youth comes from the underluminosity of 
GPS/CSO optical narrow-emission lines, 
suggesting that the Str\"omgren sphere of the recently-triggered AGN  
is still in an expansion phase in these sources, and we are thus 
witnessing the birth of their Narrow-Line Region \citep[NLR;][]{vink2006}.

The youth scenario is part of a wider evolutionary scenario 
\citep{fanti1995,snellen2000}, according to which GPS/CSOs would first 
evolve into symmetric Compact-Steep-Spectrum (CSS) sources, equally powerful but with 
sizes of about 1--15 kpc, and then further expand outside the host 
galaxy to become large-scale, powerful 
radio galaxies.
The evolutionary link between GPS and CSS sources is supported by their  
membership to the anticorrelation between linear size and peak frequency 
\citep[$LS \propto \nu_{\rm p}^{-0.65}$;][]{odea1997}.
However, it is not clear yet whether all GPS/CSOs will eventually become large-scale
radio sources, and whether the evolved sources will display a 
Fanaroff-Riley type I (FRI) or type II (FRII) morphology,
because of the frequency's observational limits in the exploration of this relation for 
super-galactic sized sources, as well as because of the 
biases in the surveys from which the distribution of the radio sources
in the power-linear size ($P-LS$) diagram is drawn \citep[e.g.,][and references therein]{snellen2000}.

The first systematic studies of the host galaxies of GPS sources, 
conducted in the optical and near-infrared (NIR) bands 
\citep{snellen1996,odea1996,devries1998a,devries1998b,devries2000} 
showed that they are, as CSS sources, characterized by luminous masses, brightness profiles, 
and optical-NIR colors more typical of passively- or non-evolving giant elliptical galaxies, 
than either spirals, small ellipticals, brightest cluster galaxies (BCGs), 
or central-dominant (cD) galaxies at similar redshifts. 
This seems to be consistent with what was found for a sample of FRIIs 
rather than with the properties of FRIs, 
which are preferentially 
hosted by cD galaxies \citep{owen1989,owen1991,zirbel1996},
and suggest that GPS sources are more likely to evolve, through the CSS phase, in 
FRIIs rather than in FRIs \citep{devries2003}.
However, the  morphologies of the majority \citep[$\sim$60\%;][]{odea1996} of the hosts 
of GPS sources show signs of recent mergers and/or interactions.
This evidence,
apparently at odds with the passively-evolving scenario,
was first interpreted as an indication that GPS sources might be 
associated with the first of a sequence of mergers \citep{devries2003}, 
possibly leading to the formation of a BCG or a cD only over time 
scales much longer than the lifetime of the radio source.
In subsequent investigations, however, the presence of young stellar 
population in these giant ellipticals
was revealed by means of 
stellar-population synthesis models applied to photometric 
\citep{devries2007} and spectrophotometric data \citep{holt2009}.
Furthermore, \citet{devries2007} showed the similarity, 
out to $z \simeq 0.55$, between the luminosities of the GPS-source hosts and 
of the Luminous Red Galaxies (LRGs), which are thought to represent the 
most massive early-type galaxies, and are often associated with BCGs.
Therefore, the properties of the hosts do not provide unambiguous indications on 
the subsequent GPS/CSO evolutionary stages.

Despite these results,  
population studies highlight the existence of far too 
many compact sources compared to the number of powerful large-scale  
objects \citep{odea1997}.    
Beside the possibility, not easy to justify, that these sources undergo a short-lived 
activity \citep{readhead1996}, two scenarios prove to be particularly attractive
to explain this statistical evidence. 
In the first scenario, the jet activity is intermittent 
on time scales comparable to the age of the small sources \citep[e.g.,][]{reynolds1997}: 
this scenario, supported by the observations of double-double radio galaxies  
\citep{lara1999,schoenmakers2000,kaiser2000},
and by the detection of candidates for dying compact sources \citep{giroletti2005,parma2007},
finds a natural explanation in the framework of 
accretion-disk instabilities \citep{czerny2009}.
An accretion disk operating above a given threshold accretion rate 
($\dot m_* \simeq 0.025\, \dot m_{Edd}$) 
is expected to experience instabilities driven by the radiation pressure, 
causing the source undergo alternate phases of high and low activity;
for moderate accretion rates, the duration of the active phases is short enough 
($\sim 10^3-10^4$ years) to make the compact source unable to grow to super-galactic
sizes, whereas accretion rates close to the Eddington limit are required for the 
development of large-scale sources \citep{czerny2009}.
In the second scenario, many sources undergo processes of jet-flow turbulent disruption  
before their lobes can grow to large sizes, either disappearing from the radio sky 
\citep{alexander2000} or developing an FRI-type morphology \citep{kaiser2007}, 
in both cases experiencing a drop in luminosity.
The recent finding that FRIs smaller than $\sim$40 kpc were 
almost absent in a sample of radio galaxies optically identified in the SDSS
\citep{best2009} seems to suggest that all radio sources begin their 
life with collimated jets, but less powerful jets are more easily disrupted, 
as the source grows, in denser environments, leading to the production of FRI sources. 
Further evidence supporting this view might derive from the confirmation that 
the large fraction of low-power, compact objects with non-FRI morphology 
discovered within a sample of FRI candidates at $1<z<2$ 
\citep{chiaberge2009}, are indeed young sources.
The two above scenarios are not necessarily in conflict with each other:
for sufficiently high accretion rates,
accretion-disk instabilities might partake  
to the radio-source evolution, otherwise driven 
by the interplay between jet power and environmental conditions.
However, this picture needs to be further explored.

It has recently become clear that instrumental clues
on the evolution of young GPS sources may come from the X-ray electromagnetic window.
Early detections of GPS sources by {\it ASCA} \citep{odea2000} triggered extensive 
searches for X-ray emission from these sources; thanks to the capabilities 
of {\it XMM-Newton} and {\it Chandra}, several GPS/CSOs are now 
known to be strong X-ray emitters 
\citep{guainazzi2004,vink2006,guainazzi2006,siemiginowska2008,tengstrand2009}.
However, the origin of this X-ray emission is not known yet.

The best spatial resolution currently available in the X-ray band 
($\sim$1$''$ with {\it Chandra}) is not sufficient to resolve the X-ray morphology 
of most GPS/CSOs: thus, the identification of the origin of the X-ray emission in 
these objects relies entirely on spectral studies.
The components of GPS/CSOs first proposed to be the source of 
the observed X-rays are the ISM of the host galaxy shocked by the 
expanding radio lobes 
\citep{heinz1998,odea2000}, and the accretion disk's hot corona 
\citep{guainazzi2004,vink2006,guainazzi2006,siemiginowska2008}: 
both components are expected to generate {\it thermal} X-ray radiation, 
but they are difficult to disentangle and do not rule out alternative scenarios
\citep[][and references therein]{siemiginowska2009}.
Furthermore, the location of GPS/CSOs on the radio/X-ray luminosity plane
does not fully elucidate the relationship between ``young'' and ``mature'' radio 
galaxies: GPS/CSOs are as X-ray luminous as FRIIs, despite their  
larger radio luminosity, but they seem to lie on the high radio-power tail  
of the radio-core/X-ray correlation discovered for 
FRIs \citep[][and references therein]{tengstrand2009}. 

A way of investigating the origin of the X-rays in GPS/CSOs,
and address, at the same time, some of the open issues 
on the physics and fate of young radio sources,
is to simultaneously model the dynamical and radiative evolution 
of GPS/CSO galaxies, and constrain the models through the wealth of 
multiwavelength data currently available.
We recently proposed a dynamical-radiative model that, for the first time, 
describes the evolution of the broad-band emission of a young GPS 
with a given jet kinetic power, as it expands through the ISM of the 
host galaxy \citep{stawarz2008}.
The model accounts for the radiative contribution of the various AGN components,
as well as for environmental absorption effects, and predicts significant and 
complex {\it non-thermal} X-ray to $\gamma$-ray emission. 

In the present paper, we apply the model to a sample of GPS/CSO galaxies,
with the manifold aim of reproducing their complex radio to X-ray spectra,
and constraining relevant source parameters as the jet kinetic 
power, the accretion rate, and the absorption mechanisms.
New observational evidence supporting the model is also discussed.
The paper is organized as follows:
in Section \ref{sec_model}, we briefly review the main features of the dynamical-radiative model
we presented in \citet{stawarz2008};
in Section \ref{sec_sample}, we describe the source sample that we chose for the application of 
the model, and the relevant broad-band data; 
in Section \ref{sec_SEDmodeling}, we show the results of the SED modeling;
in Section \ref{sec_support}, we present further evidence supporting the proposed scenario;
we discuss our results in Section \ref{sec_discussion}, and in Section \ref{sec_conclusions} 
we draw our conclusions.

A model of $\Lambda$-dominated universe \citep[$\Omega_\mathrm{{\Lambda}}=0.7$, 
$\Omega_{\mathrm{M}} = 0.3$, $\Omega_{\mathrm{k}} = 0$;][]{spergel2003}, with 
$H_{0} = 72$ km~\persec~Mpc$^{-1}$ \citep{freedman2001} will be adopted throughout 
this paper.
Source-intrinsic quantities from the literature given below were corrected for 
this cosmology, unless otherwise stated.

\section{THE MODEL}
\label{sec_model}

We recall below the most relevant features of our non self-similar 
dynamical-radiative model, presented in \citet{stawarz2008}.
The reader is referred to the original paper (and references therein)
for a more comprehensive discussion.
 
\subsection{Dynamical Evolution}
We based our description of the dynamical evolution of a young radio source in its 
``GPS phase'' on the model proposed by \citet{begelman1989} to 
explain the expansion of classical double sources in an ambient medium 
with density profile $\rho(r)$. 
The relevant equations were derived by assuming that:  
(i)   the jet momentum flux, which is 
      proportional to the jet kinetic power $L_{\mathrm{j}}$, 
      is balanced by the ram-pressure of the ambient medium spread 
      over an area $A_{\mathrm{h}}$; 
(ii)  the lobe sideways expansion velocity, $v_c$, 
      equals the speed of the shock driven by the overpressured cocoon, 
      with internal pressure $p$, in the surrounding medium; 
(iii) the kinetic energy 
      transported by the jet pair during the entire 
      source lifetime $t$ is converted into the cocoon's energy $pV$  
      (where $V$ is the volume of the cocoon) at the heads of the jets.

For a young GPS source expanding in the central gaseous core of a giant 
elliptical galaxy, and characterized by age $t$, linear 
size $LS(t) \ltsim 1$ kpc, and transverse size $l_c(t)$, 
we could constrain the model with a number of reasonable approximations: 
(i)   a constant ambient density $\rho=m_{\mathrm{p}} n_0$
      (with $m_p$ the proton mass, and $n_0 \simeq 0.1$ \percmc),
      representative of the inner King density profile of the host galaxy; 
(ii)  a constant hot-spot advance ve\-lo\-ci\-ty $v_{\mathrm{h}}$,
      as suggested by many obervations of CSOs
      (see Section \ref{sec_introduction});
(iii) a scaling law $l_{\mathrm{c}}(t)\sim t^{1/2}$, reproducing the initial, 
      ballistic phase of the jet propagation, according to 
      \citet{kawakatu2006} and \citet{scheck2002}
      \citep[see, however,][for an alternative scenario]{kawakatu2009a}.
Under the aforementioned conditions, all the lobe physical quantities become 
functions of two parameters only: the jet kinetic power $L_{\mathrm{j}}$, 
and the source linear size $LS$.

\subsection{Spectral Evolution}
In the framework of the dynamical model described above, 
we studied the evolution of the broad-band radiative output of GPS sources 
with the source expansion, for a given jet kinetic 
power $L_{\mathrm{j}}$.
The magnetic field in the expanding lobes scales as 
$B=(8\pi\eta_{\mathrm{B}}p)^{1/2} \sim L_{\mathrm{j}}^{1/4}LS^{-1/2}$,
with $\eta_{\mathrm{B}}=U_{\mathrm{B}}/p<3$, and $U_{\mathrm{B}}$ being the magnetic 
energy density. 
The electron population $Q(\gamma)$ (with $\gamma$ the electron 
Lorentz factor), injected from the terminal jet shock to the expanding 
lobes, evolves under the joint action of adiabatic and radiative 
energy losses, yielding a lobe electron population  
$N_{\mathrm{e}}(\gamma)$, characterized by a broken power-law form with 
critical energy $\gamma_{\mathrm{cr}}$ when $Q(\gamma)$ is a power law, 
and by a more complex form when $Q(\gamma)$ is a broken power law with 
intrinsic break $\gamma_{\mathrm{int}} \simeq 2 \times 10^3$ 
\citep[$\gamma_{\mathrm{int}}=m_{\mathrm{p}}/m_{\mathrm{e}}$, 
with $m_{\mathrm{p}}$ and $m_{\mathrm{e}}$ the proton and electron masses, 
respectively;][]{stawarz2007}.  
Assuming that the lobe electrons, in rough equipartition 
with the magnetic field and the protons 
\citep[e.g.,][and references therein]{orienti2008b},
provide the bulk of the lobe pressure, the electron energy density 
is $U_{\mathrm{e}}=\eta_{\mathrm{e}} p$, with $\eta_{\mathrm{e}} \ltsim 3$.

The lobe electrons emit synchrotron radiation, with luminosity 
$L_{\mathrm{syn}}$ roughly constant with time, and energy density 
$U_{\mathrm{syn}} \sim LS^{-3/2}$. 
This radiation suffers then from absorption processes, responsible for the 
characteristic spectral change across the turnover at GHz frequencies.
Because the coupling between the synchrotron-self-absorption (SSA) process 
and our model equations returns turnover frequencies $\nu_{\rm p}$ systematically 
lower than those typically observed, and reproduces neither the observed 
slopes of the optically-thick spectra, nor the $\nu_p-LS$ anticorrelation, 
the absorption process that we favoured and implemented is the free-free 
absorption (FFA).

Recent data on compact sources of different brightness appear to be in good agreement
with the predictions of the SSA model for homogeneous sources \citep{devries2009}.
However, this result does not exclude the viability of the FFA models \citep{bicknell1997,begelman1999}. 
In fact, \citet{devries2009} showed that the Bicknell et al.'s FFA model
works well on bright sources, although it does not on faint sources, unless  
the ratio between jet power and radio luminosity depends on luminosity. 
Although this scenario might be more complicated than homogeneous SSA, this solution 
was never investigated either in Bicknell et al.'s or in Begelman's FFA models in sufficient detail.
In any case, this issue is not of major relevance for our sample, because we focus on 
bright sources, for which the FFA model works properly.

In particular, we adopted the FFA scenario by \citet{begelman1999}, 
which was shown to return reliable turnover frequencies and be a promising 
candidate to account for the $\nu_p-LS$ anticorrelation, while relaxing considerably 
the constraints on the ambient conditions required by Bicknell et al.'s model.
Yet following \citet{begelman1999}, we ascribed FFA to the external 
layers of interstellar gas clouds that have been photoionized by the  
UV radiation from the active nucleus, and are actually ionization-bounded.
These gas clouds might be associated with the NLR clouds. 

The particles of the lobes also produce inverse-Compton (IC) radiation via 
up-scattering of both the synchrotron radiation (synchrotron-self-Compton 
mechanism; SSC) and the local, thermal photon fields. 
The energy density $U_{\mathrm{rad}}$ of the thermal fields was evaluated 
by taking into account the contributions by a putative accretion disk, 
producing the bulk of its luminosity at UV frequencies, 
by a dusty torus, re-radiating the disk's UV photons in the IR domain,
and by the stars of the host galaxy, mostly contributing the NIR-optical 
photons. We obtained  
$U_{\mathrm{UV}}\sim LS^{-2}$ and $U_{\mathrm{IR}}\sim LS^{-2}$, 
whereas $U_{\mathrm{opt}}$ is independent of $LS$ for the considered $LS \lesssim 1$ kpc.

The IC scattering of all the aforementioned radiation fields yields significant and 
complex high-energy emission, from X-ray to $\gamma$-ray energies. 
In GPS {\it quasars} the putative direct X-ray emission of the accretion 
disk's hot corona and of the beamed relativistic jets may overcome the X-ray 
output of the lobes. In GPS {\it galaxies} these two contributions are instead 
expected, respectively, to be obscured by the torus and Doppler-hidden; the 
lobes are expected to be the dominant X-ray source.

\section{THE SOURCE SAMPLE}
\label{sec_sample}

We chose to apply our model to the sample of eleven GPS/CSO galaxies
known as X-ray emitters up to 2008.\footnote{During the preparation of this
manuscript, a paper by \citet{tengstrand2009} appeared, reporting the detection 
of additional X-ray emitting GPS galaxies; the modeling of the broad-band spectra
of these new sources is beyond the scope of the present work.}
The source list is given in Table \ref{tab_data}.

The sources are all members of the catalogue of radio sources with flux density 
greater than 1 Jy at 5 GHz by \citet{kuehr1981}.
The GPS nature of their radio spectra was observed at several 
epochs \citep[][and references therein]{odea1998}. 
Although the {\it bona-fide} GPS nature of three of our sample's members 
(IERS B0108+388, PKS B0941-080, and IERS B1345+125) was recently questioned  
on the basis of either flux-density variability or values of the radio spectral indices 
\citep{torniainen2007}, we decided to keep these sources in our sample 
because the presence of the spectral peak at GHz frequencies was always confirmed.
A CSO morphology characterizes the radio structures of all the sources, 
although one of them (PKS B1345+125) is an unusual example of CSO, 
with one of the two jets not being detected at frequencies above 
$\sim$1 GHz.
Our sample's members have radio power $P_{\rm 5 GHz}=10^{25.5}-10^{27.9}$ \wattperhz, and X-ray 
luminosities  $L_{\rm 2-10\, \mathrm{keV}} \simeq 6\times 10^{41} - 5\times 10^{44} $ erg \persec;
they are hosted by galaxies located at redshifts $z=0.08-0.99$, i.e. at luminosity distances 
between 0.4 and 6.6 Gpc.
The linear (projected) sizes of the radio sources are in the range $\sim$$10-400$ pc, 
thus providing ``snapshots'' of different stages of the source expansion.
More details on classification and properties 
of individual GPS/CSOs and their host galaxies are given in Appendix \ref{app_A}. 
The source physical quantities  that are more relevant 
to our SED modeling are summarized in Tables \ref{tab_data} and \ref{tab_spitzer}. 

Fig.\ \ref{fig_seds} shows the SEDs of the GPS/CSO galaxies of our sample.
With the exception of some of the mid/far-infrared (MFIR) data,
which will be discussed below,
the SED data were derived from the literature 
(see Appendix \ref{app_A} for the complete reference list). 
In particular, the near-infrared (NIR) and optical magnitudes 
were converted into fluxes by means of the absolute calibrations of 
\citet{bessel1979} for standard UBVK and Cousins' RI filters, 
\citet{wamsteker1981} for standard JHL bands, and \citet{allen1973} 
for standard RI filters; 
the dereddening was performed by means of the extinction laws
given by \citet{rieke1985} and \citet{cardelli1989}, assuming as B-band 
Galactic extinctions ($A_{\rm B}$) the values by \citet{schlegel1998} 
provided by the NASA/IPAC Extragalactic Database (NED).
Source-intrinsic reddening effects were not considered in these bands,
given the dominance of the host-galaxy contribution.
The X-ray unabsorbed fluxes were derived from the absorbed fluxes by means of 
XSPEC v.\ $11.3.1$, taking both the Galactic and the source-intrinsic hydrogen 
column densities ($N_{\mathrm{H,Gal}}$ and $N_{\mathrm{H,intr}}$) into account.
The SED luminosities at the source rest-frame frequencies were then computed by 
assuming modern cosmology (see above).
No $K$-correction was applied to the data.

In our sample, literature data in the MFIR band 
were available for two sources only: IERS B1345+125 and IERS B1404+286.
Given the importance of the MFIR portion of the SED for our modeling (see 
Section \ref{sec_SEDmodeling}), we thus searched for and analysed archival 
{\it Spitzer}\footnote{http://www.spitzer.caltech.edu/} 
IRAC\footnote{Infrared Array Camera} \citep{fazio2004} and MIPS\footnote{Multi-Band 
Imaging Photometer for {\it Spitzer}} 
\citep{rie04} observations of our sample's members. 

IRAC data for IERS B1345+125 \citep[prog.\ ID 32, January 2004;][]{faziowang2004} and IERS B1404+286 
\citep[prog.\ ID 30443, July 2006;][]{rieke2006} were available in all of the four bands 
(3.6, 4.5, 5.8, and 8.0 $\mu$m). 
MIPS observations at 24 and 70 $\mu$m were performed for IERS B1345+12 
\citep[prog.\ ID 30877, July 2007;][]{evans2006} and IERS B1404+286 
\citep[prog.\ ID 30443, July 2006;][]{rieke2006} an additional 24 $\mu$m zodiacal light calibration 
image with IERS B0500+019 observed in the field (prog.\ ID 1882, April 2007) was also available.
The analysis of the aforementioned IRAC and MIPS data is described in Appendix 
\ref{app_B}; the flux densities are reported in Table \ref{tab_spitzer}.

\section{MODELING THE BROAD-BAND SPECTRA}
\label{sec_SEDmodeling}
Our dynamical-radiative model, summarized in Section \ref{sec_model}, is a powerful tool 
to study the evolution of a typical  GPS-source synchrotron SED as a function of the time-dependent 
source linear size $LS(t)$, given the kinetic power of the jets, $L_{\mathrm{j}}$, and the energy spectrum of 
the hot-spot electrons injected into the lobes; 
furthermore, assuming typical luminosities for the putative torus' and accretion-disk's 
components, as well as for the host galaxies, the model enables one to investigate the 
temporal evolution of the high-energy, comptonized SED component \citep{stawarz2008}.

By applying the model to sources with {\it measured} (projected) linear sizes, we could here 
constrain their jet kinetic powers, as well as the spectra of their hot-spot particles; 
furthermore, we could test the viability of the FFA effect as the main responsible for the 
optically-thick part of the radio spectra; 
finally, the observational constraints on the luminosities of the host galaxies, 
and, for a few sources, of the torus emission, enabled us to evaluate the 
contribution of the comptonized radiation to the high-energy emission of the source. 

In Fig.\ \ref{fig_seds}, we show the modeling of the intrinsic broad-band SEDs of our sample's sources,
and Table \ref{tab_par} reports on the values of the best-fit parameters.
More details on the modeling of the various SED components are given below.

\begin{figure*}
\hbox{
\includegraphics[scale=0.7]{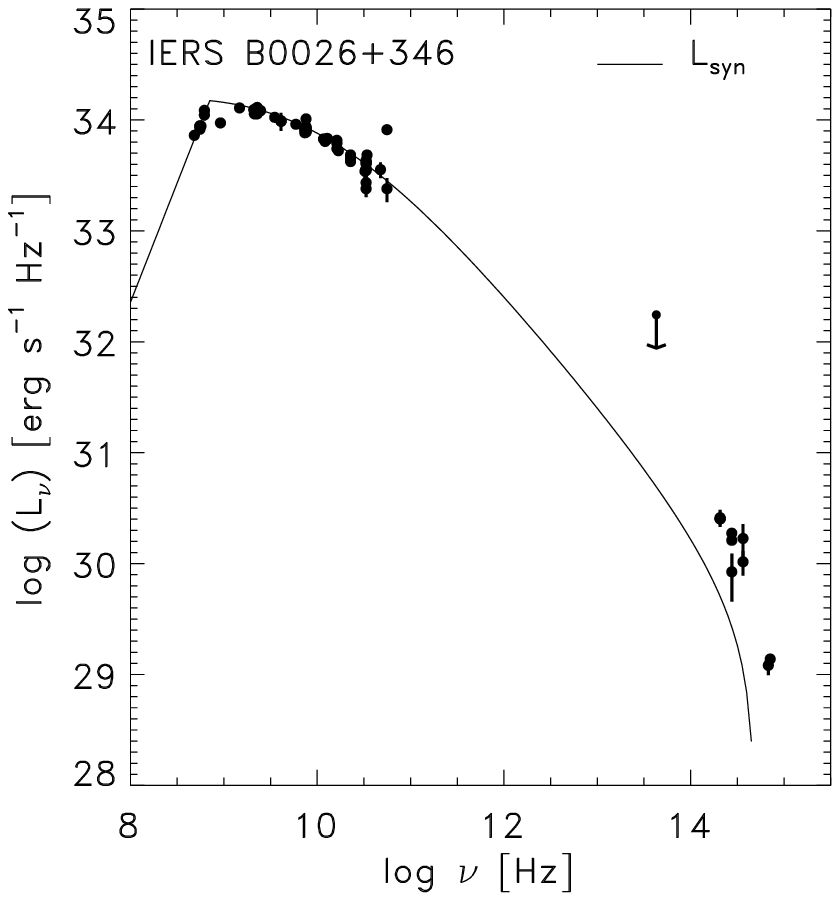}
\includegraphics[scale=0.7]{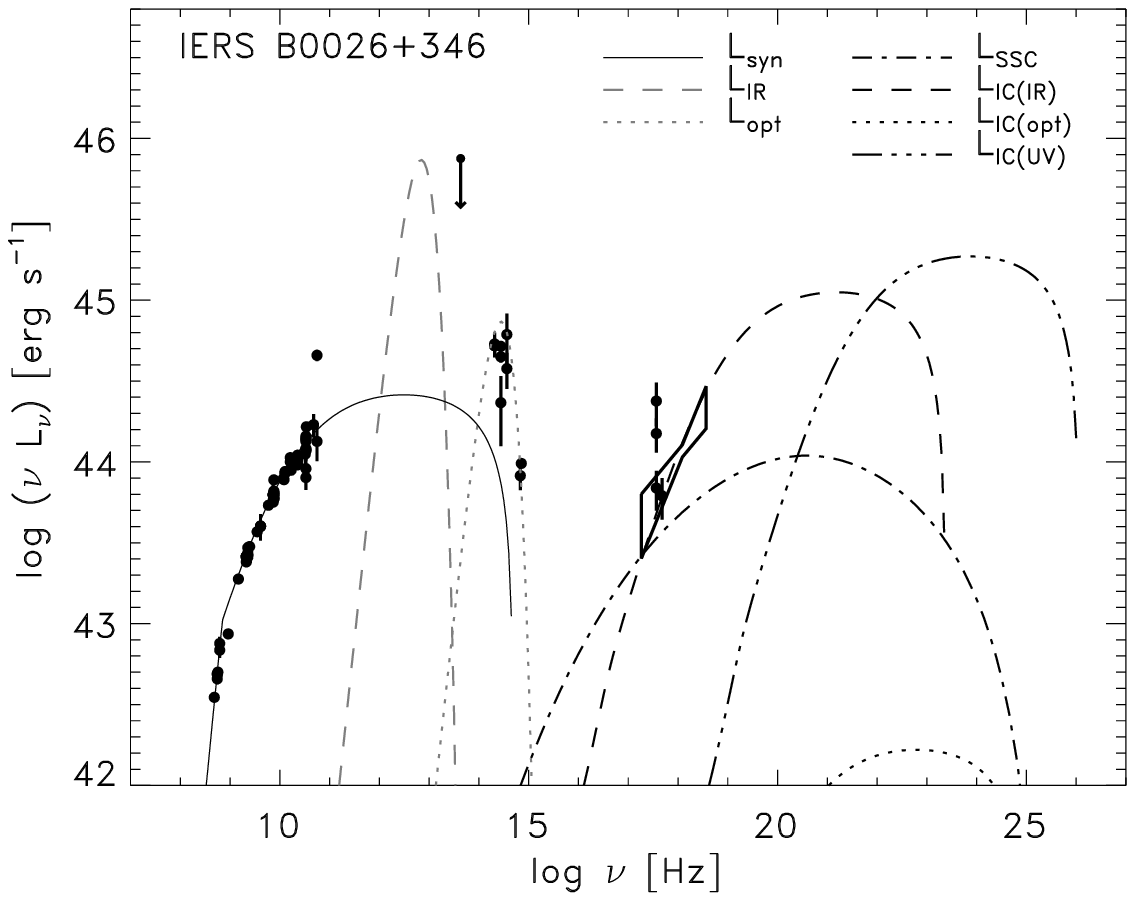}
}
\hbox{
\includegraphics[scale=0.7]{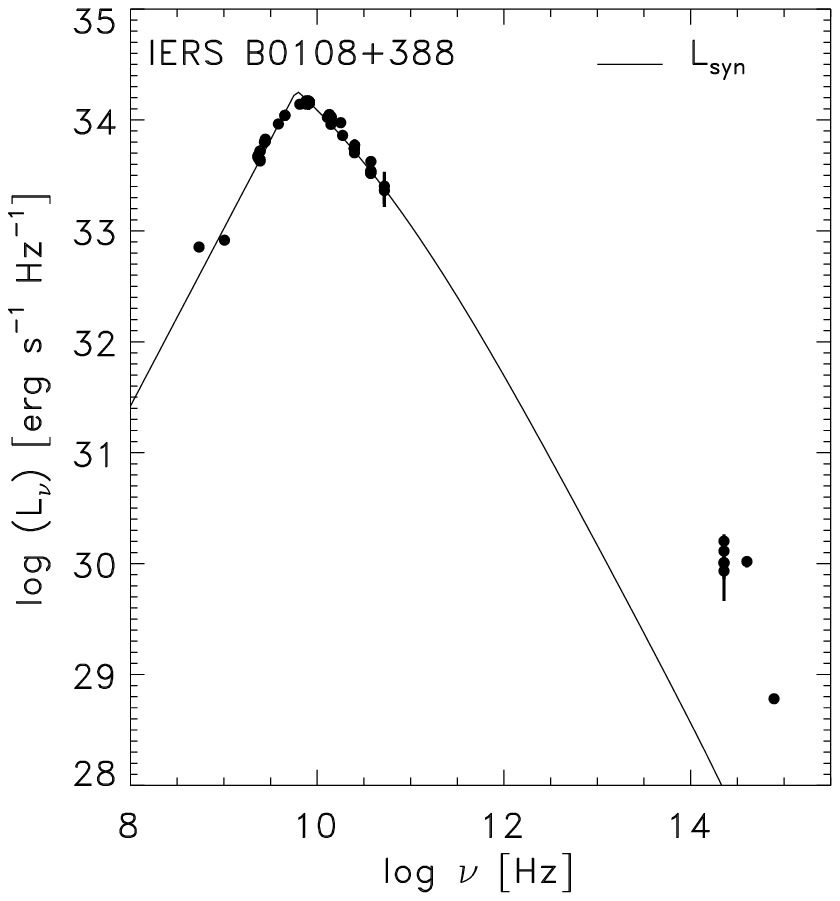}
\includegraphics[scale=0.7]{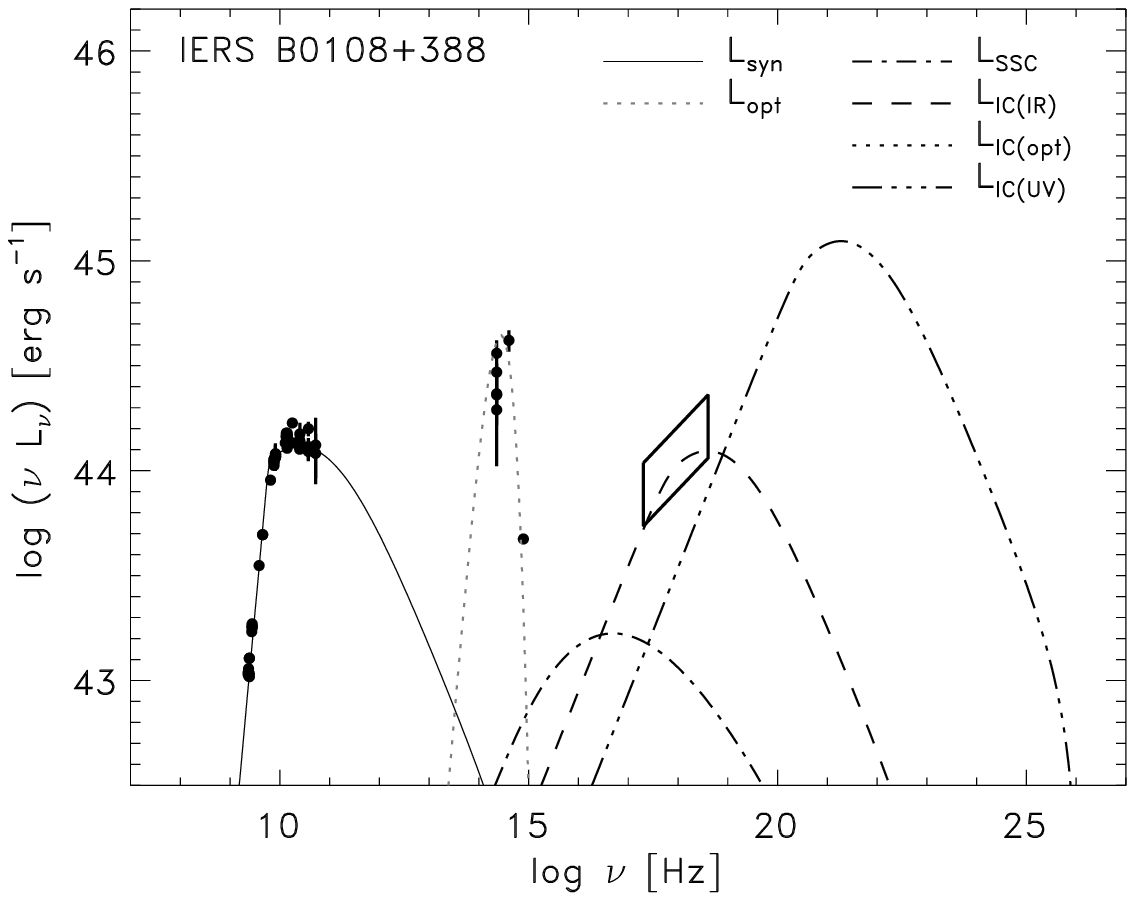}
}
\caption{Modeling of the source-frame radio spectra and SEDs of the GPS/CSO galaxies of our sample.
            Radio to X-ray data were derived from the literature (the references are 
            listed, source by source, in Appendix \ref{app_A}), excepted for the MFIR data 
            of IERS B0500+019, IERS B1345+125, and IERS B1404+286, 
            which were derived from the analysis of archival {\it Spitzer} data (Appendix \ref{app_B}).
            The model curves, obtained with the parameters listed in Table \ref{tab_par}, 
            are displayed with different-style lines. 
            {\it Left panels}: The solid-line curve shows the 
            optically-thin synchrotron emission at frequencies higher than the turnover frequency, 
            and the synchrotron free-free absorbed spectrum at frequencies below the turnover. 
            {\it Right panels}: 
            The black, solid lines represent the synchrotron emission;
            the black, dash-dotted line indicates the corresponding SSC emission; 
            the grey, dashed line shows the thermal MFIR emission from the torus; 
            the grey, dotted line shows the thermal star light; 
            the black, dashed and dash-dot-dotted lines 
            show the comptonized thermal emission from the torus and the disk, respectively;
            the black, dotted line indicates the comptonized star light. 
            Note that the shown X-ray spectra are unabsorbed, whereas the radio spectra are
            displayed as absorbed.
            When an IC spectral component does not appear in the plot, its luminosity is 
            below the scale minimum.}           
\label{fig_seds}
\end{figure*}

\addtocounter{figure}{-1}
\begin{figure*}
\hbox{
\includegraphics[scale=0.7]{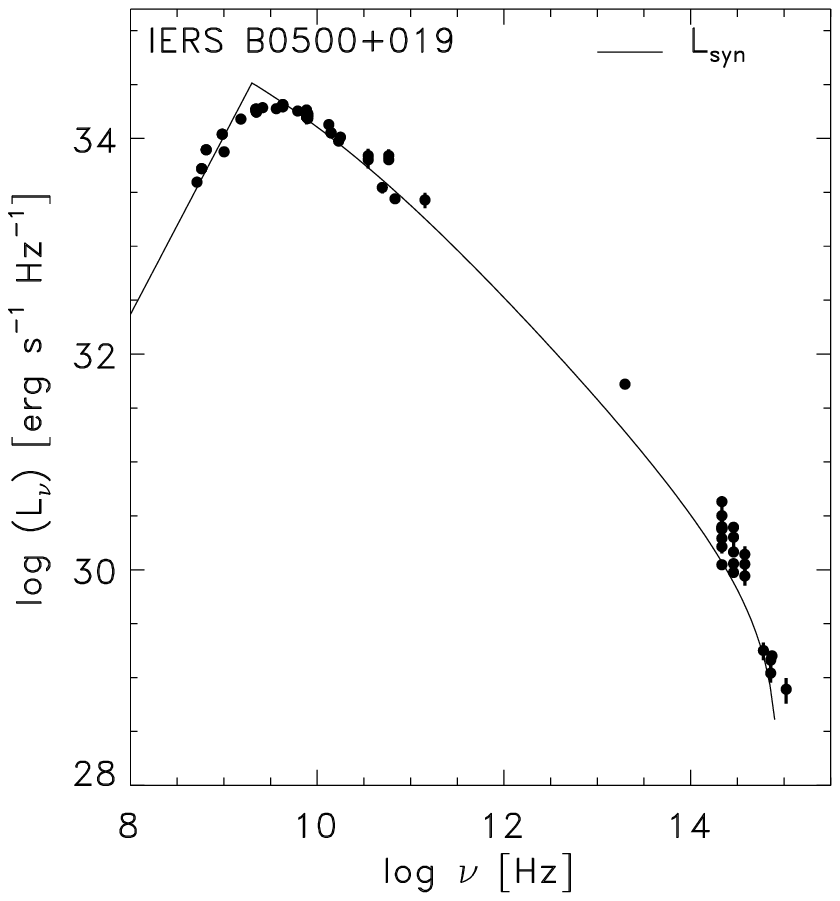}
\includegraphics[scale=0.7]{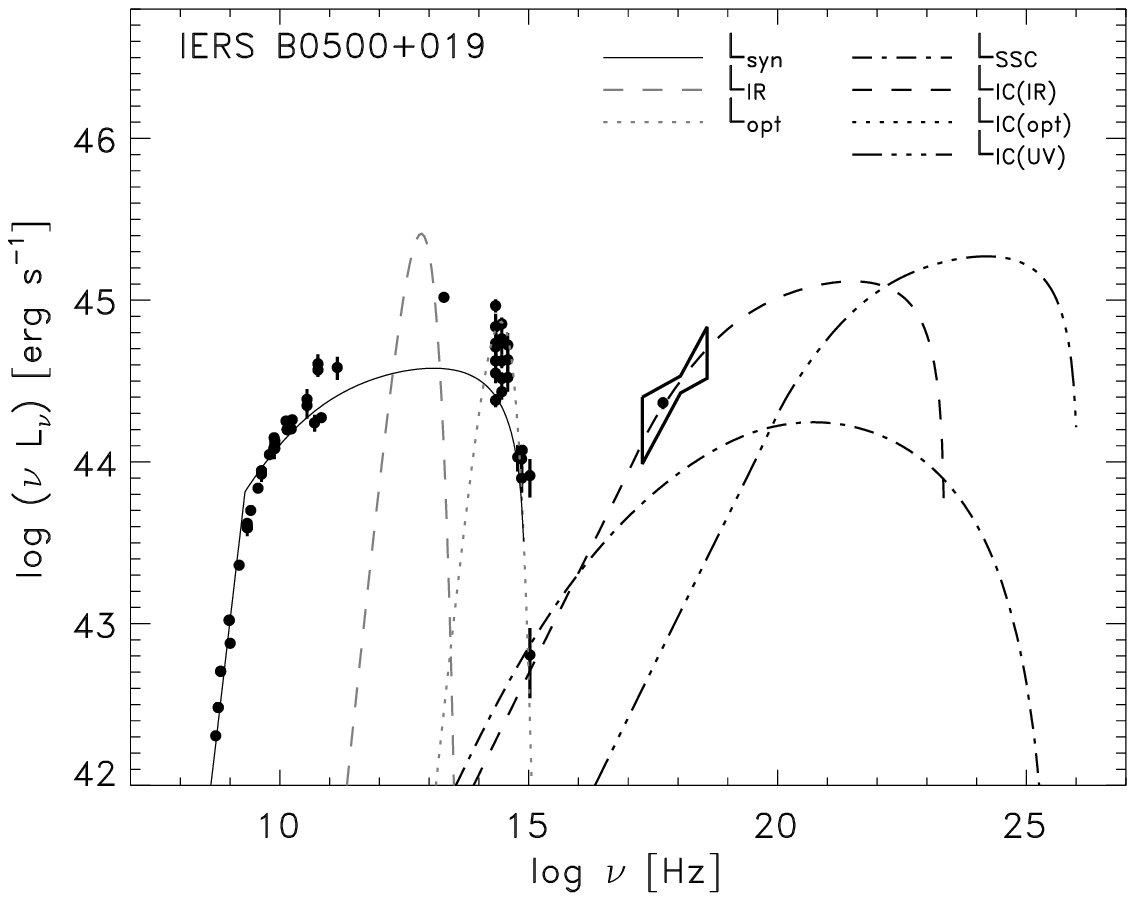}
}
\hbox{
\includegraphics[scale=0.7]{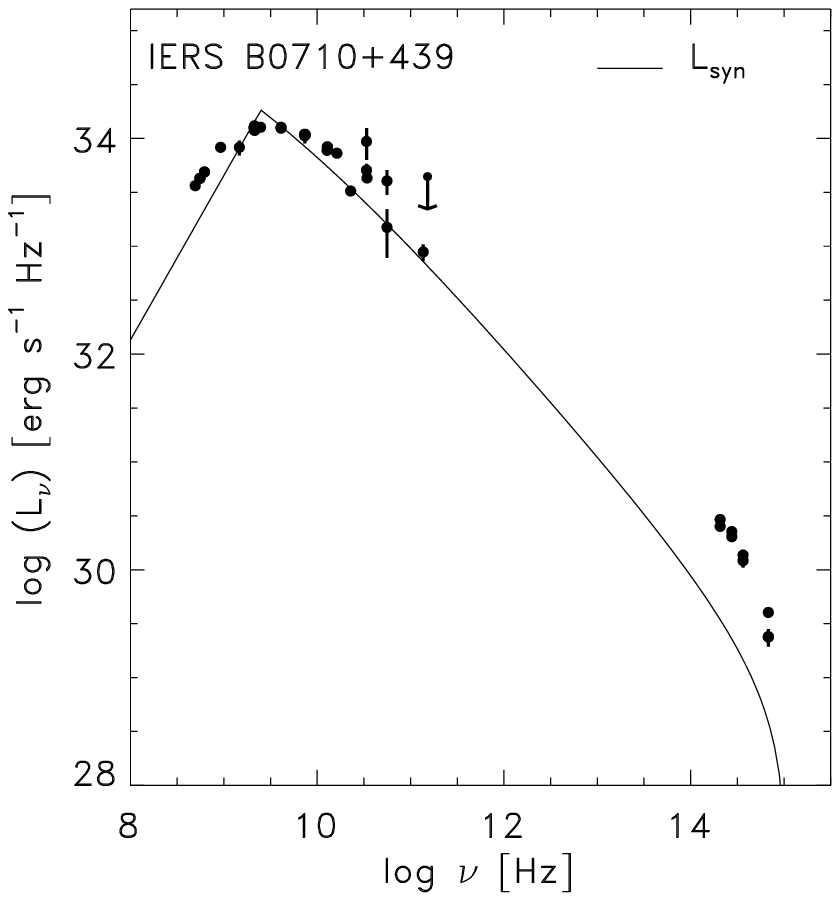}
\includegraphics[scale=0.7]{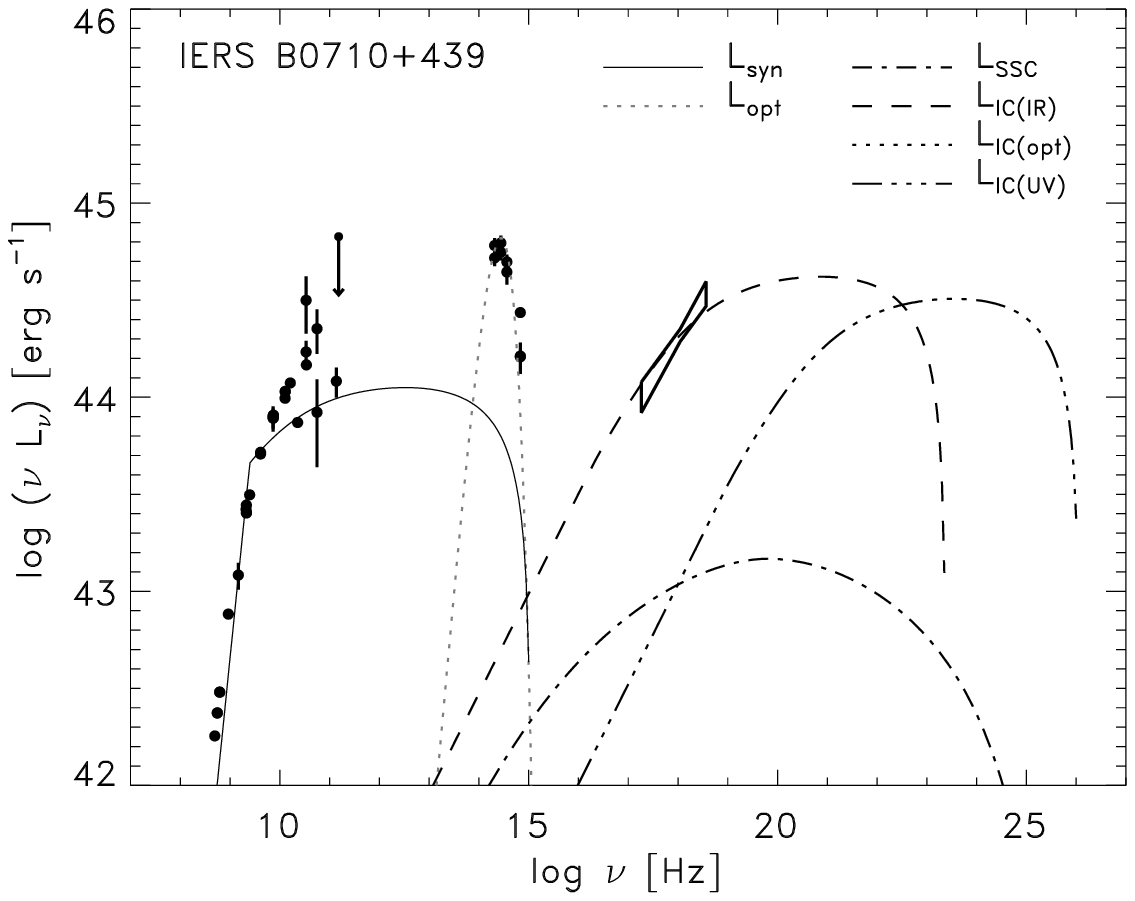}
}
\hbox{
\includegraphics[scale=0.7]{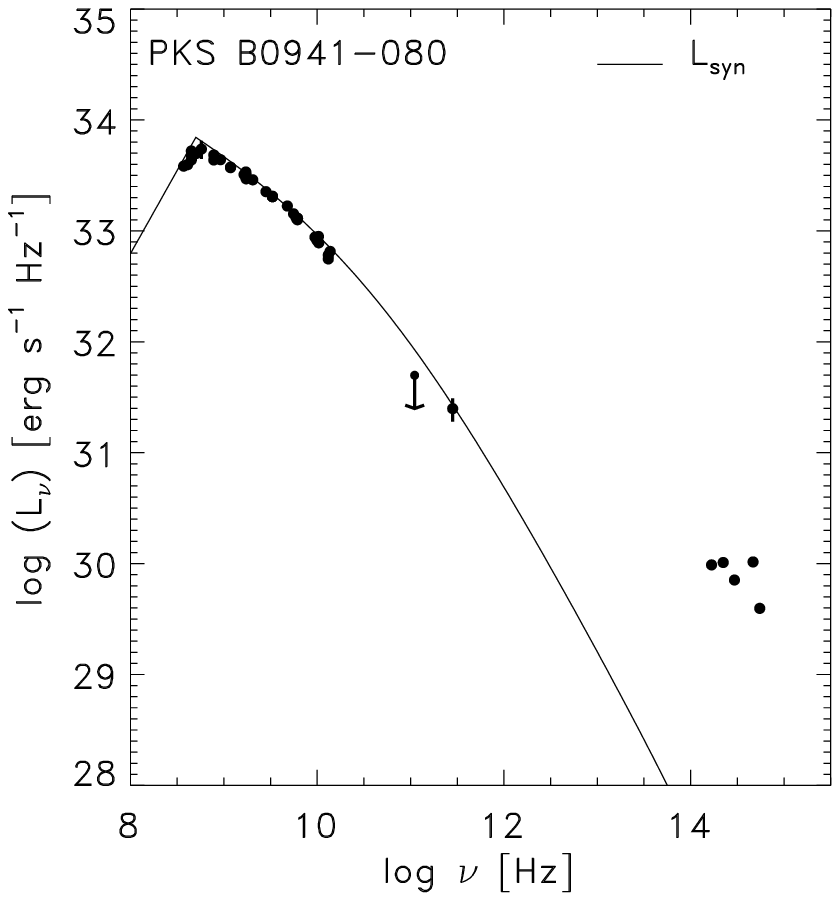}
\includegraphics[scale=0.7]{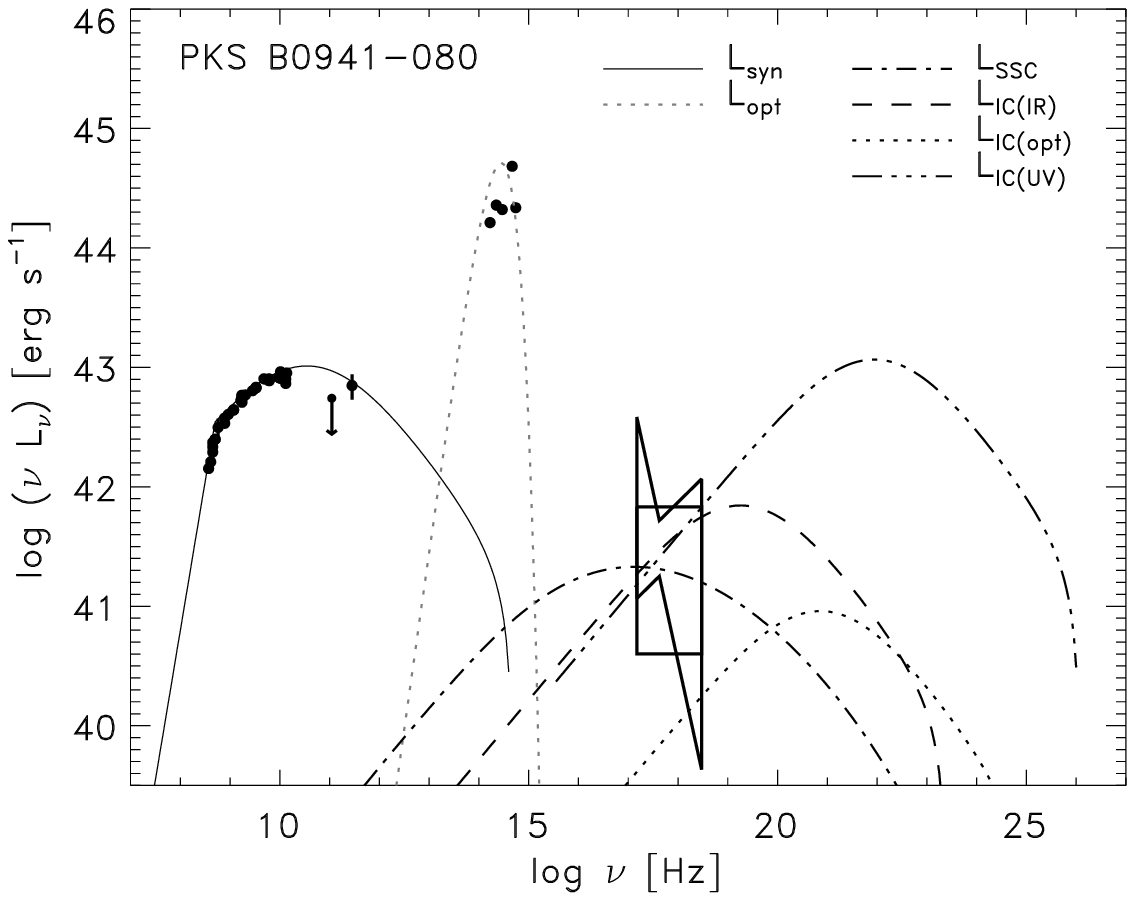}
}
\vspace{1.0cm}
\caption{{\it Continued}}
\end{figure*}

\addtocounter{figure}{-1}
\begin{figure*}
\hbox{
\includegraphics[scale=0.7]{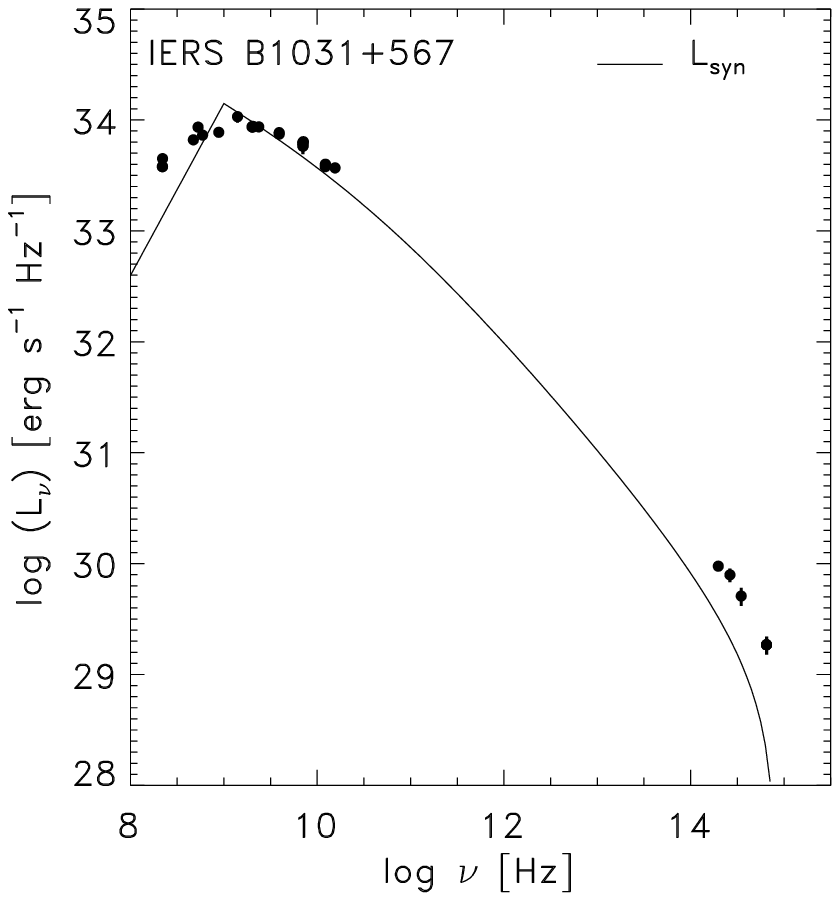}
\includegraphics[scale=0.7]{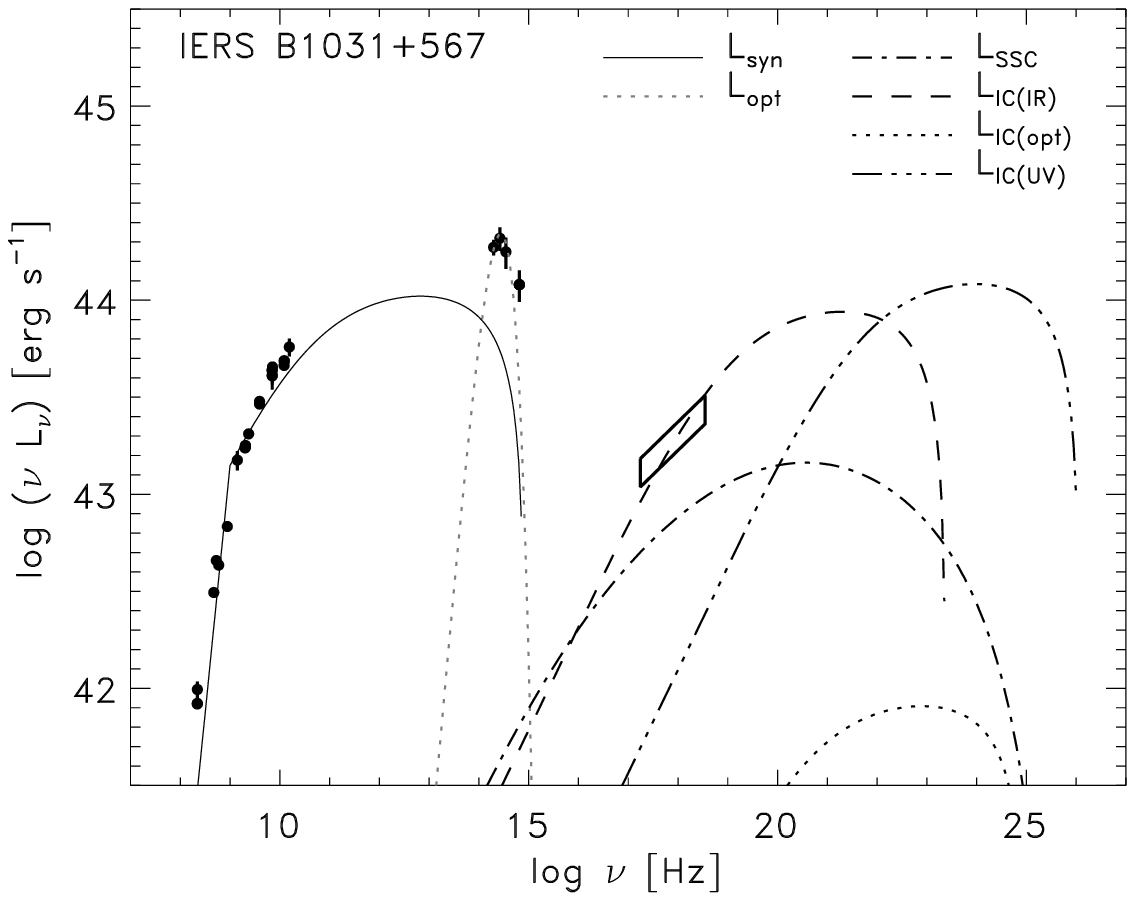}
}
\hbox{
\includegraphics[scale=0.7]{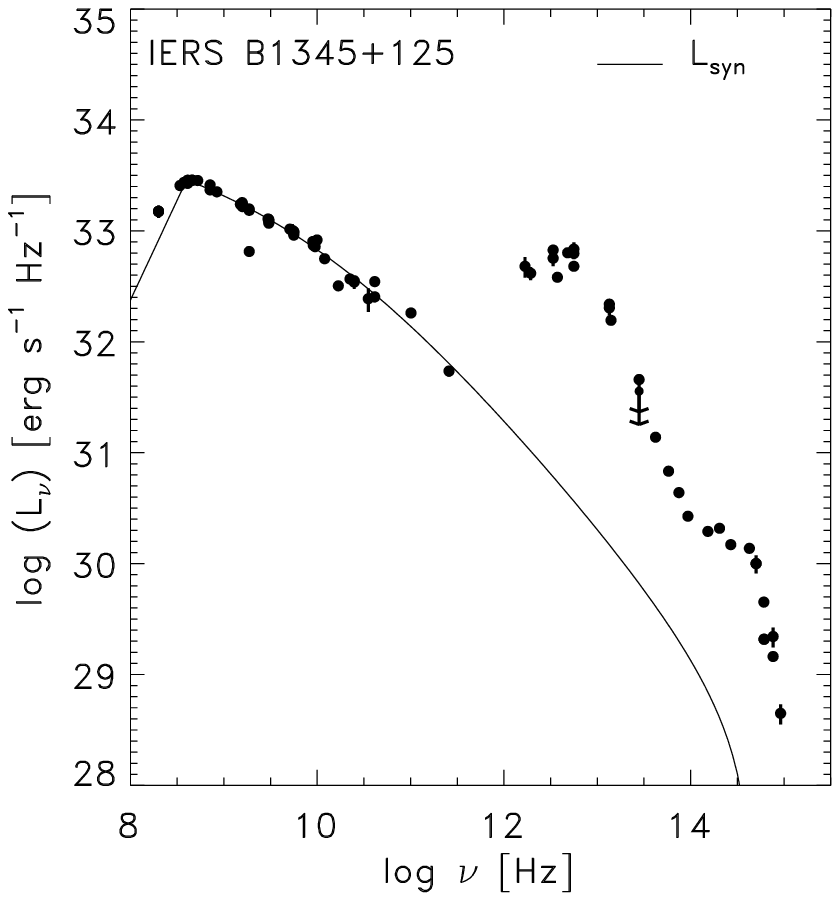}
\includegraphics[scale=0.7]{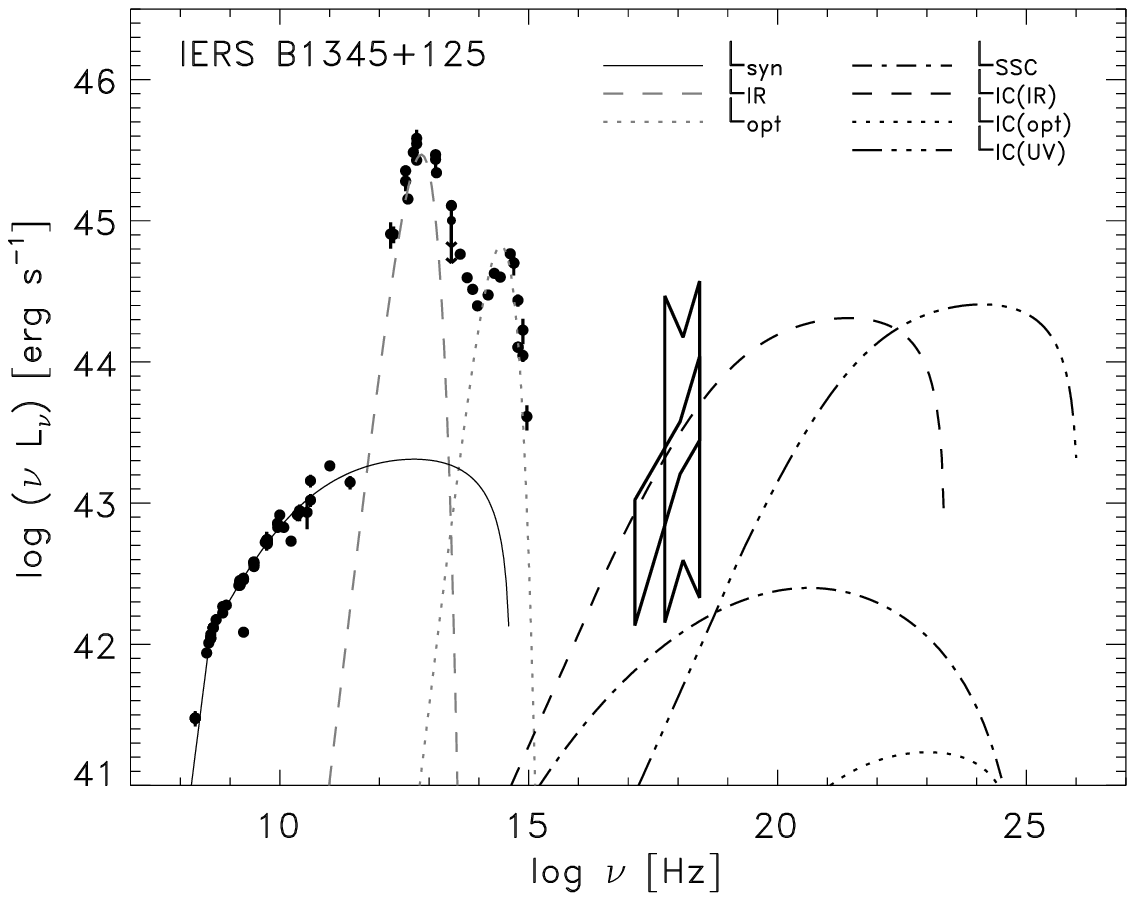}
}
\hbox{
\includegraphics[scale=0.7]{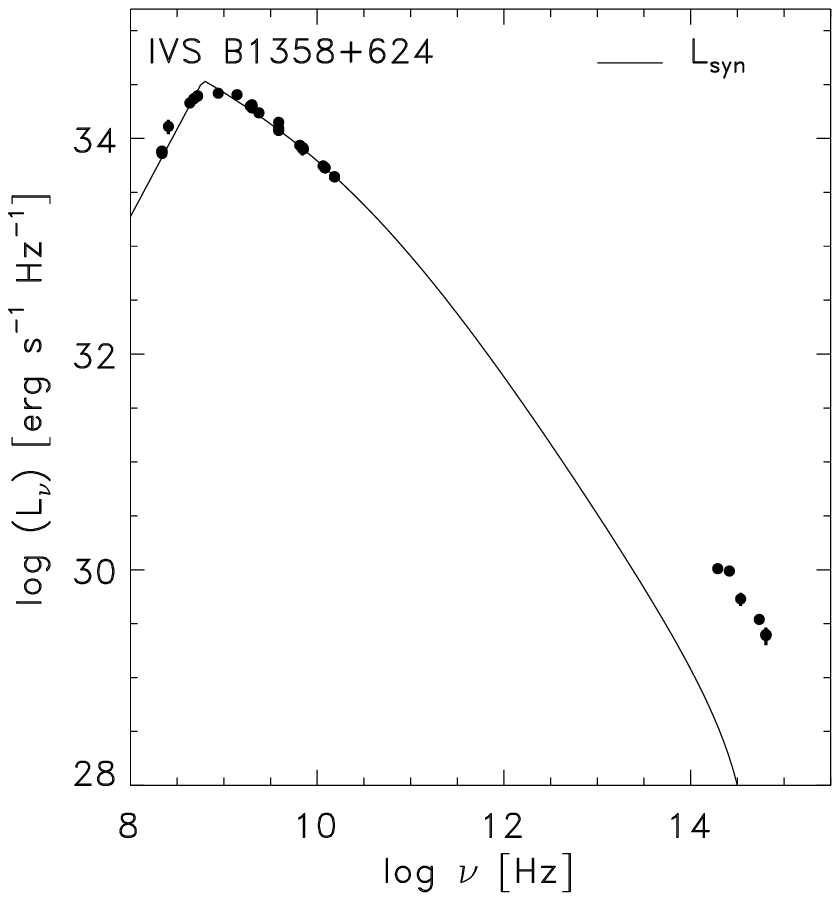}
\includegraphics[scale=0.7]{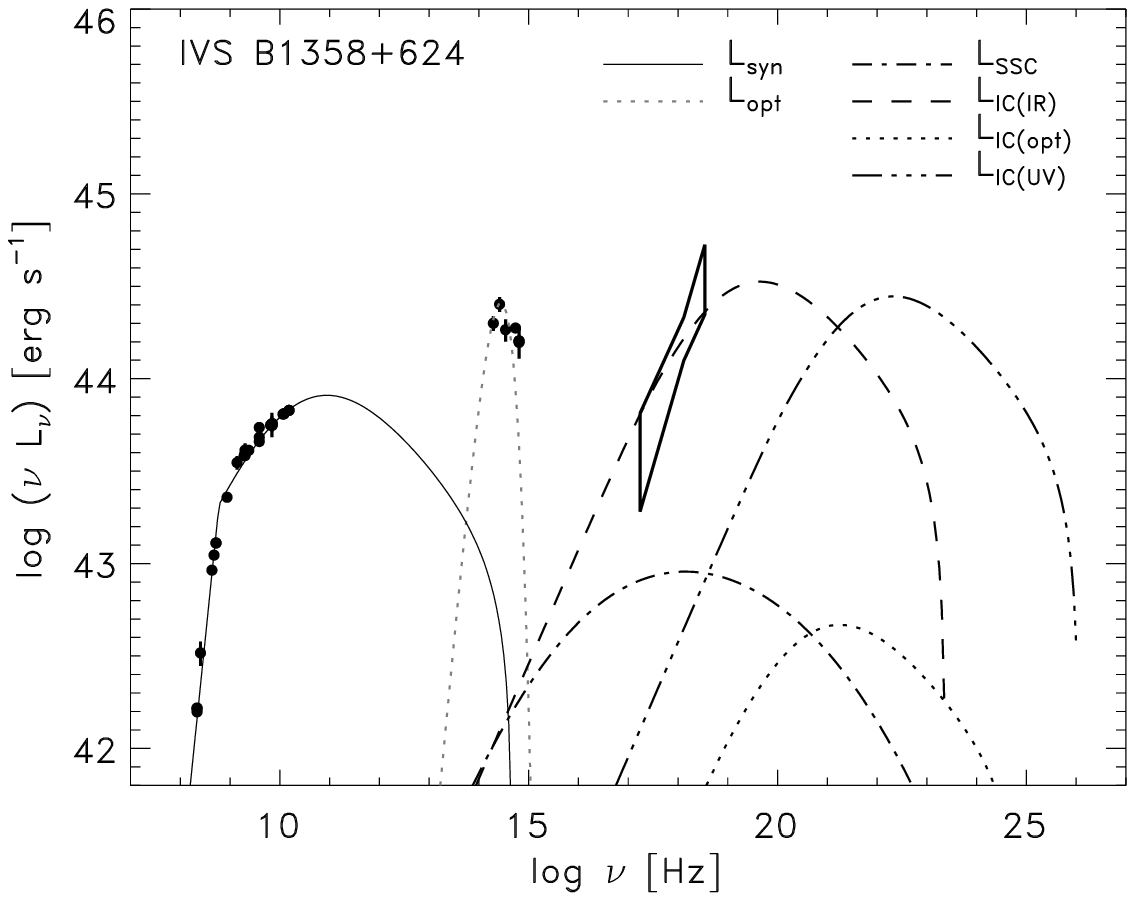}
}
\caption{{\it Continued}}
\end{figure*}

\addtocounter{figure}{-1}
\begin{figure*}
\hbox{
\includegraphics[scale=0.7]{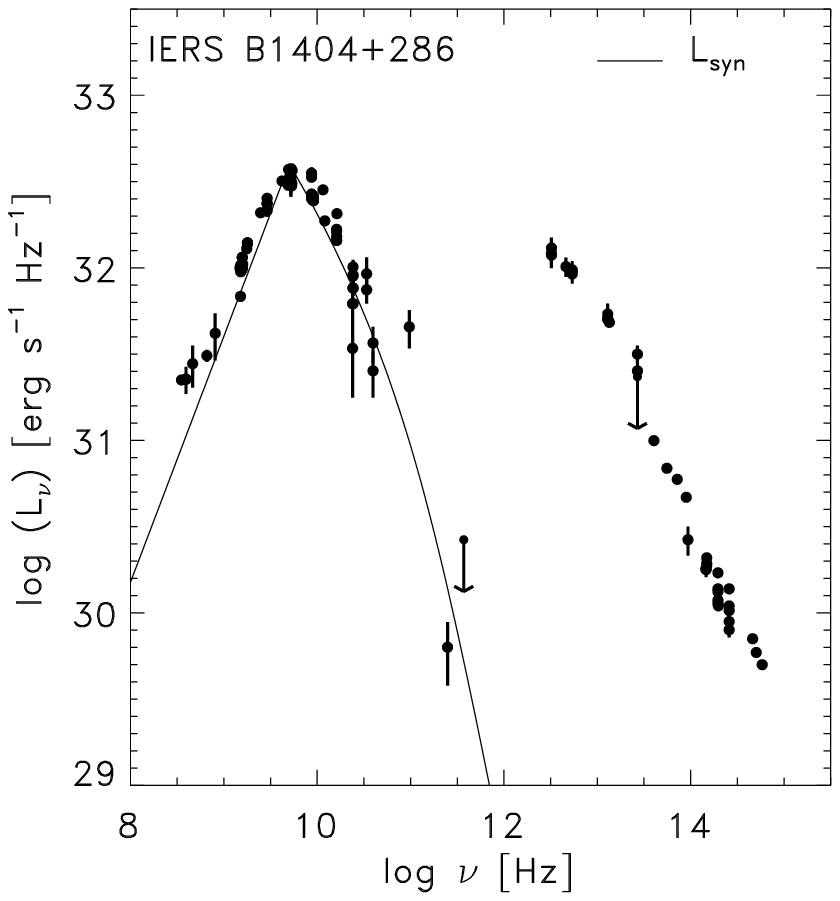}
\includegraphics[scale=0.7]{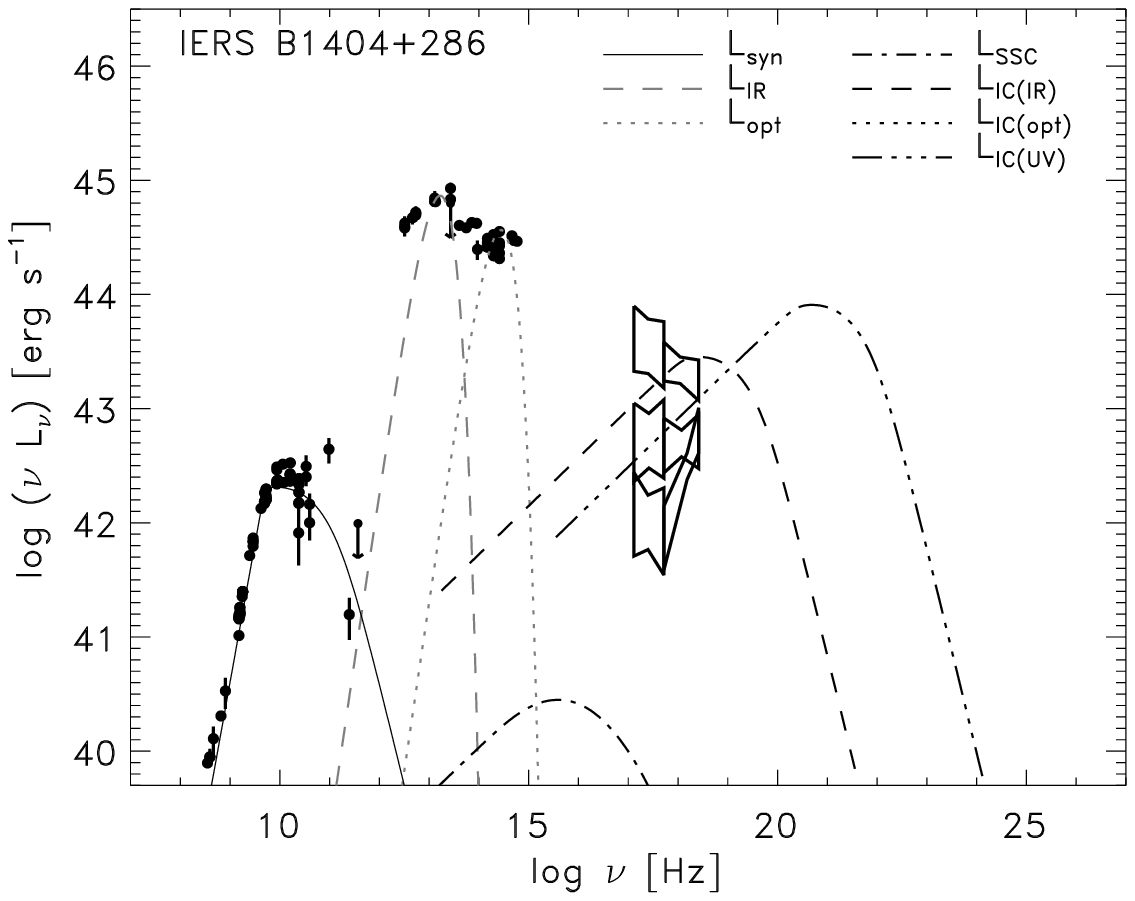}
}
\hbox{
\includegraphics[scale=0.7]{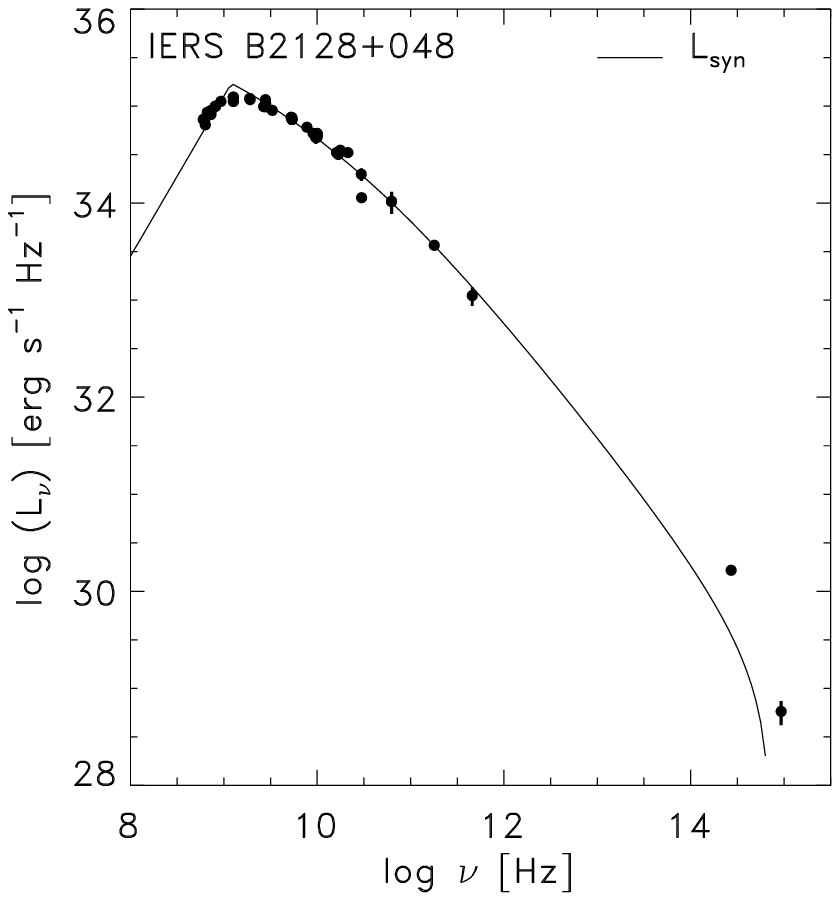}
\includegraphics[scale=0.7]{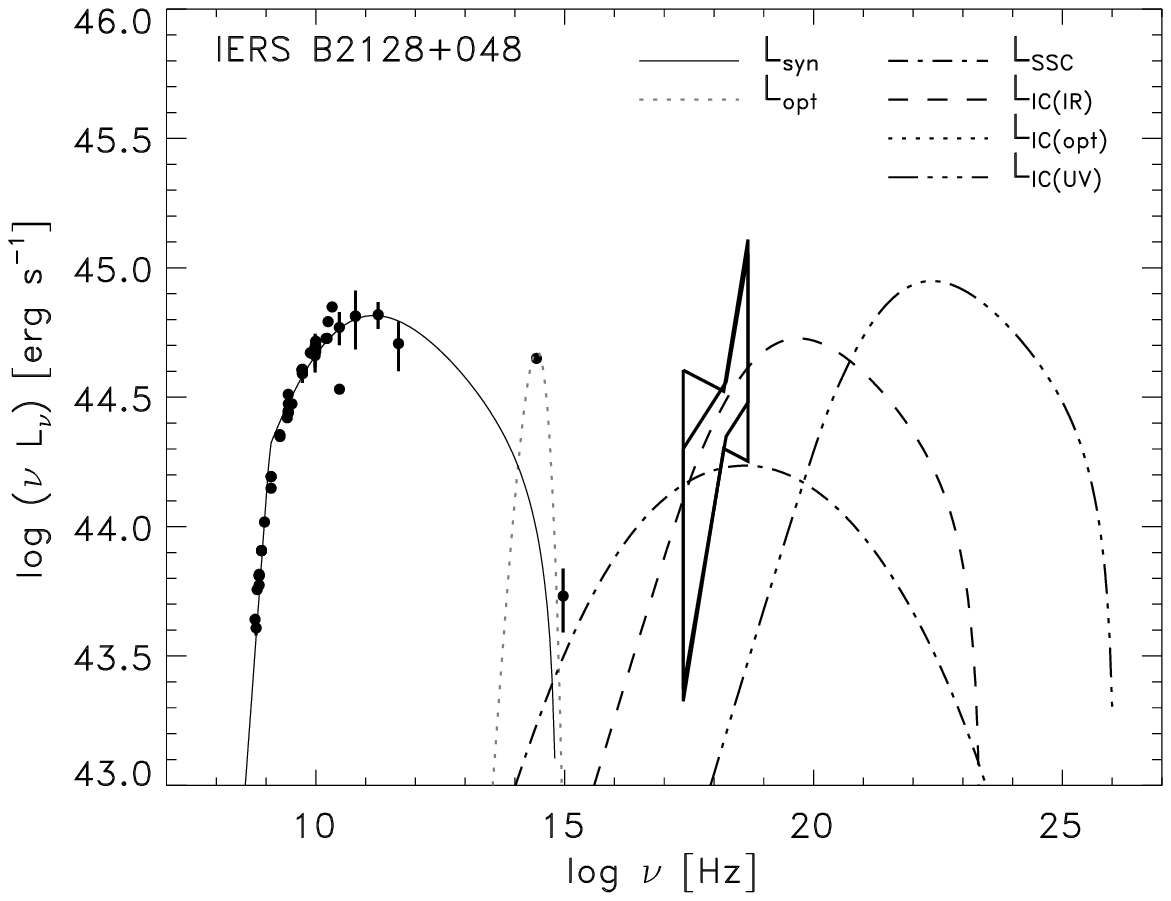}
}
\hbox{
\includegraphics[scale=0.7]{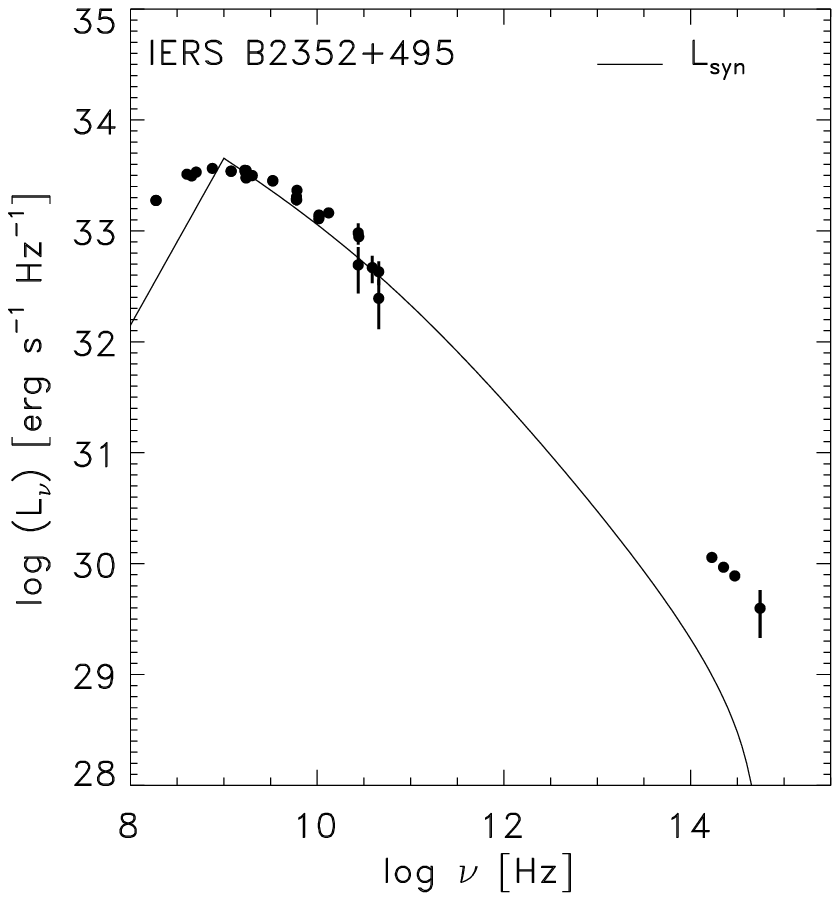}
\includegraphics[scale=0.7]{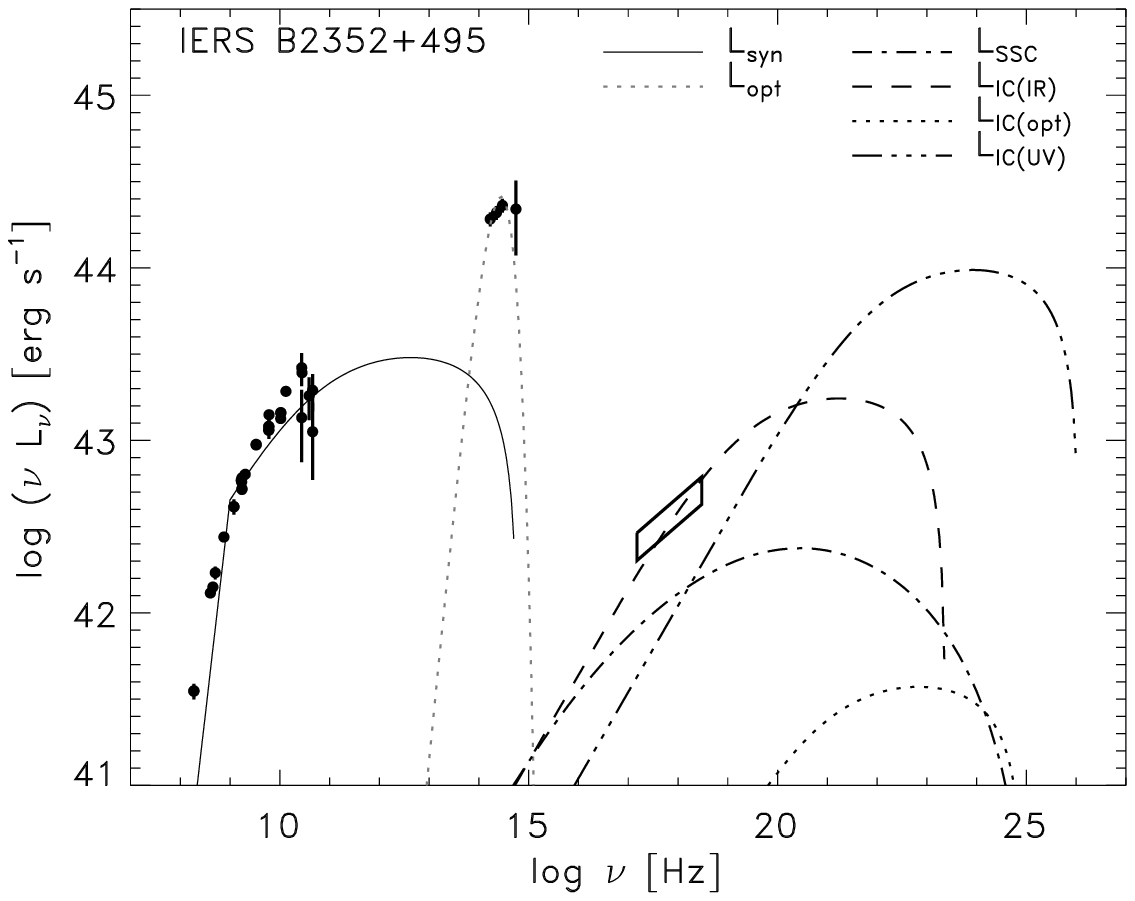}
}
\caption{{\it Continued}}
\end{figure*}

\subsection{Synchrotron emission}

In our model, the computation of the theoretical radio spectrum 
is performed under the assumptions that the radio emission is 
dominated by the contribution of the lobes, and that each lobe contributes 
the same amount of radiation. 
Although the theoretical source morphology is necessarily much simpler 
than the actual source structure, the above approximations 
reasonably fit our sample of GPS/CSO galaxies.

For the selected sources, we thus modelled the radio spectrum
as synchrotron radiation produced by the electron population of  
the lobes, $N_e(\gamma)$, which represents the evolution of 
the injected particle hot-spot population $Q(\gamma)$.
Among the possible injection functions $Q(\gamma)$ 
discussed in Section \ref{sec_model}, 
we chose the broken power law
$Q(\gamma) \sim \gamma^{-s}$, with $s=s_{\mathrm{1}}$ for 
$\gamma < \gamma_{\mathrm{int}}$, and $s=s_{\mathrm{2}}$ for 
$\gamma>\gamma_{\mathrm{int}}$, 
the break being fixed at $\gamma_{\mathrm{int}} \simeq 2\times 10^3$;
a similar spectrum was indeed inferred from the broad-band modeling 
of the hot-spot emission of the radio galaxy Cyg A, 
and interpreted as the result of a transition between two different 
acceleration mechanisms: the cyclotron resonant absorption,  
and the diffusive shock acceleration \citep[][and references therein]{stawarz2007}.
The lower and higher Lorentz factors
of the emitting particles were chosen as $\gamma_{\rm min}=1$ and 
$\gamma_{\rm max}=10^5$.
The emission of this particle population well reproduces the 
optically-thin part of the spectrum, 
i.e. the emission at frequencies higher than the turnover. 

FFA effects, as those generated by a spray of interstellar clouds engulfed 
by the expanding lobes and mingled with the synchrotron-emitting 
gas, are included in the model in order to compute the 
optically-thick portion of the spectrum, i.e. the emission at 
frequencies lower than the turnover.
The value of the turnover frequency enters the model as an 
input parameter ($\nu_{\rm p, intr}$);
the calculation of this value would indeed require 
the physical and geometrical modeling of the absorber, and thus the 
introduction of additional free parameters in the model \citep[see][]{begelman1999}.

The combination of the above synchrotron emission with the adopted 
absorption model returned a spectral shape of the form 
$L_{\nu} \propto L_{\nu,\mathrm{ua}}\, \nu^{-2}$ (with $L_{\nu,\mathrm{ua}}$ 
the luminosity unaffected by FFA), and satisfactorily fits 
the shape of most of the optically-thick spectra without the need of 
assuming source inhomogeneities.
This feature of our scenario is not shared by the SSA scenario.
SSA effects would return an optically-thick spectrum with slope 
{\it independent} of the optically-thin slope, and equal to 5/2 for a homogeneous source; 
to obtain a flatter spectrum an inhomogeneous source should be assumed.
Nevertheless, the issue of homogeneity can also play a role in our scenario.
In fact, our absorption model significantly deviates from the data 
in the case of IERS B2352+495, a source whose spectrum is much flatter than 
predicted at frequencies below the turnover; in this spectral region,
the source structure is indeed more complex than we assumed 
\citep[][and references therein]{araya2010}.

The spectral fits described above enabled us to constrain $L_{\mathrm{j}}$, $s_1$, and $s_2$
(see Table \ref{tab_par}).

\subsection {Thermal radiation fields}
The thermal radiation fields that we considered as soft-photon sources for the 
IC emission are the host galaxy's stars, the circumnuclear torus, and the accretion 
disk. 

We are interested in evaluating the relevance of the contribution of these  
radiation fields in the IC process, rather than in carefully modeling the SED of 
these components.
Therefore, for the sake of simplicity, we modelled the spectra of the above 
thermal photon fields as black-body spectra with the appropriate frequency peaks 
and bolometric luminosities. 

\subsubsection{Host galaxy}
\label{sec_galaxy}
As mentioned in Section \ref{sec_introduction}, the properties of GPS-source host galaxies 
are consistent with those of non-passively evolving giant ellipticals: they
often display signs of recent star formation as well as morphological disturbances.
Nevertheless, departures from our assumption of a constant-density medium for the inner kpc-scale 
core (see Section \ref{sec_model}) are negligible for any GPS-source host galaxy.

As shown by, e.g., \citet{silva1998}, the typical SED of an evolved giant elliptical
displays a dominant optical-NIR hump peaking about 1--2 $\mu$m, with peak luminosity 
of a few $\sim$10$^{44}$ erg \persec, and a secondary MFIR hump peaking about 
$\sim$100 $\mu$m, and a factor $\sim$300 fainter.
Starburst, interacting, and ULIRG galaxies can however show a MFIR hump with a peak luminosity 
up to a factor $\sim$50 higher than the optical-NIR peak.

We could satisfactorily model all of our galaxy optical-NIR SEDs with a black-body spectrum peaked 
at $\lambda=1.5\, \mu$m ($\nu=2\times10^{14}$ Hz), and a bolometric luminosity in the range  
$L_{\rm opt}=3\times 10^{44}-10^{45}$ erg \persec (see Table \ref{tab_par}).
MFIR data were available for three objects of our sample: 
in all of these cases, we did not associate the MFIR emission with the galaxy, 
but with circumnuclear dust (see Section \ref{sec_torus}).

\subsubsection{Torus}
\label{sec_torus}
Besides the emission from the host-galaxy dust accompanying possible star-formation 
activity, a putative circumnuclear dusty torus re-radiating part of the 
UV radiation absorbed from the accretion-disk would also contribute to the total emission in the 
MFIR domain.

Most of our sample's sources lack MFIR observations and/or detections. 
In the three objects detected at MFIR frequencies, the radio-to-MFIR spectra 
clearly show that the MFIR data lie well above the extrapolation of the synchrotron 
spectrum, ruling out the jet as the source of this radiation,
and suggesting a thermal origin of the MFIR spectral component.

These sources are not spatially resolved in this energy window, preventing us to 
locate the site of the MFIR emission.
Nevertheless, as we already mentioned in \citet{stawarz2008}, 
hints for the presence of a dusty torus around their AGN come from 
numerous observations of samples of GPS and CSS sources.
\citet{heckman1994} showed, by means of IRAS data (12--100 $\mu$m), 
that the MFIR luminosity of GPS and CSS sources is comparable to that of extended 
sources with similar radio power and redshift, and is consistent with 
thermal emission from dust heated by the UV emission of the central AGN, rather than
with radiation from a circumnuclear starburst.
Analysis of ISO data by \citet{fanti2000} also proved that FIR (60--100 $\mu$m) 
luminosities of GPS and CSS radio galaxies are not significantly different from those of 
extended objects, and associated the observed luminosities (on average 
$\ge 6 \times 10^{11}$ L$_{\sun}$)
to dust with temperatures from  25--30 K (warm component) up to 60--100 K (hot component),
whose masses are $\sim 5\times10^5$ M$_{\sun}$ 
and $2\times10^8$ M$_{\sun}$ respectively.
Finally, combining ISO data at 2--200 $\mu$m with millimetric and sub-millimetric observations
of 3CR radio galaxies and quasars, \citet{haas2004} demonstrated that both classes show
MIR-to-FIR luminosity ratios typical of powerful AGNs;
the MFIR emission can be ascribed to dust with temperatures between $\sim$30 and 100 K;
quasars and galaxies preferentially display $L_{\rm MIR}/L_{\rm FIR} >1$ and $<1$, 
respectively, according to the idea 
that the thermal dust emission is located in the torus, whose internal, 
hotter component cannot be observed by edge-on observers.
The above findings were confirmed by {\it Spitzer}/MIPS ($24-160\, \mu$m) 
observations of FRII radio galaxies and quasars: \citet{shi2005} showed the dominance 
of thermal emission from AGN-heated dust in their MFIR spectra, and no difference between 
small- and large-scale sources; CSS sources with sizes of 2--400 kpc displayed 
MFIR luminosities between $10^{10}-10^{14}$ L$_{\sun}$, with no size-luminosity correlation. 

On the other hand, 
\citet{polletta2008} modelled, by means of clumpy-torus models,  the 
IR SEDs of a sample of MIR-luminous AGNs at $z=1.3-3$ discovered by {\it Spitzer} (IRAC and MIPS, 
$\lambda=3.6-160 \mu$m): the MIR data could be reproduced by a single-temperature
($\sim$300 K) dust component; for sources detected at FIR frequencies as well, 
the additional, FIR spectral component could be modelled either with an AGN-heated 
colder dust component detached from the torus, or with an exceptionally 
powerful ($L>3.3\times10^{12}$ L$_{\sun}$) starburst, thus leaving the issue of the origin 
of FIR emission open.

Finally, {\it Spitzer} data of FRI radio galaxies \citep{leipski2009} recently proved that star formation
is actually present in these sources, with a contribution to the MIR emission ranging from minor to dominant;
for sources in which the nuclear MIR component could be identified, this component is dominated 
by non-thermal emission, although some FRIs do exist with clear indication of AGN-heated 
dust emission.
The debate on MFIR emission is thus clearly open.

As discussed in Section \ref{sec_sample},
MFIR data were available for three of our sample's members only. 
For the sake of simplicity, we chose to model the MFIR emission 
of these GPS/CSO galaxies with single-temperature black-body spectra, 
which we associated with torus' dust. 
The spectral fit yielded 
black body temperatures $T\sim 50-125$ K, and thus peak  
wavelengths $\lambda_{\mathrm{IR}}=23-60\, \mu$m.
For the remaining sources, with no MFIR data available, we {\it assumed} the torus' dust 
to radiate as a black body with temperatures $T\sim 50$ K; its spectrum thus  
peaks at wavelengths $\lambda_{\mathrm{IR}}\sim 60\, \mu$m.
The broad-band spectral fit procedure yielded MFIR luminosities  
$L_{\mathrm{IR}}=8 \times 10^{43}-6 \times 10^{45}$ erg \persec (see Table \ref{tab_par}).

Even though our analysis of the {\it Spitzer}/IRAC observations suggests 
the presence of a hotter, MIR-emitting dust component in two sample's sources
(IERS B1345+125 and IERS B1404+286), 
these objects are brighter in the FIR than in the MIR domain.
Therefore, our single-temperature black-body spectral model accounts for the 
dominant IR component.
Furthermore, the exact shape of the IR spectrum does not play a crucial role in determining 
the resulting X-ray spectrum, which is more heavily affected by the peak frequency of the 
IR spectrum and by the electron energy distribution.

\subsubsection{Accretion disk}
\label{sec_disk}
Both the direct UV emission from the disk and the possible X-ray emission from a hot corona 
\citep[e.g.][]{koratkar1999,cao2009}
are expected to be largely obscured by the dusty torus, in edge-on AGNs.
Emission lines are thus used to investigate the accretion properties of obscured sources.  

According to the findings by \citet{baum1989} and \citet{rawlings1991} for extended sources, 
the [\OIIIline]$\lambda$5700 luminosity was claimed to be positively 
correlated with the radio luminosity
also for a sample of  GPS and CSS sources \citep{labiano2008}:
this would support the traditional scenario according to which the AGN 
powers both the ionized gas and the radio emission.

Additional sources of ionization might however be plausible.
Early studies by \citet{morganti1997} 
showed that no significant differences in line luminosities and ratios were present between 
compact (CSS) and extended radio sources with comparable radio power and redshift,
and only tentative evidence for a lower \OIII luminosity in CSS sources was found. 
More recently, however, \citet{labiano2008} claimed 
the existence of a possible positive correlation between \OIII luminosity and 
the size of the radio source, suggesting that the jet expansion  contributes to 
enhance the line emission.

\citet{kawakatu2009b} recently compared the observed  properties of some narrow emission lines
of a sample of young radio-loud galaxies (average size of $\sim$3 kpc) 
with those of a sample of radio-quiet Seyfert 2 galaxies, and found that young radio-loud 
galaxies have systematically larger low- vs. high-ionization emission line ratios 
([\OIline]$\lambda$6300/[\OIIIline]$\lambda$5700). 
They concluded that powerful, young AGNs favour accretion disks without a strong 
big blue bump (hereafter BBB), 
whereas a strong BBB would be present in Seyfert 2s.
The BBB is actually missing in geometrically thick and optically thin 
accretion disks.
However, such disks would be characterized by a 
radiatively inefficient accretion flow \citep[RIAFs; e.g.,][]{narayan1995}, unable to generate the 
bolometric luminosities necessary to produce the \OIII emission observed 
in young radio galaxies \citep{kawakatu2009b}, and in conflict with the high 
accretion rates found by \citet{wu2009a} for a sample of young radio sources.

As acknowledged by \citet{kawakatu2009b}, an alternative explanation for the observed 
line ratios might be a scenario in which young radio galaxies do have a disk SED with 
BBB, but they are characterized by systematically different combinations of the 
hydrogen column density \NH and the ionization parameter $U$ 
(generally, higher \NH and lower $U$).
Alternatively, following our model, the hard spectral component needed 
to explain the \OIline/\OIII line ratio might come from the UV photons produced through IC
in the expanding lobes (see Sect. \ref{sec_IC}).

The above evidence highlights that the actual SED of the accretion disk in GPS sources is 
still not well constrained, and might be different from that of both standard disks
and RIAFs.

In our broad-band SED modeling, we assumed that the disks of GPS sources do  
produce the BBB \citep{koratkar1999} characteristic of Seyfert 2s and quasars:
we fixed the emission peak at $\sim$10 eV ($\nu_{\mathrm{UV}}=2.45\times10^{15}$ Hz; 
$T\simeq 1.16\times 10^5$ K ) 
and found luminosities $L_{\mathrm{UV}} = 10^{45}-10^{46}$ erg\persec 
(see Table \ref{tab_par}).
This UV disk emission would be partly 
reprocessed by the dusty torus and re-radiated in the MFIR band, partly absorbed 
by the NLR gas clouds, and partly upscattered off the lobe high-energy particles 
to $\gamma$-ray energies.
In the modeling procedure, the disk luminosities were constrained
by the MFIR luminosities 
(see Table \ref{tab_par}); 
the MFIR luminosity values, in turn, were constrained either by MFIR observations 
(when available) or by the X-ray spectral fitting (see Section \ref{sec_IC}).
As we showed in \citet{stawarz2008}, possible detections of (or upper limits on)
GeV $\gamma$-ray emission of GPS/CSO galaxies by the Large Area Telescope 
on board the {\it Fermi Gamma-Ray Space Telescope} 
\citep[hereafter {\it Fermi}/LAT;][]{atwood2009}
would constrain the disk luminosity in the UV band, thus enabling a test 
of the BBB assumption for the disk spectrum (see Section \ref{sec_IC}).

\subsection {Inverse-Compton emission}
\label{sec_IC}

The high-energy spectral components were computed through 
inverse-comptonization of the aforementioned synchrotron and thermal radiation fields
off the lobe electron population responsible for the synchrotron radio spectrum.
As discussed in \citet{stawarz2008}, we assumed a magnetic field only a factor 
of a few below the equipartition value ($\eta_{\mathrm{B}}=0.3$, $\eta_{\mathrm{e}}=3$); 
this assumption is roughly consistent with the finding by \citet{orienti2008b} 
that equipartition holds in the lobes of GPS sources.

As Fig.\ \ref{fig_seds} shows, in most of our sources the modelled X-ray emission
is dominated by the inverse-comptonization of the MFIR photon field (IC(IR)). 
The contributions of the inverse-comptonized UV radiation (IC(UV)) and of the SSC emission
reaches comparable magnitude in a few objects;
the starlight does not play, as IC seed photon, a relevant role in any of our sources 
\citep[see, however,][for the complete theoretical case study]{stawarz2008}.
Our model well reproduces the observed X-ray spectra for reasonable assumptions on the 
thermal emission of the circumnuclear dust and of the accretion disk. 

The model also predicts significant $\gamma$-ray emission up to the GeV-TeV 
domain, not observed so far in any GPS galaxy.
Even though, in some of our sample's sources, the high-energy tail of the 
IC(IR) spectral component reproducing the observed X-ray spectrum extends to  
the soft $\gamma$-ray energy band, the modelled $\gamma$-ray emission 
can be mostly identified as the result 
of the IC scattering of the UV photons produced by the accretion disk with 
characteristic BBB and appropriate luminosity, which we assumed to operate 
in the source.

A comparison of our modelled SEDs with the {\it Fermi}/LAT sensitivity 
curves \citep{atwood2009} shows that none of our GPS/CSO galaxies could be detected by 
{\it Fermi}/LAT in less than one year: this is consistent with the current 
lack of any of these sources in the {\it Fermi}/LAT First Source Catalog \citep{abdo2010}.
The modeling also predicts that five out of eleven sample's members 
(i.e., IERS 0026+346, IERS B0108+388, IERS B0500+019, IERS B1345+125, IERS B1404+286) 
could be detected by {\it Fermi}/LAT at $2\sigma$ level after an integration 
time of about one year. 
This prediction should however be taken with caution: as mentioned in 
Section \ref{sec_disk}, the 
$\gamma$-ray fluxes depend upon the assumed accretion disk luminosities 
$L_{\mathrm{UV}}$, as well as upon the high-energy tail of the lobe electron 
distribution.  
A non-detection by {\it Fermi}/LAT might thus imply either UV accretion disk 
luminosities lower than we assumed, or a different electron distribution 
in the lobes.

\section{FURTHER SUPPORT TO THE X-RAY LOBE SCENARIO: THE \NHmath-\NHI CORRELATION} 
\label{sec_support}
\subsection{Context}
As we recently pointed out \citep{ostorero2009}, 
the prediction of the X-ray--emitting lobes that characterizes 
our model may be supported by further observational evidence.

The X-ray emission of GPS galaxies was traditionally 
mostly interpreted as thermal radiation from the accretion disk, absorbed by a 
gas component associated with the AGN and characterized by an equivalent hydrogen 
column density \NH \citep{odea2000,guainazzi2004,guainazzi2006,vink2006,siemiginowska2008},
rather than as non-thermal emission from the jet or the lobe.
Non-thermal emission from the jet/lobe, although not always completely ruled out,
was never considered to be significant.
The main argument brought in support of a disk-dominated X-ray emission scenario 
is the apparent discrepancy between the equivalent total-hydrogen column density 
\NH derived from the X-ray spectral analysis and the neutral hydrogen column 
density \NHI derived from the radio measurements of the 21-cm absorption line,
when the latter is available. 
The authors concluded that, because \NH always exceeds \NHI of 1--2 orders 
of magnitudes, the X-rays must be produced in a source region that is more 
obscured than the region where the bulk of the radio emission comes from, and 
thus located {\it within} the radio lobes;
if this was not the case, an unreasonably high fraction of ionized hydrogen 
(\HIIline) should be assumed to account for the above difference 
\citep{guainazzi2006,vink2006}.
Such a scenario would also be consistent with the observed anticorrelation 
between \NHI and linear size found by \citet{pihlstroem2003}, being the fraction 
of ionized gas likely low in a young radio source with still expanding 
Str\"omgren sphere \citep{vink2006}.

The discrepancies between the \NH and \NHI values mentioned above
should in fact be regarded with caution.
The \NHI estimate is derived, from the measurements of the hydrogen 
21-cm absorption lines, as a function of the ratio between the gas 
spin temperature $T_{\rm s}$ and its covering factor $c_f$, representing 
the fraction of the source covered by the \HI screen \citep[e.g.][]{gupta2006}.
The common assumption $T_{\rm s}/c_f = 100$ K 
refers to the case of complete coverage ($c_f=1$) of the emitting 
source by a standard cold ($T_{\rm k}\simeq 100$ K) ISM cloud
in thermal equilibrium, and thus with spin temperature $T_{\rm s}$ equal 
to the kinetic temperature $T_{\rm k}$ ($T_{\rm s}=T_{\rm k})$.
However, this assumption 
returns a value of \NHI that represents a lower limit
to the actual column density of both the total and the neutral hydrogen 
\citep[e.g.][]{pihlstroem2003,vermeulen2003,gupta2006}.
In fact, when $T_k\simeq100$ K, the expected high abundance of the molecular 
hydrogen, $H_2$ \citep{maloney1996}, might account for the difference 
between \NH and \NHI.
On the other hand, in the AGN environment,
illumination by X-ray radiation might easily raise $T_{\rm k}$ up to $10^3-10^4$ K 
\citep{maloney1994,conway1995}, making $T_{\rm s}$ raise accordingly 
\citep{davies1975,liszt2001};
a source covering factor smaller than unity would also increase 
the $T_{\rm s}/c_f$ ratio; 
both the above effects would potentially lead to \NHI values fully 
consistent with the \NH estimates.
Finally, temperatures as high as several $10^3$ K would likely 
imply the presence of a non-negligible fraction of ionized 
hydrogen \citep[\HIIline;][]{maloney1996,vink2006}, also contributing to
relax possible residual column-density discrepancies.

Evidence is mounting that, in GPS and CSS sources, the \HI absorption 
lines are not generated by a screen covering the source uniformly: instead,
they originate in clouds of neutral hydrogen connected with (at least one 
of) the jet/lobe radio structures, and possibly interacting with it
\citep{morganti2004b,labiano2006,vermeulen2006,araya2010}.
The association of the bulk of the \HI absorption with the optical emission line 
currently supports the identification of the absorbers with the atomic cores of the 
NLR clouds, although the presence of \HI elsewhere is not ruled out 
\citep{labiano2006,vermeulen2006}.

In our GPS-source model, the NLR clouds are gradually engulfed by the expanding 
radio lobes.
Moreover, the cloud density is assumed to decrease with the distance $r$ from 
the AGN as $r^{-n}$ (with $1<n<2$), according to the observations of the NLR 
in Seyfert galaxies \citep{kraemer2000a,kraemer2000b,bradley2004}, and consistently 
with the anticorrelation between 
\NHI and linear size found by \citet{pihlstroem2003}.
Because in our model the contribution of the lobes to the X-ray output of the source 
is significant, becoming stronger and stronger with increasing source compactness,
the NLR clouds might thus play an important role in the absorption of the lobe X-ray
radiation.

\subsection{\NHmath-\NHI Correlation} 
\label{sec_NH_sample}

\begin{figure}
\includegraphics[scale=0.5]{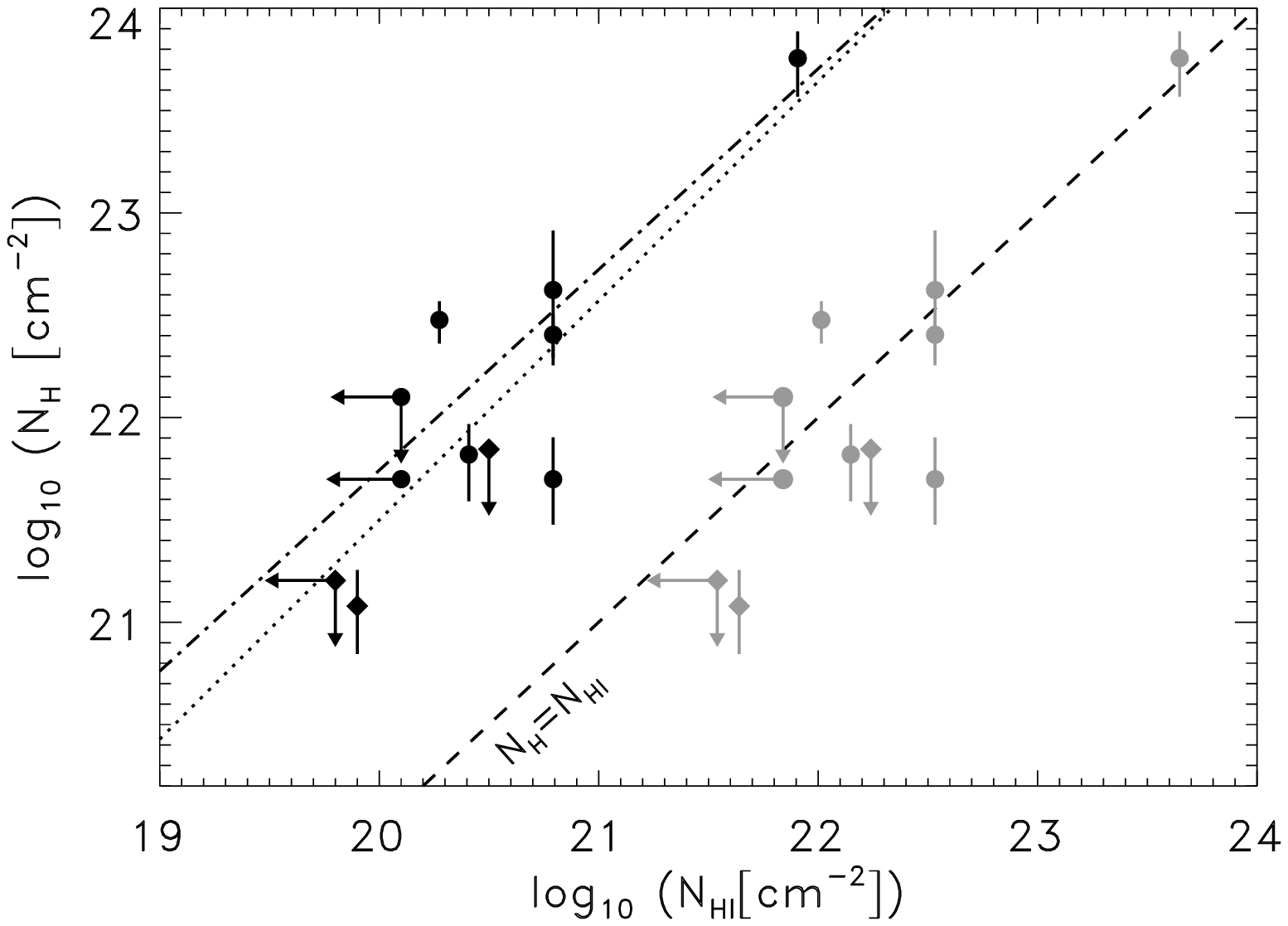}
\caption{X-ray column densities (\NH) as a function of radio column densities (\NHI) 
for the seven GPS/CSO galaxies of our sample ({\it sub-sample D5+U2; bullets}), 
IERS B0108+388, IERS B0500+019, PKS B0941-080, IERS B1031+567, 
IERS B1345+125, IVS B1358+624, and IERS B2352+495, and the three 
additional GPS galaxies reported by \citet{tengstrand2009} 
({\it sub-sample ADU; diamonds}): 4C +32.44, PKS 0428+20, and 4C +14.41.
Black symbols: \NHI was computed by assuming $T_{\mathrm{s}}=100$ K and $c_f=1$; 
arrows represents upper limits. 
Grey symbols: as an example, the same sources with $T_{\mathrm{s}}/c_f=5.5 \times 10^3$ K.
Dash-dotted line: linear fit to {\it sub-sample D5+AD} 
of \NHmath/\NHI {\it detections} (with $T_{\mathrm{s}}=100$ K);
dotted line: linear fit to {\it sub-sample D5+U2+ADU} including both 
{\it detections} and {\it upper limits}.
Data are from: 
\citet{guainazzi2006}; \citet{mirabel1989}; \citet{odea2000};
\citet{pihlstroem2003}; \citet{siemiginowska2008}; Siemiginowska (priv.\ comm.); 
\citet{tengstrand2009}; \citet{vermeulen2003}; \citet{vink2006}.} 
\label{fig_nh}
\end{figure}

As discussed above,
besides the modeling of the broad-band SEDs, a way of discriminating 
among different scenarios, and unveil the actual X-ray production site, is 
to compare the properties of the X-ray and
radio absorbers, i.e.  compare the column densities derived from the analysis of 
the X-ray data, \NH, and those obtained from radio measurements at the redshifted 21-cm 
wavelength, \NHI.
Such a comparison can be performed on a source-by-source basis: 
an {\it ad hoc} increase of either the $T_{\mathrm{s}}$ parameter 
or the $T_{\mathrm{s}}/c_f$ ratio can always remove possible \NH and \NHI discrepancies.
Alternatively, one can compare the column densities of a sample of sources: the existence of 
a positive, significant \NHmath-\NHI correlation would suggest that the X-ray
and radio absorbers coincide, thus supporting the co-spatiality of the X-ray and
radio source.

We investigated the existence of a connection between \NH and \NHI 
in our sample. 
For a positive correlation, we searched the source sub-sample
for which both \NH and \NHI estimates (either detections or upper limits) 
are available (see Table \ref{tab_xrad}).
The results of the correlation analysis are reported in Tables \ref{tab_correlation1},
\ref{tab_correlation2}, and Fig.\ \ref{fig_nh}, and are discussed below.

Estimates of \NH were available for ten out of eleven sample's members,
and an upper limit was available for the remaining source;
\NHI was instead measured in 6 sources only, and \NHI upper limits
were derived for 2 additional objects.
Estimates of both \NH and \NHI were available for a sub-sample of five sources;
taking the upper and lower limits to \NH and \NHI into account, the sub-sample 
extended to eight sources.

Note that, due to the poor photon statistics, the upper limit to \NH for source 
PKS B0941-080 was derived 
(A. Siemiginowska, priv.\ comm.) from a re-analysis of the {\it Chandra} data 
\citep{siemiginowska2008} in two different ways: (i) by fixing the absorption equal 
to the Galactic value, and letting the photon index $\Gamma$
free to vary; (ii) by letting the intrinsic absorption free to vary, and 
fixing the photon index to a value varying in the range $\Gamma$=1.7$-$1.9,
the upper-limit of this range being reported as the mean $\Gamma$ of both radio 
galaxies \citep{brinkmann1995} and GPS sources \citep{brinkmann1997}. In our analysis, 
we adopted the more conservative estimate (i).
For source IERS B1345+125, estimates of \NH were derived from two observations
carried out by different instruments ({\it ASCA} and {\it Chandra}) in different epochs 
(1996 and 2000, respectively). Given the fact that we cannot rule out long-term 
column-density variations, we included both \NH estimates in our data set; however, 
including the average \NH does not change our results significantly.
Finally, the physics of IERS B1404+286 appears to be so complex 
that a variety of models can satisfactorily fit the X-ray data 
\citep{guainazzi2004}.
As a consequence, the inferred values of \NH can vary by a factor $\gtsim$100
(see Table \ref{tab_xrad}).
This forced us to remove this source from our correlation analysis.

Our final correlation sample thus consisted of seven sources. 
Hereafter, the sub-sample of objects for which column density estimates 
did exist will be referred to as {\it sub-sample D5}, wherease {\it sub-sample U2}
will indicate the sources for which upper limits to \NH and/or \NHI were estimated; 
our full correlation sample will be indicated as {\it sub-sample D5+U2}.
The column-density detections and upper limits that entered the above sub-samples 
were indicated with boldface characters in Table \ref{tab_xrad}.

A Pearson correlation analysis applied to {\it sub-sample D5} revealed a  
significant (probability of the null-hypothesis of no correlation being true: 
Prob. $\simeq 1.7\times 10^{-5}$) \NHmath-\NHI positive correlation.
The statistical significance of this correlation, however, is admittedly 
driven by source IERS B0108+388, characterized by the highest \NHmath-\NHI values. 
Indeed, in the same sub-sample, the evaluation of the correlation by means 
of more robust, {\it non-parametric} (or {\it rank}) methods 
\citep[i.e.\ Spearman's and Kendall's correlation coefficients;][]{press1992}, 
decreased the significance of the correlation.

In the full correlation sample, i.e., {\it sub-sample D5+U2}, the correlation was 
investigated by means of survival analysis techniques. 
In particular, we made use of the software package ASURV Rev.\ $1.2$ \citep{lavalley1992}, 
which implements the methods for bivariate problems presented in \citet{isobe1986}.
The sub-sample {\it D5+U2} was shown to display a positive correlation, 
with significance level about 8\%.
The results of our correlation tests are reported in the first raw of 
Tables \ref{tab_correlation1} and \ref{tab_correlation2}.

It is rather promising that the significance of the correlation substantially improved
when the sample size was increased to ten sources by including the three new X-ray 
GPS galaxies, among those reported by \citet{tengstrand2009}, 
for which both \NH and \NHI estimates were 
available.\footnote{Detection: 4C +32.44; upper limits: PKS~0428+20 and 4C~+14.41 .} 
In this case, the probability of the null-hypothesis of no correlation being true 
decreased to $\sim$2\%.
The results on the full composite sample, including the additional sub-samples 
{\it AD} (additional detection) and {\it ADU} (additional detection and upper limits),
are given in the second raw of Tables \ref{tab_correlation1} and \ref{tab_correlation2}.
We wish to stress that dropping source IERS B0108+388 
(which drove the correlation in {\it sub-samples D5 and D5+U2}) 
from the full sample {\it D5+U2+ADU}, returned a correlation with still 
good significance (5\%: Spearman; 6\%: Kendall).

We note that our tentative correlation, that we anticipated for a smaller source 
sample in \citet{ostorero2009}, is in agreement with the tentative anticorrelation 
between \NH and linear size reported by \citet{tengstrand2009}, given the known 
GPS-source \NHI-$LS$ anticorrelation \citep{pihlstroem2003}.
We also note that a comparison of \NHI and \NH in a sample of spiral-hosted Seyfert 
galaxies did not reveal any correlation \citep{gallimore1999}.

As far as the law describing the relationship between \NH and \NHI is concerned, 
according to Pearson's test we could fit a linear relation to the log(\NHmath)-log(\NHI) 
sub-samples {\it D5} and {\it D5+AD}; however, this relation is not a good description 
of the data ($\chi^2_{\mathrm{red}}=9.63$ and $9.80$, respectively).
As for the censored sub-samples, {\it D5+U2} and {\it D5+U2+ADU}, we applied 
the ASURV Schmitt's linear regression to estimate the best-fit straight-line 
parameters; however, this procedure did not enable us to evaluate the 
goodness of the fit.
The results of our linear regression analysis are reported in the last two columns
of Tables \ref{tab_correlation1} and \ref{tab_correlation2}, and are displayed 
in Fig.\ \ref{fig_nh}.

\section{DISCUSSION} 
\label{sec_discussion}
Our systematic SED analysis provided us with constraints on some interesting 
physical parameters of our sample's members, as the spectrum of the electrons 
injected from the hot spots into the lobes, the luminosities of the accretion 
disk and the torus, the accretion efficiency, and the jet kinetic power, 
as discussed below.

The broken power-law energy distribution assumed for the electron population injected 
into the lobes is characterized by a break energy 
$\gamma_{\mathrm{int}} \simeq m_{\mathrm{p}}/m_{\mathrm{e}}$;
furthermore, the lower-energy segment ($\gamma<\gamma_{\mathrm{int}}$) 
of the spectra have indices $0.7< s_1 \lesssim 2$, 
and are thus mostly flatter than the canonical spectra generated by 
diffusive (1st order Fermi) shock acceleration.
These findings are in good agreement with the results inferred from the modeling of the 
broad-band spectra of both the hot spots of some extended radio galaxies 
\citep{stawarz2007,godfrey2009} and luminous blazars \citep{sikora2009}.
Therefore, as the aforementioned cases, our sources seem to fit into a scenario
in which two different acceleration processes are at work in mildly-relativistic shocks: 
a pre-acceleration process responsible for the lower-energy spectra, 
as e.g.\ the cyclotron resonant absorption \citep{hoshino1992,amato2006}, 
and an acceleration process acting at higher energies, as e.g.\ the Fermi-type mechanism.
The transition between the two acceleration regimes, marked by the energy break 
$\gamma_{\mathrm{int}}$, would reflect the dominant role of the protons in the 
jet's dynamics.

The torus luminosities yielded by our modeling ($\sim$$10^{44}-10^{45}$ erg \persec) 
are as large as the ones observed in powerful quasars and FRII radio galaxies 
\citep[e.g.,][]{shi2005};
moreover, the accretion disk luminosities ($10^{45}-10^{46}$ erg \persec) 
are as large as the ones estimated for powerful flat-spectrum radio quasars (FSRQs) 
\citep[e.g.,][]{sambruna2006,ghisellini2009}.
However, the jet kinetic powers that we found ($\sim 2\times 10^{44}- 4\times10^{45}$ erg \persec) 
are smaller than the values claimed for FSRQs
\citep[$10^{46} - 10^{48}$ erg \persec; e.g.,][]{sambruna2006,xu2009,ghisellini2009};
they rather seem to be comparable to the jet powers estimated for steep-spectrum radio quasars (SSRQs)
with similar BH mass, and to be higher than those of BL Lacertae objects and FRI 
radio galaxies \citep[e.g.,][]{xu2009}.

As noted by \citet{czerny2009}, assuming a standard 10\% radiative efficiency for 
both the accretion disk and the radio jet, the jet kinetic power sets 
a lower limit to the accretion luminosity: $L_{\mathrm{acc}}> L_{\mathrm{j}}/10$.
The results of our modeling yielded accretion-disk luminosities $L_{\mathrm{UV}}$ 
and jet kinetic powers $L_{\mathrm{j}}$ such that 
$L_{\mathrm{UV}}> L_{\mathrm{j}}/10$ (see Table \ref{tab_par}, col.\ 10):
this confirms the radiative efficiencies typically assumed, and characterizes  
our sources with jet/disk luminosity ratios of $0.01-0.1$, 
a factor $\sim$100 lower than the ratios typically found in powerful blazar sources    
\citep[$\sim$1-10; e.g.,][]{celotti2008}.

\begin{figure}
\includegraphics[scale=0.5]{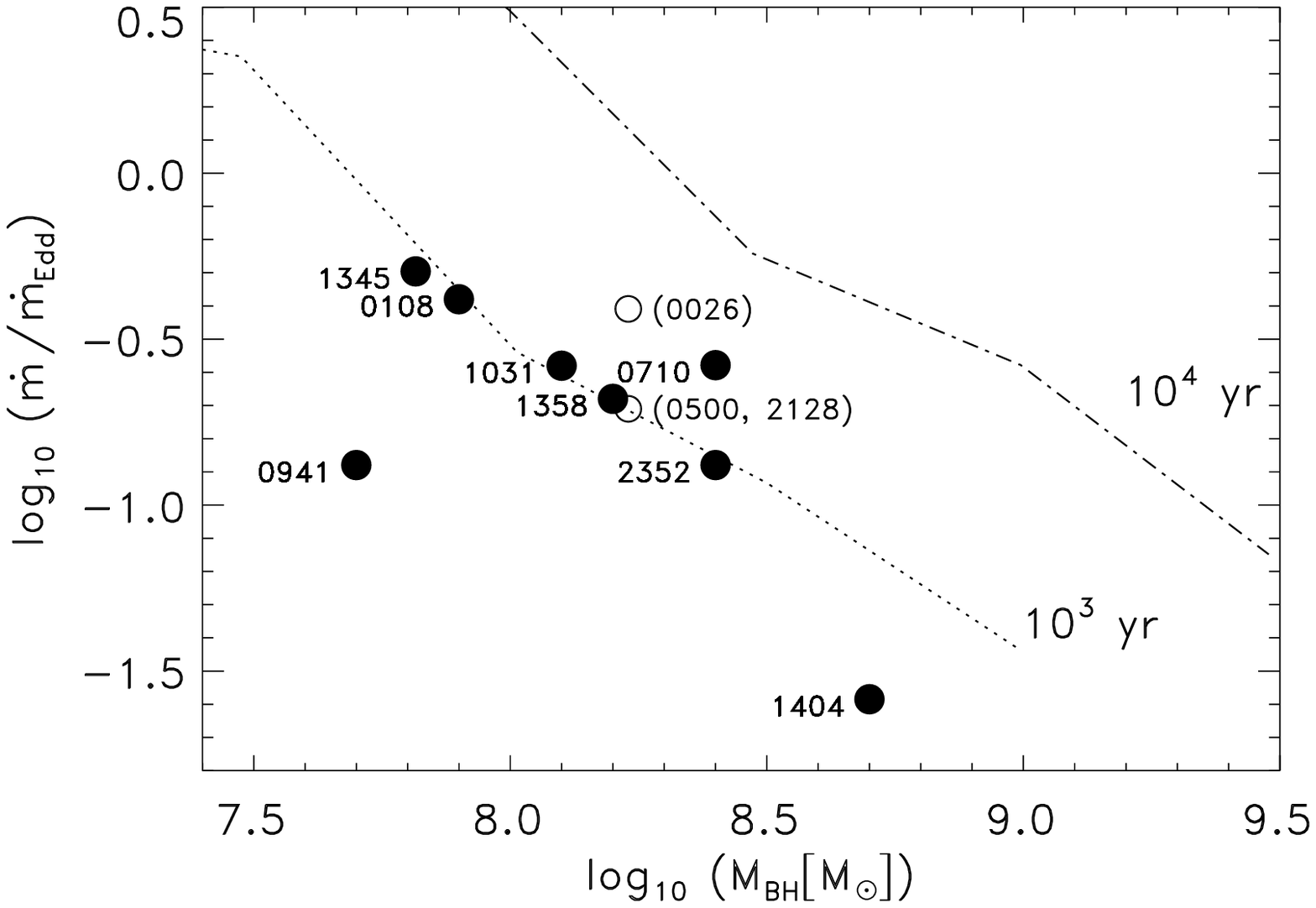}
\caption{The Eddington ratios of the GPS/CSO galaxies of our sample, inferred from the SED 
model parameters, are displayed as a function of their BH mass (data are from Table \ref{tab_par}), 
and compared to the theoretical curves of the accretion disk instability model
by \citet{czerny2009}. Source names are indicated in a short format in the plot.
Filled circles indicate the sources for which $\dot m/\dot m_{\mathrm{Edd}}$ was computed with the 
estimate of the BH mass; open circles show the sources (names are 
in parenthesis) for which the BH mass was assumed 
as the average BH mass of the GPS-source sample by \citet{wu2009b}.
Dotted and dashed-dotted lines represent the theoretical curves of constant duration
($10^3$ and $10^4$ years, respectively) for accretion disk outbursts 
triggered by the radiation pressure instability, assuming a disk viscosity 
$\alpha=0.02$ \citep[][their Fig.\ 4]{czerny2009}.}
\label{fig_mdot}
\end{figure}

The best-fit values of the accretion-disk luminosities enabled us to 
make a reasonable guess on the source accretion rates.
BH mass estimates were available for eight of our sample's members;  
by using the estimated BH mass values for these sources, and the 
average BH mass of  $1.698\times 10^8$ M$_{\sun}$  
derived by \citet{wu2009b} for a sample of GPS sources in the remaining three sources, 
we inferred for our GPS/CSO galaxies Eddington ratios of  
$\dot m / \dot m_{\mathrm{Edd}}=L_{\mathrm{UV}}/L_{\mathrm{Edd}}=0.026-0.506$
(see Table \ref{tab_par}, col.\ 11). 

These ratios correspond to accretion rates that are comparable to (for source IERS B1404+286)
or much higher than (for all the other sources) the critical accretion rate 
$\dot m_{*}\simeq 0.025\, \dot m_{\mathrm{Edd}}$ 
proposed by \citet{czerny2009} as the threshold above which the radiation pressure 
instability operates in accretion disks, giving rise to intermittent source activity.
Therefore, the results of our modeling are consistent with all 
of our sample's members being intermittent radio sources, with IERS B1404+286 being a 
border-line case.

In this context, according to the prescription by \citet{czerny2009}, for 
values of the disk viscosity $\alpha$ in agreement with the observational 
constraints \citep[$\alpha \simeq 0.02$; e.g.,][]{starling2004},
the sources characterized by $\dot m \gtsim 0.1 \dot m_{\mathrm{Edd}}$ 
are potentially able to undergo outburst phases lasting $\sim$10$^4$ years, and thus 
grow to super-galactic scales after a few of these outbursts, provided that
they have the appropriate BH mass.
Higher viscosity values would imply higher accretion rate thresholds.  
Except for IERS B1404+286, our sample's members are characterized by accretion 
rates higher than this limit;
however, their BH masses imply outburst durations shorter than $10^4$ years.
This can be seen in Fig.\ \ref{fig_mdot}, where our sample's members are plotted in the 
$M_{\mathrm{BH}}$ vs.\ $\dot m/\dot m_{\mathrm{Edd}}$ diagram, 
and are compared to the lines of constant outburst duration
predicted by the radiation pressure disk instability model by \citet{czerny2009}: 
all the sources lie below the line representing the outburst duration 
$T=10^4$ years, and are thus expected to undergo shorter-lived outbursts which
will prevent them to reach super-galactic scales.
Interestingly, most of our sources follow the $10^3$ year outburst model curve.
Note that for three of our sample's members (open symbols in Fig.\ \ref{fig_mdot}) 
the BH mass was {\it assumed} (see above);
the availability of a BH mass estimate might change the location 
of these objects in the diagram.

\section{CONCLUSIONS}
\label{sec_conclusions}
For the first time, we systematically compared the broad-band spectra of a 
sample of GPS/CSO galaxies with a model.
Specifically, we showed that the dynamical-radiative model we proposed in \citet{stawarz2008}
can reproduce the observed emission, from the radio to the X-ray energy range, 
of eleven  GPS/CSO galaxies characterized by different linear sizes, and 
thus sampling different stages of the source expansion.

The radio spectra were modeled as synchrotron emission by 
a lobe electron population that represents the evolution of an injected hot-spot electron 
population suffering from adiabatic and radiative energy losses.
At frequencies below the turnover, the shape of most of the spectra could be 
satisfactorily accounted for by an absorption scenario dominated 
by the FFA mechanism. 
 
The X-ray emission could be interpreted as non-thermal radiation 
produced through IC scattering of the local thermal radiation fields 
off the lobe electron population.
Among the possible seed photon fields, the dominant role in the generation of the 
X-ray radiation via IC scattering is played by the MFIR radiation, which we associated 
to the putative circumnuclear dusty torus; dust associated with circumnuclear 
star-forming regions might also contribute to this emission, but this 
would not affect our results significantly.
The UV radiation emitted by the accretion disk gives a contribution comparable to 
that of the MFIR photons in a few sources only, but provides the bulk of the 
seed photons responsible for the $\gamma$-ray emission.
Our model proved to be a viable alternative to the thermal, accretion-disk 
dominated scenario for the interpretation of the X-ray emission of GPS/CSO galaxies.

Finally, from a more observational perspective, our radiative model 
finds further support in the comparison of the measurements  
of the radio and X-ray hydrogen column densities of a sub-sample of our 
source ensemble: the data suggest a positive correlation, which, if confirmed, 
would point towards the co-spatiality of the radio and X-ray emission regions.
New radio measurements, necessary to improve the statistics of our correlation sample,
as well as the quality of the available data, are currently being performed.

\acknowledgments

We are indebted with A.\ Siemiginowska and E.\ Ferrero for stimulating 
conversations on GPS sources, with R.\ Morganti for fruitful discussions 
on \HI absorption, and with L.\ Costamante, A.\ Siemiginowska, and M.\ Guainazzi, 
for clarifying many issues on the analysis of the X-ray data of GPS sources.
We thank an anonymous referee for a careful revision of the paper, 
and for constructive suggestions.
We gratefully thank A. Siemiginowska for estimating the intrinsic \NH for 
sources PKS B0941-080.
L.O.\ started this work while she was supported by a fellowship of the 
University of Torino; she currently holds a 2009 National Fellowswhip 
``L'OR\'EAL Italia Per le Donne e la Scienza'' of the L'OR\'EAL-UNESCO 
program ``For Women in Science''.
L.O.\ and A.D.\ gratefully acknowledge partial support from the INFN grant PD51.
This work is part of a project that started when L.O., {\L}.S., and S.W.\ were 
partly funded by EC under contract HPRN-CT-2002-00321 (ENIGMA).
R.M.\ acknowledges support from MNiSW grant N N203 301635.
\L.S.\ acknowledges support from Polish Ministry of Science and 
Higher Education within the project N N203 380336.
C.C.C.\ was supported by an appointment to the NASA Postdoctoral Program
at Goddard Space Flight Center, administered by Oak Ridge Associated 
Universities through a contract with NASA.
S.W. acknowledges partial support by BMBF/DLR through grant 50OR0303. 
We acknowledge financial support by the Department of Energy 
contract to SLAC no.\ DE-AE3-76SF00515.
L.O., A.D., and \L.S.\ warmly thank M.J.\ Geller and A.\ Siemiginowska 
for their kind hospitality at the CfA, where part of this work was performed.
This work is partly based on observations made with the {\it Spitzer Space Telescope}, 
which is operated by the Jet Propulsion Laboratory (JPL), California Institute 
of Technology (CalTech) under a contract with the National Aeronautics 
and Space Administration (NASA).
This research has made use of the NASA/IPAC Extragalactic Database 
(NED), which is operated by the JPL, CalTech, under contract with NASA.

\appendix

\section{NOTES ON INDIVIDUAL SOURCES}
\label{app_A}

\subsection{IERS B0026+346 (COINS J0111+3905; OB +343)}
The object, quite  powerful in the radio band  
($P_{\rm 5\,GHz}=10^{27.1}$ \wattperhz), was observed repeatedly at many 
radio frequencies since the 1980's.
Because of its convex radio spectrum \citep[e.g.,][]{jauncey1970,kuehr1981}, 
it was identified as a GPS source, although the spectral turnover did not 
look dramatic. The GPS nature of the source was also confirmed, on the basis of 
flux variability and spectral shape, by \citet{torniainen2007}.
Despite the detection of a $\sim$4.7$c$ apparent speed, atypical 
for GPS sources, for the brightest source component \citep{kellermann2004},
recent flux-density monitoring measurements by \citet{torniainen2007}
detected only moderate variability and a spectrum steep enough to confirm the 
classification of the object as a GPS source on the basis of their criteria. 
The source is seen as an unresolved core by the VLA \citep{ulvestad1981};
however, VLBA maps enabled \citet{kellermann1998} to classify the object as a  
double source, by assuming that the core of one of the weaker, intermediate components 
was the core of a CSO, although they could not exclude the core as being actually 
located at one of the extremities. 
The overall source structure extends over $\sim$35 mas ($\sim$211 pc),
and is hosted by a galaxy located at $z=0.517$ \citep{zensus2002}, whose optical 
$R$-band magnitude is $m_{\rm R}\sim 21$ \citep[][and references therein]{snellen1996}.

In the X-ray energy range, the source was firstly successfully detected 
(0.54--3.9 keV) by the {\it Einstein Observatory} in 1980 
\citep[and references therein]{bregman1985,kollgaard1995}.
The target was then re-observed by {\it ROSAT} 
(0.1--2.4 keV), during both the {\it ROSAT} All Sky Survey (RASS) 
and a pointed observation in 1992 \citep[and references\ therein]{kollgaard1995}, 
and was detected in a possibly lower flux level.
Conclusive spectral information could be extracted neither from 
the {\it Einstein} nor from the {\it ROSAT} data.
Finally, the source was the target of a pointed observation by {\it XMM-Newton} 
(0.5--10 keV) in 2004 \citep{guainazzi2006}: the source was detected with a 
$2-10$ keV luminosity of $2.3\times10^{44}$ erg \persec.
the data spectral fitting revealed a spectrum with slope 
$\alpha=0.43^{+0.20}_{-0.19}$ and a relatively high column density 
($N_H=1.0^{+0.5}_{-0.4} \times 10^{22}$~\percmq), whereas the timing analysis
showed the existence of a flux variable of a factor $\sim$3 on time-scales 
of a few ks.
From the comparison of the {\it Einstein}, {\it ROSAT}, and {\it XMM} 
1-keV flux densities
a soft--X-ray flux decline of 
a factor $\gtsim$ 3 over $\sim$10 years emerged: however,  due to the quality 
of the X-ray data, it was not possible to ascribe those variations to 
a change of the column density of the intervening absorber \citep{guainazzi2006}.
Based on the detection of rapid X-ray variability in the XMM data set, 
\citet{guainazzi2006} constrained, through the light-travel argument, 
the size of the X-ray emitting region to $\ltsim$10 $\mu$pc,
concluding that the X rays are produced either at the base of the jet or 
in the accretion disk; the lack of excess soft X-ray emission is also brought by the 
authors in support of a scenario in which the hotspots are X-ray silent.

\subsubsection{References for spectrum and SED of Fig. 1} 
\citet{jauncey1970}; 
\citet{kapahi1981};  
\citet{ulvestad1981}; 
\citet{peacock1981b}; 
\citet{perley1982}; 
\citet{rudnick1982}; 
\citet{hutchings1994}; 
\citet{heckman1994};  
\citet{taylor1994}; 
\citet{kollgaard1995};  
\citet{kovalev1999}; 
\citet{terasranta2001}; 
\citet{zensus2002}; 
\citet{hirabayashi2000};  
\citet{waldram2003};
\citet{terasranta2005}; 
\citet{beichman1981}; 
\citet{stickel1994}; 
\citet{snellen1996}; 
\citet{guainazzi2006}; NED.


\subsection{IERS B0108+388 (COINS J0029+3456; OC +314)}

This GHz-peaked radio source, very powerful in the radio band  
($P_{\rm 5\,GHz}=10^{27.3}$ \wattperhz), has been classified as a CSO. It  
displays a nucleus with inverted spectrum, and twin jets with steeper 
spectra extending 3 mas ($\sim$20 pc) from the nucleus
toward the NE and SW \citep[][and references therein]{carilli1998}.
High-resolution imaging of the source was first performed
at 5 GHz as part of the PR VLBI survey \citep{pearson1988},
showing the source to have a simple double structure.
\citet{conway1994} and \citet{taylor1996}, with more sensitive 
and higher resolution images at higher frequencies, revealed  
a steeply inverted core connected to both outer components 
by a faint chain of components. Overall, IERS B0108+388 has an ``S''
symmetry, a relatively common feature among CSOs \citep{readhead1996}.
Unlike most CSOs, IERS B0108+388 also displays a large-scale 
jet with steep spectrum, extending about 25$''$ ($\sim$170 kpc) to the east 
of the nucleus \citep{baum1990}.
Proper motions enable an estimate of the kinematic age of $\sim$400 
yr \citep{owsianik1998,polatidis2003}.
The radio flux density is weakly polarized ($0.30\pm0.08$ at 4.8 GHz);
although it was claimed not to vary significantly \citep{aller1992},
a study by \citet{torniainen2007} revealed flux variations 
greater than a factor of $\sim$6
at some radio frequencies, 
and led the authors to not confirm the true GPS nature of the source.
The radio source is associated with a narrow emission line galaxy 
at z=0.66847 with R=22.0 mag 
\citep{stanghellini1993,stickel1996a,carilli1998},
and is located (in projection) within $0.5''$ ($\sim$3.4 kpc) of the galactic center.
Optical images of the host reveal a very red, diffuse, 
and slightly asymmetric galaxy (perhaps a face-on spiral), 
although the faintness of the galaxy makes the classification 
difficult \citep{stanghellini1993}.
There is no evidence for a strong point-source
contribution in the R-band image. The I-band
image is more compact, leading \citet{stanghellini1993} to
propose an increased contribution from the active nucleus
in the NIR, possibly indicating the existence of nuclear
obscuration toward the AGN in IERS B0108+388. 
Variability in the near IR, with a very red observed color ($\alpha_{IR-opt}\le -3$) 
during an IR maximum, was revealed by \citet{stickel1996a}.

By means of the $M_{\mathrm{R}}-M_{\mathrm{BH}}$ relation, 
\citet{wu2009b} estimated a BH mass of $7.94\times10^{7}M_{\sun}$.

In the X-ray band, the source was detected for the first time 
by {\it XMM-Newton} in 2004 \citep{vink2006}, with a luminosity of 
$1.18\times10^{44}$ erg \persec.
The source counts were not sufficient 
to independently estimate, in the fitting procedure, spectral index $\alpha$
and column density \NH. By fixing $\alpha=0.75$, \NHmath$=(57\pm 20)\times10^{22}$ \percmq 
was obtained. 

The detection of a strong \HI 21-cm absorption line, with a width of 100 km \persec 
and optical depth $\tau=0.44$, yields a neutral hydrogen column density 
\NHImath$= 80.7\times10^{20}$ \percmq, under the standard assumption 
of $T_{\mathrm{s}} \sim 100$ K and $c_f=1$ \citep{carilli1998}.
VLBI opacity maps recently allowed to prove that the mechanism responsible for 
the turnover of the radio spectrum is SSA in the core, 
whereas in the lobes FFA by ionized gas with a temperature 
$T\sim2000$ K \citep{marr2001}.

\subsubsection{References for spectrum and SED of Fig. 1} 
\citet{tinti2005}; \citet{dallacasa2000}; 
\citet{stickel1996a}; \citet{vink2006}; NED.


\subsection{IERS B0500+019  (PKS 0500+019; OG +003)}
Observed in the radio band since the 1970's, and well known as a convex-spectrum
radio source \citep[e.g.,][]{jauncey1970}, 
this very powerful ($P_{\rm 5\,GHz}=10^{27.3}$ \wattperhz)
object was classified as a GPS source by \citet{odea1991}.
The GPS nature of the source was recently confirmed, on the basis of 
flux variability and spectral shape, by \citet{torniainen2007}.
VLBI, VLBA and VSOP images show a radio structure displaying a symmetric, 
S-shaped morphology, as other GPS sources do, dominated by components similar 
to jets and/or micro-lobes, with the possible presence of a weak core, 
and with the northern part brighter than the southern part 
\citep{stanghellini1997,stanghellini2001,fey2000,fomalont2000}.
The total extension of the radio sources is $\sim$15 mas ($\sim$96 pc).
The total fractional polarization is below 0.1\% at 5 GHz \citep{stanghellini1998},
although polarized emission at the $\sim0.3\%$ level is detected in the brightest 
component \citep{stanghellini2001}.

The optical host of the radio source was the subject of a long debate:
originally identified with an unresolved QSO lying 2$''$ ~north of a 
galaxy with $m_r=21.5$ and with extremely red optical spectrum \citep[$\alpha_{RB}= 7.1$, 
with $F_{\nu}\propto \nu^{-\alpha}$;][]{fugmann1988a,fugmann1988b}, it was subsequently 
associated with the southern, asymmetric galaxy, with total magnitude $m_R=20.7$
and dominated by a strong point-like source in the NIR band \citep{stickel1996a}.
Because this galaxy is part of an apparent group of more than 10 galaxies, the 
observed asymmetry might be generated by gravitational interactions with companion 
objects \citep{stickel1996b}.
The galaxy produces moderate-ionization narrow emission lines \citep{stickel1996b}, and is
located at $z=0.583$ \citep{devries1995}.
Based on the detection of a further, unidentified emission line, \citet{stickel1996b}
proposed that a background, reddened quasar is strongly aligned with the 
foreground galaxy.  
This scenario, however, was not confirmed by subsequent observations by
\citet{devries1998a,devries1998b,devries2000}. 
Besides confirming the galaxy at $z=0.583$ to be the host of the radio source, 
these authors revealed a degree of source nucleation much higher in the K band 
than in the J and H bands, contrary to the behavior of both the GPS sources of the same 
redshift and the FRIIs with the same degree of J-band nucleation;
this led the authors to suggest that the very red color of the nucleus 
is produced by an absorption excess in the host galaxy.
\citet{jackson2002} too identified the host as a narrow-line radio galaxy, 
providing a revised redshift of $z=0.584$ \citep{hook2003}.  

Whereas the radio emission of the source is fairly stable 
\citep{stanghellini1997,torniainen2007}, 
moderate variability was detected in both the optical and NIR bands 
\citep{stickel1996a}.

The source is also a MFIR emitter: we detected it in the field of an archival 
{\it Spitzer}/MIPS calibration observation at 24 $\mu$m (see Section \ref{sec_sample}, 
and Appendix \ref{app_B} for details).

In the X-ray regime, the {\it Einstein Observatory} observed the source 
in 1980, 
yielding however only a flux upper limit in the $0.33-4.64$ keV energy range 
\citep{ledden1983}.

The source was successfully detected by {\it ROSAT} 
during the RASS as well as in two pointed observations in 1993: although a unique 
spectral fit was not possible due to the paucity of low-energy counts, the spectrum 
was found to be consistent with either a significant excess absorption or a hard 
slope, even though thermal spectral models (however with poorly constrained temperature) 
could not be ruled out \citep[and refs.\ therein]{kollgaard1995}.
The source spectrum was better determined through the {\it XMM-Newton} observations in 
2004 \citep{guainazzi2006}: with a $2-10$ keV luminosity of
$5\times 10^{44}$ erg \persec, the source spectrum was well fitted with a 
slope $\alpha=0.62^{+0.21}_{-0.19}$ and \NHmath$=5.0^{+3.0}_{-2.0}\times 10^{21}$~\percmq.

An independent estimate of the column density of neutral hydrogen comes from the 
measurements of the redshifted 21-cm absorption lines, 
yielding \NHImath$=6.2\times10^{18}(T_{\mathrm{s}}/f)$ \percmq ~\citep{carilli1998}.
Based on the standard assumption of $T_{\mathrm{s}} \sim 100$ K and $c_f=1$, 
yielding \NHImath$=6.2\times10^{20}$ \percmq, \citet{guainazzi2006} conclude that 
a difference of a factor $\sim$10 between \NHI and \NH supports a source scenario 
in which the absorbed X-rays come from a region located well within
the lobe radio hotspots, completely X-ray silent.
Assuming a non-thermal radio-to-optical quasar spectrum for the source, 
\citet{carilli1998} derived, from the difference between the optical data and 
the extrapolation of the radio-to-infrared spectrum into the optical band, a lower 
limit to the rest-frame visual extinction, $A_V \ge 3$, implying $T_{\mathrm{s}} \ge 500$ K, 
and thus \NHImath$\ge 3.1\times10^{21}/f$ \percmq.
This limit would be largely consistent with the above-mentioned \NH value provided 
by the X-ray data analysis. 

\subsubsection{References for spectrum and SED of Fig. 1} 
\citet{devries1998b}; \citet{stickel1996a}; \citet{devries1995};   
\citet{fugmann1988b}; \citet{guainazzi2006}; this work ({\it Spitzer}/MIPS data); NED.


\subsection{IERS B0710+439 (COINS J0713+4349; OI +417)}
This source is a well known CSO \citep{taylor1996}, with radio power  
$P_{\rm 5\,GHz}=10^{27.1}$ \wattperhz.
The flux density is very weakly polarised ($< 0.15\% \pm 0.11\%$ at 5 GHz), 
and the observed variations of the flux are not statistically significant 
\citep{aller1992,torniainen2007}.
The known GPS nature of the source was recently confirmed, on the basis of 
flux variability and spectral properties, by \citet{torniainen2007}.

The total angular size of the source is 24.1 mas \citep{owsianik1998},
which corresponds to a projected linear size of $\sim$145 pc.  
VLBI maps at several frequencies (1.6, 5, 10.7, and 15 GHz) showed the
overall triple structure of the source. Despite having three components,
this source was provisionally classified as a compact double, based on 
the fact that more than 80\% of the emission came from two almost equally 
bright components \citep{pearson1988}. 
\citet{conway1992} argued that the two outer components were hotspots 
and minilobes, whereas the centre of activity was associated with the 
middle component, based on its compactness, spectrum and weak flux 
density variability. 
However, more recent multi-frequency observations reveal a
compact component with a strongly inverted spectrum at the
southern end of the middle component, suggesting that the true
centre of activity lies there \citep{taylor1996}.

\citet{readhead1996} gives age estimates of 1200-1800 yr based 
on synchrotron ageing, and 1500-7500 yrs from energy supply arguments;
more recently, the first detection of hotspot advance velocities 
in a CSO was measured in this source by \citet{owsianikconway1998};
these measurements, together with those by \citet{polatidis2003}, 
yielded an age smaller than 1000 yr.

In the optical band, IERS B0710+439 has been identified with a galaxy \citep{peacock1981a}
of magnitude $r=19.7\pm0.2$ \citep{peacock1981a}.
The galaxy displays a very irregular morphology, with large differences in the two bands,
and has a very red color \citep[$r-i=1.9$;][]{stanghellini1993}. 
There is evidence of strong interaction, and presence of a large amount of 
obscuring matter. The numerous objects around the source suggest that 0710+439 is the 
dominant galaxy of a cluster of galaxies \citep{stanghellini1993}. 
The emission-line redshift of the galaxy is z=0.518 \citep{lawrence1996};
the optical spectrum shows absorption lines characteristic of an evolved stellar
population, and an optical continuum shape typical of an elliptical
galaxy without any evidence for a non-stellar component.

By means of the $M_{R}-M_{\mathrm{BH}}$ relation, 
\citet{wu2009b} estimated a BH mass of $2.512\times10^{8}M_{\sun}$.

At high energies, the source was detected as an X-ray emitter by {\it XMM-Newton} in 2004 
\citep{vink2006}. Its $2-10$ keV luminosity is $2.16\times 10^{44}$ erg \persec,
and the spectral analysis yielded a slope $\alpha=0.59\pm0.06$ for the energy spectrum, 
and a source-intrinsic column density \NHmath$=(0.44\pm0.8)\times10^{22}$~\percmq.

\subsubsection{References for spectrum and SED of Fig. 1} 
\citet{stanghellini1993}; \citet{snellen1996}; \citet{odea1996};
\citet{devries1998a}; \citet{vink2006};  NED.


\subsection{PKS B0941-080}
The source has a radio power  $P_{\rm 5\,GHz}=10^{26.1}$ \wattperhz,
and an overall size of $\sim$50 mas ($\sim$ 177 pc) \citep{odea1998}. 
Imaging with the VLA shows a slightly resolved secondary component 20$''$ ($\sim$71 kpc)
east of the main one \citep{stanghellini1998}.
The VLBI morphology of the main component is that of a compact double \citep{dallacasa1998}
typical of many CSS/GPS radio galaxies. 
The source was classified as a GPS source \citep[][and references therein]{odea1998}, 
although recent investigation by \citet{torniainen2007} 
did not confirm the GPS-source classification, indicating the object as a steep-spectrum source.

Ground-based optical-NIR observations showed the host galaxy of the main radio source 
as an elliptical envelope, with magnitude $r=17.9$ and 
color $r-i=1$, including two cores \citep{stanghellini1993}. 
HST-NICMOS (JK) imaging revealed that the system is actually composed of two galaxies, 
interacting with each other, and characterized by blue nuclei \citep{devries2000}.
The GPS source is associated with the core of the northern, larger, and brighter 
galaxy \citep{stanghellini1993,devries2000}.
The prominent emission line spectrum allows one to derive, from seven identified 
spectral features, a redshift $z=0.228$.

By means of the $M_{BH}-\sigma$ relation, \citet{wu2009a} recently estimated 
a central BH mass of $5.01 \times 10^{7} M_{\sun}$.

In the X-ray regime, the source was pointed at by {\it Chandra} in 2002
\citep{guainazzi2006,siemiginowska2008}: 
the target was detected, and subsequent, independent analyses first 
failed in determining a significant spectral fit \citep{guainazzi2006}
(fixed $\alpha=1.0$; \NH undetermined), 
and then derived a steep-spectrum source ($\alpha=1.62^{+1.29}_{-1.03}$)
not absorbed by any excess hydrogen column \citep{siemiginowska2008}.
The latter analysis also reported the detection of the eastern source component
revealed by the VLA, although with very low source counts.
The 2--10 keV X-ray luminosity of PKS B0941-080 is the lowest of our GPS-galaxy
sample \citep[$9\times10^{41}$ erg \persec;][]{guainazzi2006}.

\subsubsection{References for spectrum and SED of Fig. 1} 
\citet{devries1998a};  \citet{stanghellini1993}; 
\citet{guainazzi2006}; \citet{siemiginowska2008}; NED.


\subsection{IERS B1031+567 (COINS J1035+5628; OL +553)}

This source, with radio power $P_{\rm 5\,GHz}=10^{26.9}$ \wattperhz,  
is a known GPS source \citep[][and references therein]{odea1998}, and was 
also included in the sample of bona-fide GPS sources assembled by 
\citet{torniainen2007}. 

The source was found to have a compact double morphology in 5 GHz observations by 
Pearson \& Readhead \citep{taylor1996}. VLBA observations at 8.4 and 15.4 GHz 
showed the source to be dominated by two leading edge-brightened, steep-spectrum  
components  \citep{taylor1996}, proposed by the authors to be the working 
surfaces and lobes of two oppositely directed jets. 
Based on these data, the authors concluded that the source was likely to be a CSO, 
with the center of the activity located in the source center.
The CSO morphology was later confirmed by VSOP observations \citep{fomalont2000}.
The overall source size is $\sim$40 mas ($\sim$224 pc). 

Measurements of the hotspot separation velocity, based however only on two epochs,
yield estimates of $\sim$600--1800 yr \citep{taylor2000,polatidis2003}.

In the optical band, the source is identified with an elliptical-like galaxy
with magnitude $r=20.2$, showing
a possible asymmetric morphology at low brightness levels; the presence of 
weak resolved objects in the field suggest that this galaxy is the dominant 
member of a galaxy cluster.

By means of the $M_{\mathrm{R}}-M_{\mathrm{BH}}$ relation, 
\citet{wu2009b} estimated a BH mass of $1.259\times10^{8}M_{\sun}$.

In the X-ray band, the source was detected for the first time in 2004 
by {\it XMM-Newton} \citep{vink2006}, with a $2-10$ keV luminosity of $2.2\times10^{43}$  
erg \persec. 
The source counts were not sufficient to independently estimate, in the fitting 
procedure, spectral index $\alpha$ and column density \NH. By fixing 
$\alpha=0.75$, \NHmath$=(0.5\pm0.18)\times10^{22}$~\percmq was obtained. 

\subsubsection{References for spectrum and SED of Fig. 1} 
\citet{stanghellini1993},\citet{odea1996}; \citet{snellen1996}; 
\citet{devries1998a}; \citet{vink2006}; NED.


\subsection{IERS B1345+125 (PKS 1345+12; 4C +12.50)}
This is a luminous, compact 
($\sim$170 pc) 
radio source hosted by an elliptical-like galaxy at $z=0.12174$,
classified as an ultra-luminous infrared galaxy (ULIRG).

The main host-galaxy body displays the same shape from NIR 
to near-UV wavelengths, is $\sim$15$''$ ($\sim$32 kpc) wide and rather 
asymmetric, and appears to be associated with 
a group of 15 fainter nebulous objects, many of which are located very close to its 
distorted halo \citep{heckman1986}.
Ground-based and HST observations clearly reveal two nuclei separated 
by $\sim$1.8$''$ ($\sim$3.8 kpc)
embedded in a common, distorted envelope, including a strongly curved 
tidal tail \citep[e.g.,][]{gilmore1986,heckman1986,smith1989,labianoetal2008}.

The consistency of the inner ($10''$) NIR brightness profile with 
that of a merger remnant, and the little evidence of extinction, indicative 
of ISM stripping, in the very central region \citep{shaw1992}, together with 
the detection of a rich young stellar population (YSP) \citep{zaurin2007}, 
confirmed that this complex has a merger origin, as firstly proposed by 
\citet{gilmore1986} to explain the double nucleus and the asymmetry of 
the large-scale structure.

The south-eastern (SE) nucleus is believed to belong to an elliptical galaxy
\citep{gilmore1986}, and its NIR color is consistent with reddened starlight 
\citep{evans1999}. 
The north-western (NW) nucleus displays a narrow-line Seyfert-2 
optical spectrum \citep{gilmore1986},
although the presence of a buried quasar is suggested by the detection of NIR broad 
Pa$\alpha$ lines \citep{veilleux1997,veilleux1999}, by the similarity 
of the extreme NIR color 
with that of quasars with a warm (500$-$1000 K) dust component 
\citep{evans1999,surace1999}, and by the highly-polarized UV emission 
\citep{hurt1999}; stellar velocity dispersion derived from NIR VLT spectra 
indicate the presence of a $6.54 \times 10^{7} M_{\sun}$ BH \citep{dasyra2006}.

The source, detected by {\it IRAS} and {\it ISO} 
as a bright MFIR emitter \citep{golombek1988,fanti2000}, was
recently observed in the MFIR domain by {\it Spitzer}, 
with both IRAC and MIPS (see Section \ref{sec_sample}, and Appendix \ref{app_B} 
for details).

Although the optical counterpart of the radio source was originally identified with
the SE nucleus \citep{gilmore1986}, newer astrometry led to the 
association of the radio activity to the NW nucleus 
\citep{stanghellini1993,evans1999,axon2000,batcheldor2007}. 
The latter identification is adopted throughout this paper.

The powerful \citep[$P_{\rm 5\,GHz}=10^{25.94}$ \wattperhz ;][]{odea1998} 
radio source is resolved neither by 
MERLIN (at 18 cm) nor by VLA (at 6 and 2 cm) observations by \citet{spencer1989}, 
which constrained the (projected) linear size to $<0.1''$ ($<213$ pc).
VLBI images at 5 GHz show a single jet disrupting at $\sim$30 mas ($\sim$60 pc) 
south of the core, and then expanding in a diffuse $\sim$40-mas-sized lobe reaching a 
distance of $\sim$80 mas ($\sim$170 pc) from the core \citep{stanghellini1997,shaw1992}. 
The puzzling lack of any counter-lobe at these frequencies \citep{stanghellini1997,
stanghellini2001}, together with the observation of highly-polarized components and 
superluminal motion \citep{lister2003} make the source an unusual example of CSO.
A more symmetric structure is however detected by the VLBI at lower frequencies 
($\sim 1266$ MHz), where a fainter counter-jet/lobe extending toward the 
North for $\gtsim$40 mas can be identified \citep{morganti2004b}.
The radio structure is misaligned by 
$\sim$45$\deg$ with respect to the UV light distribution \citep{axon2000}.

The source was shown to be non-variable at 318 and 430 MHz over a 14-year 
period of flux-density monitoring with the Arecibo telescope \citep{salgado1999}.
Although classified as a GPS source \citep[][and references therein]{odea1998}, a recent study
by \citet{torniainen2007} did not confirm the GPS nature of the object, 
classifying the source as a steep-spectrum source.

In the X-ray domain, IERS B1345+125 was first revealed by {\it ASCA} (2--10 keV) in 1996 
\citep{odea2000}, after a previous non-detection by {\it ROSAT} 
\citep[0.2--2 keV;][]{odea1996}, and was the first GPS source to be 
reported as an X-ray emitter.
The {\it ASCA} image does not resolve the two nuclei. However, a comparison of 
the X-ray power with the H$\alpha$ emission-line luminosity makes 
\citet{odea2000} conclude that the X rays come from the western nucleus, 
associated with the GPS source.
Highly uncertain spectral index ($\alpha=0.6^{+1.2}_{-0.8}$) and 
column density (\NHmath$=4.2^{+4.0}_{-2.4} \times 10^{22}$ \percmq) 
are derived from the modeling of these X-ray data.
Higher-quality X-ray data were obtained in 2000 with {\it Chandra} \citep{siemiginowska2008},
which also revealed the presence of extended emission on $\sim$10 arcsec (20 kpc) scale.

The detection of CO(1$\rightarrow$0) emission implies a mass of molecular 
hydrogen $M(H_2)\sim(3.3-6.5)\times10^{10}M_{\sun}$ 
\citep{mirabel-etal1989,evans1999},
concentrated in the central $\sim$2 kpc of the galaxy.

H{\sc i} absorption was detected against the radio source
by \citet{mirabel1989}, showing a narrower (FWHM$\sim$200 km\persec) and 
a broader (FWHM$\sim$700 km\persec) component,
and implying an average \NHImath=$6.2\times10^{18}(T_{\mathrm{s}}/f)$ \percmq.
More recent measurements by \citet{morganti2004a,morganti2004b} revealed
a much wider ($\sim$2000 km\persec) broad line, blueshifted with respect 
to the systemic velocity of the galaxy. 
\citet{morganti2004b} also assessed, thanks to the spatial resolution of the VLBI,  
the actual location of the H{\sc i} absorber, 
showing that a single H{\sc i} cloud of (projected) size $\gtsim 22\times 65$ pc, 
with mass $M\sim (10^5-10^6) M_{\sun}$, column density \NHImath$\sim 10^{22}(T_{\mathrm{s}}/100K)$ \percmq, 
and located at the edge of the northern jet/lobe (projected distance of 40--100 pc from 
the core), can account for the whole unresolved H{\sc i} absorption. This would prove the clumpy 
nature of the ISM in the nuclear region of this source, making the scenario of 
a strong jet-cloud interaction a viable interpretation of the observations of not only 
this galaxy, but also of other young radio sources \citep{morganti2004b}.
\citet{holt2003} found complex optical emission line profiles at the 
position of the nucleus with line width of $\sim$2000 km\persec and blueshifted with respect
to the narrow HI absorption component.

Whether the GPS source is truly young is still a matter of debate.
A model-dependent estimate by \citet{odea2000} implies the source to be 
younger than $3\times10^4$ yr. However, \citet{zaurin2007} find no clear 
evidence of a time delay of the AGN activity with respect to the major 
merger-induced starburst ($<$6 Myr ago); this would also be in agreement 
with the apparent connection between the extended 
\citep[$\gtsim$100 kpc;][]{stanghellini2005} and compact 
radio emission, indicating recursive AGN activity on a time-scale longer than 
previously thought.
\citet{lister2003} speculate the existence of a precessing BH to explain the 
S-shaped jet; if confirmed, this would point to an age of the radio source 
$<$10$^5$ yr.

\subsubsection{References for spectrum and SED of Fig. 1} 
\citet{pushkarev2005}; \citet{stanghellini1993};         
\citet{snellen1996}; \citet{odea1991};               
\citet{golombek1988}; \citet{devries1998a};                            
\citet{fanti2000}; \citet{scoville2000}; 
\citet{odea2000}; \citet{siemiginowska2008};
this work ({\it Spitzer}/IRAC, and {\it Spitzer}/MIPS data); NED.
                   

\subsection{IVS B1358+624 (COINS J1400+6210; 4C +62.22)}
This source has radio power $P_{\rm 5\,GHz}=10^{26.9}$ \wattperhz;
it is one of the 11 low-frequency variable sources mapped with global VLBI 
at 608 MHz by \citet{padrielli1991}, who showed the object to be a compact double 
separated by 46 mas with a possible bridge of emission. 
\citet{dallacasa1995} confirm that the two lobes are connected by a 
knotty, broad jet; \citet{taylor1996} provide new evidence of the
source structure: they resolve the two lobes, and detect a long, one-sided 
jet with a faint, compact component at the base; based on the compactness 
of the component, on its inverted spectrum, and on its location, the authors 
confirmed the source to be a member of the CSO class, with the central 
component being the center of the activity. No evidence of hotspots 
on either sides of the central engine was found \citep{taylor1996}.
The overall source size is $\sim$70 mas ($\sim$380 pc).

The convex, stable, and GHz-peaked spectrum enabled the classification of the source as 
a GPS source \citep{odea1998,torniainen2007}.

An optical image of this source \citep{stanghellini1993} shows that the host is a galaxy 
of $r=19.8$ mag  with a blue color ($r-i=0.3$) and a boxy morphology; there is evidence  
of obscuring material in a direction (NE-SW) roughly perpendicular 
to the VLBI axis, but on larger scale \citep{dallacasa1995}.

In the X-ray band,  {\it XMM-Newton} detected the source in 2004
as an X-ray emitter with $2-10$ keV luminosity of $1.67\times 10^{44}$ erg \persec
\citep{vink2006}. 
The fitting procedure yielded a spectral index $\alpha=0.24\pm0.17$ and a total-hydrogen column 
density \NHmath$=(3.0\pm0.7)\times10^{22}$~\percmq. 

By means of the $M_{\mathrm{R}}-M_{\mathrm{BH}}$ relation, 
\citet{wu2009b} estimated a BH mass of $1.585\times10^{8}M_{\sun}$.

An independent estimate of the amount of neutral hydrogen comes from the 21-cm observations 
by \citet{vermeulen2003}, who detected a column density \NHImath$=1.88\times10^{20}$~\percmq,
under the usual assumption of $T_{\mathrm{s}}=100$ K and uniformly covered source.

\subsubsection{References for spectrum and SED of Fig. 1} 
\citet{devries1998a}; \citet{stanghellini1993}; 
\citet{odea1996}; \citet{snellen1996}; \citet{vink2006}; NED.


\subsection{IERS B1404+286 (MRK 0668; OQ +208)}
Observed in the radio band since the 1970's, this compact radio source is one 
of the closest bright ($P_{\mathrm{5 GHz}}=10^{25.4}$ \wattperhz) 
GPS galaxies.
Its radio spectrum is convex, stable, and GHz-peaked, and the classification of the 
object as a GPS source \citep{odea1998} was confirmed also by \citet{torniainen2007}.

The radio source displays a CSO morphology \citep{stanghellini2002}: 
it has a weak core at 15 GHz, two-sided faint and short jets, and 
two mini-lobes located 10 mas apart, giving a projected size of less 
than 10 pc.
The two micro hot-spots of this radio source seem to increase their
distance at a rate of about $0.3c$, comparable to what already found
in other CSOs, yielding a kinematical age of a few centuries
\citep{stanghellini2002}.

Based on VSOP observations, \citet{kameno2000} reported a very inverted
spectrum below the peak frequency of the weaker SW region
of this source. They considered the extremely inverted spectrum 
inconsistent with SSA, and propose a process of FFA to explain the 
radio spectral shape. 
\citet{xiang2002} confirmed the incompatibility of the spectrum with 
pure SSA result, on the basis of ground-based VLBI observations.

The host galaxy of the radio source, located at $z=0.077$, has $m_R=14.6$
and color $r-i=0.2$;
companions in the galactic envelope, and a 
tail of low brightness
emission in N-S direction, suggest that the galaxy is
dynamically disturbed \citep{stanghellini1993}.
The optical spectrum is that of a Seyfert 1; however, given the power of 
the associated radio source, it was classified as a broad line
radio galaxy \citep[e.g.,][and references therein]{stanghellini1997}.

{\it IRAS} detected IERS B1404+286 as a bright MFIR emitter \citep{golombek1988};
recently, the target was re-observed in the MFIR domain by {\it Spitzer}, 
with both IRAC and MIPS (see Section \ref{sec_sample}, and Appendix \ref{app_B} 
for details).

By means of the $M_{\mathrm{R}}-M_{\mathrm{BH}}$ relation, 
\citet{wu2009b} estimated a central BH mass of $5.012\times10^{8}M_{\sun}$.

In the X-ray domain, the source was observed by ASCA 
and subsequently by {\it XMM-Newton} \citep[][and references therein]{guainazzi2004}:
it displayed the typical features of obscured AGN, despite the optical classification 
as a Seyfert 1; the detection of a prominent iron $K_{\alpha}$ emission line 
strongly suggested a Compton-reflection scenario for the X-ray emission, but the spectrum 
could be satisfactorily fitted by several different models, and a contribution from 
upscattered IR radiation off the lobes could not be ruled out by the authors.  
Depending on the assumed model for the X-ray spectrum, the luminosity of the source
in the 2--10 keV range varies from $4\times10^{42}$ erg \persec to $>9\times10^{43}$ erg \persec.

\subsubsection{References for spectrum and SED of Fig. 1} 
\citet{tinti2005}; \citet{dallacasa2000}; 
\citet{knapp1990}; \citet{stanghellini1993}; \citet{guainazzi2004};
this work ({\it Spitzer}/IRAC, and {\it Spitzer}/MIPS data); 
NED.


\subsection{IERS B2128+048 (PKS 2127+04; OX +046)}
This is the most radio powerful source of our sample, with $P_{\mathrm{5 GHz}}=10^{27.8}$ 
\wattperhz.
Early VLBI observations at 18 cm were made by \citet{hodges1984}:
they found the radio source composed of two resolved components separated
by 29 mas (225 pc). More recent VLBI mapping at 6 cm by \citet{stanghellini1997}
revealed a radio emission extending for $\sim$35 mas in NW-SE direction, 
with the typical scaled-down structure of a Classical Double. At the edges,
there are two hotspots/lobes; the central, weak component is assumed to be 
the core; a jet connects the core to the northern lobe. 
At 15 GHz, the VLBA image shows an aligned structure with four main
components, the outer of which are more resolved than the inner ones, 
although the southernmost is rather weak \citep{stanghellini2001}:
based on morphology, the source is classified as a CSO, with the 
hotspots being associated with the outer components.

The source integrated radio spectrum is convex, fairly stable, and GHz-peaked,
and the classification of the object as a GPS source \citep[][and references therein]{odea1998} 
was confirmed also recently by \citet{torniainen2007}.

The optical counterpart of this radio source is a very faint ($r=23.3$) and red 
($r-i=1.85$) galaxy \citep{biretta1985,devries2000}, 
located at $z=0.99$ \citep{stickel1994}. The redness of the spectrum is confirmed by
\citet{devries2000}, who also reported the remarkable brightness in the NIR K band.

By means of the $M_{\mathrm{R}}-M_{\mathrm{BH}}$ relation, 
\citet{wu2009b} estimated a mass of $1.698\times10^{8}M_{\sun}$ for the central BH.

In the X-ray regime, the source was pointed at by {\it Einstein} in 1979 \citep{ledden1983},
but the low photon counts only enabled the estimate of an upper limit to its flux 
in the  0.23-3.77 keV band.
The source was then successfully observed by {\it Chandra} in 2002 \citep{guainazzi2006,siemiginowska2008}
with a 2-10 keV luminosity of $4.4 \times 10^{44}$ erg \persec;  the data spectral fit gave a slope 
$\alpha=0.5^{+0.6}_{-0.7}$ and a column density \NHmath=$3.0^{+8.1}_{-3.0}\times10^{21}$~\percmq.

\subsubsection{References for spectrum and SED of Fig. 1} 
\citet{devries2000};  \citet{snellen1996}; \citet{guainazzi2006}; \citet{siemiginowska2008}; NED.


\subsection{IERS B2352+495 (COINS J2355+4950; OZ +488)}

This radio galaxy has a radio power $P_{\rm 5\,GHz}=10^{26.3}$ \wattperhz.
It was imaged at 4.99 GHz  by \citet{pearson1988} and \citet{conway1992}, 
who reported on a complex structure surrounding a compact core:
two distinct components were found, one with a simple structure, and the 
other with a complex structure.
Imaging at 1.67 GHz by \citet{wilkinson1994} also showed a two-sided morphology, 
which led the authors to classify the source as a CSO.
No radio structure on scales larger than $0.2''$ (7 kpc) has been detected for
this source \citep{perley1982}.
Consistently with \citet{pearson1988} and \citet{conway1992},
VLBI imaging at 4.99 GHz by \citet{fey1996} confirmed the complexity of the 
source structure, revealing multiple components dominated by a compact core,
a region of $\sim$2 mas ($\sim$7 pc);
two components within the central core, with flux ratio 2:1,  
contain $\sim$75\% of the total flux density at 4.99 GHz. 

The source exhibits weak variability (15\%) on time scales of a few 
months \citep{seielstad1983}
and 0.2\%-0.7\% polarization at 5 GHz \citep{perley1982}
The radio spectrum is convex and GHz-peaked, and the object's classification as 
a GPS source \citep{odea1998} was confirmed also by \citet{torniainen2007}.

The source is hosted by a galaxy with redshift $z=0.237$ \citep{odea1991}.
In the NIR band, the source is one of the few displaying a higher 
nucleation (lower extended emission) in the J band than in the H and K bands 
\citep{devries1998a}.

\citet{wu2009a} and \citet{wu2009b} recently estimated 
a central BH mass of $1.58\times 10^{8} M_{\sun}$ and $1.698\times10^{8}M_{\sun}$
by means of the $M_{BH}-\sigma$ and $M_{\mathrm{R}}-M_{\mathrm{BH}}$, respectively.

IERS B2352+495 was firstly observed in the X-rays by {\it ROSAT} and {\it ASCA}, being detected 
by neither of them \citep{odea2000}.
Observations by {\it XMM-Newton} in 2004 did reveal the source as a $2-10$ keV emitter 
with a moderate luminosity of $4.6\times 10^{42}$ erg \persec \citep{vink2006} .
The photon statistics was not good enough to independently estimate spectral index $\alpha$
and column density \NH. By fixing $\alpha=0.75$, \NHmath$=(0.66\pm0.27)\times10^{22}$~\percmq 
was obtained. 

A search for 21-cm absorption successfully detected two features, yielding  
\NHI of 0.28$\times10^{20}$~\percmq  and 2.56$\times10^{20}$~\percmq \citep{vermeulen2003},
assuming $T_{\mathrm{s}}=100$ K and unitary covering factor.
Recently, \citet{araya2010} performed high angular resolution \HI measurements of the source, 
detected the two above absorption features against the radio jet, and associated the broader 
of them with a circumnuclear gas cloud. 

\subsubsection{References for spectrum and SED of Fig. 1} 
\citet{devries1998a}; \citet{snellen1996}; \citet{vink2006}; NED.


\section{ANALYSIS OF THE {\it SPITZER} DATA}
\label{app_B}

\subsection{Analysis of the IRAC data} 
We analysed the IRAC archival data of IERS B1345+125 and IERS B1404+286 in the 
following way.
The pipeline calibrated BCD files were recombined with MOPEX rebinned to 1/5th
the native pixel size of 1.2$''$  in an attempt to separate out 
contribution from surrounding diffuse emission seen in higher resolution 
NIR images. The smallest practical aperture ($r=2.4''$) was used to extract fluxes 
most indicative of the central point source, with aperture corrections applied. 
An uncertainty of roughly 3$\%$ in the absolute
calibration \citep{rea05} was added in quadrature with measurements of the
uncertainties due to background fluctuations and the Poisson noise.

Special care was taken for a proper background evaluation.
In the case of IERS B1404+286, in order to avoid the diffraction spike 
from a nearby point source $\sim$83$''$ to the SW of the GPS source,  
half of an $r=14.4''-24''$ annulus was used to measure the background.
In the case of IERS B1345+125, due to obvious contamination from field sources, 
the background was estimated from circular apertures in source-free regions.

To quantify contributions from extended emission surrounding the point
source in the IRAC images, corresponding fluxes in a larger ($r=12''$)
aperture were measured. 
In IERS B1345+125, this formally resulted in 30$\%$, 16$\%$,
6$\%$, and 6$\%$ larger fluxes than those in the original $r=2.4''$
aperture at 3.6, 4.5, 5.8, and 8.0 $\mu$m, respectively. This confirmed the
presence of diffuse emission surrounding the point source (already
noticeable by eye) in the two shortest wavelength
(highest-resolution) bands, but we consider any similar evidence 
in the longer wavelength images to be marginal at best due to 
the large aperture corrections ($37\% -57\%$) applied to
the small aperture fluxes.

A similar test of IERS B1404+286 hints at an additional (5$\%$) 
extended flux in the 3.6 $\mu$m image, but the contribution is $< 2-3 \%$ 
in the other images.

\subsection{Analysis of the MIPS data} 
We analysed the MIPS archival data of IERS B1345+125, IERS B1404+286, and IERS B0500+019
as follows.

Using a standard $r=35''$ ($r=30''$) circular aperture and a background 
annulus of 40$''$--50$''$ (40$''$--60$''$) for the 24 $\mu$m (70 $\mu$m) 
image, we measured aperture-corrected source fluxes on the pipeline calibrated PBCD
mosaic images. The flux conversion utilized the latest
(S18)\footnote{http://ssc.spitzer.caltech.edu/mips/calib/conversion.html}
conversion factors of 0.0454 and 702 MJy/sr per instrument unit, with
absolute uncertainties of 4$\%$ and 7$\%$, respectively. As in the IRAC
data, these systematic uncertainties were added in quadrature with the
background and counting noise. 
Additionally, the 
aforementioned point source 83$''$ from IERS B1404+286 is unresolved from
the GPS source in the
lower resolution IRAS images, so presumably contaminates the older
measurements (we measured 117 $\pm$ 5 mJy at 24 $\mu$m).



\clearpage
\begin{landscape}
\begin{deluxetable}{lcccccccccc}
\tabletypesize{\scriptsize}
\tablewidth{0pt}
\tablecaption{Physical parameters of the radio sources and X-ray luminosities\label{tab_data}}
\tablehead{
\colhead{Source} & 
\colhead{$z$\tnm{a}} & 
\colhead{Angular} & 
\colhead{Linear} & 
\colhead{Scale} &
\colhead{log($P_{\rm 5 GHz}$)} & 
\colhead{$\nu_{\mathrm{p,obs}}$\tnm{c}} & 
\colhead{~~~~~~$\nu_{\mathrm{p,intr}}$\tnm{d}~~~~~~} &
\colhead{v$_{\mathrm{h,sep}}$\tnm{e}} & 
\colhead{Age\tnm{f}} & 
\colhead{$L_{X,\, 2-10\, \mathrm {keV}}$}\\ 
\colhead{name} & 
\colhead{~~} & 
\colhead{Size\tnm{b}} & 
\colhead{Size} & 
\colhead{~~} &
\colhead{~~} & 
\colhead{~~} & 
\colhead{~~} &
\colhead{~~} & 
\colhead{(kin./spec.)} & 
\colhead{~~}\\ 
\colhead{~~} & 
\colhead{~~} & 
\colhead{[mas]} & \colhead{[pc]} & 
\colhead{[pc/mas]} & 
\colhead{log(W Hz$^{-1}$)} &
\colhead{[MHz]}  & 
\colhead{~~~~~~[MHz]~~~~~~}  & 
\colhead{[$c$]}  & 
\colhead{[yr]}   & 
\colhead{[$10^{44}$ erg s$^{-1}$]} 
}
\startdata
IERS B0026+346  & 0.517   & 35 & 211.4 & 6.04 & 27.1\tnm{g} &  630.96\tnm{1} & 957.17 & \nodata &  \nodata &  1.84\tnm{2}\\
IERS B0108+388 & 0.66847 & 6\tnm{3} & 40.9 & 6.82 & 27.2\tnm{3} & 3900$-$4760\tnm{4,5;6} & 6510$-$8340 & $0.25\pm0.04$\tnm{7} & $367\pm48$\tnm{8}(k) & 2.00\tnm{9}\\
~~               & ~~      &  ~~   & ~~   & ~~   & ~~          & ~~                     & ~~        &  ~~~         &   $310\pm 70$\tnm{7}(k) &  ~~\\
~~               & ~~      &  ~~   & ~~   & ~~   & ~~          & ~~                     & ~~        &  ~~~         &   417\tnm{10}(k) &  ~~\\
IERS B0500+019  & 0.58457 & 15\tnm{3} & 96.3 & 6.42 & 27.1\tnm{3} & 2000\tnm{4} & 3169.14 &  \nodata & \nodata & 4.95\tnm{2}\\ 
IERS B0710+439  & 0.518 & 25\tnm{3} &  151.1 & 6.04 & 27.0\tnm{3} & 1500$-$1900\tnm{4,5} & 2277$-$2884 & $0.37\pm0.17$\tnm{7} & $550\pm160$\tnm{7}(k) &  2.15\tnm{9}\\
~~               & ~~      &  ~~      & ~~   & ~~   & ~~          & ~~                     & ~~        &  ~~~                      & 932\tnm{10}(k) &  ~~\\
~~               & ~~      &  ~~      & ~~   & ~~   & ~~          & ~~                     & ~~        &  ~~~                      & 1200$-$1800\tnm{11}(s) &  ~~\\
PKS B0941-080   & 0.228 & 50\tnm{3} & 177.4  & 3.55 & 26.0\tnm{3} & 500\tnm{4,5} & 614   &  \nodata & \nodata  & 0.00578\tnm{2}, 0.005\tnm{12}\\
IERS B1031+567   & 0.450 & 40\tnm{3} & 224.0 & 5.60 & 26.8\tnm{3} & $1300-1400$\tnm{5,4} & 1890$-$2030 & $0.68\pm0.15$\tnm{7} & $620\pm140$\tnm{7}(k)  & 0.33\tnm{9}\\
~~               & ~~    & ~~        & ~~    & ~~   & ~~          & ~~                      & ~~          & ~~                    & 1836\tnm{10}(k)         & ~~ \\
IERS B1345+125    & 0.12174 & 80\tnm{h,3} & 170.4 & 2.13 & 25.9\tnm{3} & 600\tnm{5,13}  & 673 &  \nodata & \nodata  &  0.40\tnm{14}, 0.41\tnm{12}\\
IVS B1358+624  & 0.4310 & 70\tnm{3} & 382.3 & 5.46 & 26.8\tnm{3} & 500\tnm{4,5}         & 715 & \nodata & \nodata  & 2.538\tnm{9}\\
IERS B1404+286  & 0.07658 & 7\tnm{3} & 9.9  & 1.41 & 25.4\tnm{3} & 4900$-$5340\tnm{15,6}  & 5280$-$5750 & $0.28\pm 0.11$\tnm{16} & $320\pm210$\tnm{17}(k)  &  0.038 -- $>$0.9\tnm{18} \\
~~~             & ~~ & ~~ & ~~ & ~~ & ~~ & ~~~ & ~~ & ~~ & 224\tnm{10}(k)  &  ~~ \\
IERS B2128+048  & 0.990 & 35\tnm{1} & 271.9 & 7.77 & 27.8\tnm{3} & 600$-$800\tnm{4,5} &  1194$-$1592 & \nodata & \nodata  & 4.20\tnm{2}, 3.21\tnm{12}\\

IERS B2352+495   & 0.23790 & 50\tnm{3} & 183.2 & 3.66 & 26.2\tnm{3} & 700$-$900\tnm{5,4} & 870$-$1110  &  $0.40\pm0.13$\tnm{7} & $1200\pm400$\tnm{7}(k) & 0.0488\tnm{9}\\
~~               & ~~      &  ~~      & ~~   & ~~   & ~~          & ~~                      & ~~        &  ~~~                  & 3003\tnm{10}(k) & ~~\\
\enddata
\tablenotetext{a}{Redshift, taken from NED.}
\tablenotetext{b}{Overall angular size of the radio source, unless otherwise stated.} 
\tablenotetext{c}{Observed turnover frequency (or frequency range) of the radio spectrum}
\tablenotetext{d}{Intrinsic (source-frame) turnover frequency (or frequency range) of the radio spectrum}
\tablenotetext{e}{Hotspot separation velocity, derived from the projected angular velocities.} 
\tablenotetext{f}{Source age, estimated by means of kinematic (k) and/or spectral ageing (s) methods.}
\tablenotetext{g}{Derived from $F_{\rm 5 \, GHz}$.}
\tablenotetext{h}{The size refers to the one-side, southern radio jet.} 
\tablerefs{(1) \citet{snellen1996}; (2) \citet{guainazzi2006}; (3) \citet{odea1998}; (4) \citet{devries1997};
           (5) \citet{stanghellini1998}; (6) \citet{tinti2005}; (7) \citet{taylor2000}; (8) \citet{owsianik1998};
           (9) \citet{vink2006}; (10) \citet{polatidis2003}; (11) \citet{readhead1996}; (12) \citet{siemiginowska2008};
           (13) \citet{stanghellini2001}; (14) \citet{odea2000}; (15) \citet{dallacasa2000}; (16) \citet{stanghellini2002};
           (17) \citet{liu2000};  (18) \citet{guainazzi2004}.}
\end{deluxetable}
\clearpage
\end{landscape}


\clearpage

\begin{deluxetable}{lccccccc}
\tablecaption{Flux densities of the sample's sources observed by the 
             {\it Spitzer Space Telescope}\label{tab_spitzer}}
\tablewidth{0pt}
\tablehead{
\colhead{Source name} &
\multicolumn{4}{c}{IRAC} & 
\colhead{~~} &
\multicolumn{2}{c}{MIPS} \\
\colhead{~~~}     & 
\colhead{$F_{3.6\, \mu\mathrm{m}}$} & 
\colhead{$F_{4.5\, \mu\mathrm{m}}$} & 
\colhead{$F_{5.8\, \mu\mathrm{m}}$} & 
\colhead{$F_{8.0\, \mu\mathrm{m}}$} & 
\colhead{~~~}     & 
\colhead{$F_{24\, \mu\mathrm{m}}$}  & 
\colhead{$F_{70\, \mu\mathrm{m}}$} \\
\colhead{~~~}   & 
\colhead{(mJy)} &  
\colhead{(mJy)} &
\colhead{(mJy)} &
\colhead{(mJy)} &
\colhead{~~}    &
\colhead{(mJy)} &
\colhead{(mJy)} 
}
\startdata
IERS B0500+019  & \nodata  & \nodata  & \nodata  & \nodata  &  ~~ & $6.3\pm 0.3$  & \nodata\\
IERS B1345+125  & $8.2\pm 0.6$ & $13.4\pm 0.9$ & $20.9\pm 1.1$ &  $42.3\pm 1.9$ & ~~ &  $479\pm 21$ & $1947\pm 186$ \\
IERS B1404+286  & $37.0\pm 1.7$& $47.0\pm 2.0$ & $54.5\pm 2.3$ &  $78.7\pm 3.1$ & ~~ &  $382\pm 17$ & $804\pm 102$ \\
\enddata
\tablecomments{Details on the analysis of the {\it Spitzer} data can be found in Appendix \ref{app_B}.}
\end{deluxetable}


\clearpage

\begin{landscape}
\begin{deluxetable}{lcccccccccc}
\tabletypesize{\scriptsize}
\tablecaption{SED modeling: Fit parameters and derived quantities \label{tab_par}}
\tablewidth{0pt}
\tablehead{
\colhead{Source name} & 
\colhead{$L_{\mathrm{j}}$} & 
\colhead{$L_{\mathrm{UV}}$} & 
\colhead{$L_{\mathrm{IR}}$} & 
\colhead{$L_{\mathrm{opt}}$} & 
\colhead{$s_1$} & 
\colhead{$s_2$} & 
\colhead{$\nu_{\mathrm{p}}$} & 
\colhead{$\nu_{\mathrm{IR}}$} & 
\colhead{$L_{\mathrm {j}}/(10\, L_{\mathrm {UV}})$} & 
\colhead{~~~$L_{\mathrm {UV}}/L_{\mathrm{Edd}}\,\,[M_{\mathrm {BH},8}]$\tnm{(a)}} \\
\colhead{ ~~}  & 
\colhead{($10^{45}$ erg \persec)} & 
\colhead{($10^{45}$ erg \persec)} &
\colhead{($10^{45}$ erg \persec)} & 
\colhead{($10^{45}$ erg \persec)} & 
\colhead{~~} & 
\colhead{~~} & 
\colhead{(GHz)} & 
\colhead{(THz)} & 
\colhead{~~} & 
\colhead{~~} 
}
\startdata
IERS B0026+346  & 0.70 & 10 & 6.00  & 1.00 & 0.70 & 2.0 & 0.7 &  5  & 0.007 & 0.390 [1.698$^{\ast}$] \\
IERS B0108+388  & 2.05 & 5  & 0.50  & 0.60 & 1.80 & 3.2 & 6.0 &  5  & 0.041 & 0.417 [0.794$^{\ddagger}$]\\
IERS B0500+019  & 1.80 & 5  & 3.50  & 1.00 & 1.70 & 2.0 & 2.0 &  5  & 0.036 & 0.195 [1.698$^{\ast}$] \\
IERS B0710+439  & 2.80 & 10 & 4.30  & 0.90 & 2.00 & 2.1 & 2.5 &  5  & 0.028 & 0.264 [2.512$^{\ddagger}$] \\
PKS  B0941-080  & 0.40 & 1  & 0.08  & 0.80 & 2.00 & 2.5 & 0.5 &  5  & 0.040 & 0.132 [0.501$^{\dagger}$] \\
IERS B1031+567  & 0.87 & 5  & 1.40  & 0.30 & 1.90 & 2.0 & 1.0 &  5  & 0.017 & 0.263 [1.259$^{\ddagger}$] \\
IERS B1345+125  & 0.17 & 5  & 6.00  & 0.90 & 1.40 & 2.0 & 0.4 &  5  & 0.003 & 0.506 [0.654$^{\diamond}$] \\
IVS  B1358+624  & 2.10 & 5  & 6.00  & 0.35 & 1.75 & 2.7 & 0.6 &  5  & 0.042 & 0.209 [1.585$^{\ddagger}$] \\
IERS B1404+286  & 0.30 & 2  & 0.70  & 0.50 & 2.15 & 5.6 & 5.0 & 13  & 0.015 & 0.026 [5.012$^{\ddagger}$] \\
IERS B2128+048  & 4.00 & 5  & 3.00  & 0.65 & 1.70 & 2.5 & 1.2 &  5  & 0.080 & 0.195 [1.698$^{\ast}$]\\
IERS B2352+495  & 0.45 & 5  & 0.60  & 0.35 & 2.00 & 2.0 & 1.0 &  5  & 0.009 & 0.132 [2.512$^{\ddagger}$] \\
\enddata
\tablecomments{The fit parameters (cols.\ 2--9) are described in Sections \ref{sec_model} and \ref{sec_SEDmodeling};
                   the derived quantities (cols.\ 10 and 11) are discussed in Section \ref{sec_discussion}. 
                   Parameters common to all the SED fits, not included in this Table, are as follows: $n_0=0.1$ \percmc, 
                   $\gamma_{\mathrm {min}}=1$,
                   $\gamma_{\mathrm{max}}=100\,m_{\mathrm{p}}/m_{\mathrm{e}}$, $\eta_{\mathrm{e}}=3$, $\eta_{\mathrm{B}}=0.3$,
                   $\nu_{\mathrm{opt}}=2.0\times10^{14}$ Hz, 
                   $\nu_{\mathrm{UV}}=2.45\times10^{15}$ Hz.
                   We assumed hotspot propagation velocity $v_{\mathrm{h}}=0.1c$ when no hotspot 
                   separation velocity $v_{\mathrm{h,sep}}$ was available 
                   (see Table 1); from the known $v_{\mathrm{h,sep}}$, we derived $v_{\mathrm{h}}$ by using the  
                   appropriate composition law for velocities, yielding in most cases 
                   $v_{\mathrm{h}}\simeq0.5v_{\mathrm{h,sep.}}$.}

\tablenotetext{(a)}{The Eddington luminosity was computed according to the relation: 
                    $L_{\mathrm{Edd}}=1.51\times 10^{46} \, M_{\mathrm{BH}}/(10^8 M_{\sun})$ erg \persec 
                    \citep[e.g.,][]{krolik1998}.
                    The assumed BH mass is given in square brackets in units of  $10^8 M_{\sun}$: it is 
                    the most recent estimate of the source BH mass when available (from \citet{wu2009a} 
                    when marked with $^{\dagger}$, from \citet{wu2009b} when marked with $^{\ddagger}$, 
                    and from \citet{dasyra2006} when marked with $^{\diamond}$),
                    and the average value of the BH mass estimated by \citet{wu2009b} for his GPS-source
                    sample ($M_{\mathrm {BH},8}=1.698$) otherwise (i.e., for masses marked 
                    with $^{\ast}$).}
\end{deluxetable}

\clearpage
\end{landscape}


\clearpage

\begin{deluxetable}{lcccccc}
\tabletypesize{\scriptsize}
\tablecaption{X-ray spectral parameters and radio \HI column densities.\label{tab_xrad}} 
\tablewidth{0pt}
\tablehead{
\colhead{Source name} & 
\colhead{$N_{\mathrm{H,Gal}}$} &   
\colhead{\NH\tnm{(a)}} & 
\colhead{$\Gamma$} & 
\colhead{Ref.} & 
\colhead{\NHI\tnm{(b)}} & 
\colhead{Ref.}\\
\colhead{~~} &
\colhead{($10^{20}$\percmq)} &
\colhead{($10^{22}$\percmq)}& 
\colhead{~~} & 
\colhead{~~} &
\colhead{($10^{18} \times (T_{\mathrm{s}}/c_f)$ \percmq)}& 
\colhead{~~} 
}
\startdata
IERS B0026+346 & 5.6   & $1.0^{+0.5}_{-0.4}$ & $1.43^{+0.20}_{-0.19}$   & 2     & \nodata           & \nodata\\
\vspace{0.1cm}
IERS B0108+388 & 5.8   & $\mathbf{57\pm20}$          & 1.75\tnm{(c)}          & 3     & $\mathbf{80.7}$    & 8\\
\vspace{0.1cm}
IERS B0500+019 & 8.3    & $\mathbf{0.5^{+0.3}_{-0.2}}$ & $1.62^{+0.21}_{-0.19}$   & 2     & $\mathbf{6.2}$  (4.5,2.5)    & 8  \\
\vspace{0.1cm}
IERS B0710+439 & 8.11  & $0.44\pm0.08$      & $1.59\pm0.06$           & 3     & \nodata            & \nodata  \\   
\vspace{0.1cm}
PKS  B0941-080 & 3.7   & \nodata            & 2\tnm{(c)}              & 2     & $<$0.80\tnm{(d)}$\,$, $\mathbf{<}$$\mathbf{1.26}$\tnm{(e)} & 6, 10 \\ 
\vspace{0.1cm}
~~             & 3.7   & \nodata            & $2.62^{+1.29}_{-1.03}$   & 9     &  ~~               & ~~  \\
\vspace{0.1cm}
~~            & 3.67    & $\mathbf{<1.26}$\tnm{(e)} & $2.28^{+0.67}_{-0.61}$    & 9b &  ~~    & ~~  \\
\vspace{0.1cm}
~~             & 3.67    &$<0.53$\tnm{(e)} & $(1.7-1.9)$\tnm{(c)}    & 9b    &  ~~               & ~~  \\
\vspace{0.1cm}
IERS B1031+567 & 0.56   & $\mathbf{0.50\pm0.18}$      & 1.75\tnm{(c)}           & 3     & $<$0.87\tnm{(d)}$\,$, $\mathbf{<}$$\mathbf{1.26}$\tnm{(e)}  & 6, 10 \\   
\vspace{0.1cm}
IERS B1345+125 & 1.1    & $\mathbf{4.2^{+4.0}_{-2.4}}$ & $1.6^{+1.2}_{-0.8}$    & 1   & $\mathbf{6.2}$\tnm{(f)} &  7 \\ 
\vspace{0.1cm}
~~             & 1.9    & $\mathbf{2.54^{+0.636}_{-0.580}}$ & $1.10^{+0.29}_{-0.28}$ & 9   & ~~               &  ~~ \\ 
\vspace{0.1cm}
IVS  B1358+624 & 1.96   & $\mathbf{3.0\pm0.7}$        & $1.24\pm0.17$         & 3      & $\mathbf{1.88}$\tnm{(g)}      & 6 \\   
\vspace{0.1cm}
IERS B1404+286\tnm{(h)} & 1.4  & $0.13^{+0.12}_{-0.10}$\tnm{(i)}  &   $2.1^{+0.6}_{-0.3}$\tnm{(i)}, $0.7^{+0.3}_{-0.4}$\tnm{(j)} & 4   & 1.83\tnm{(g)}  & 6 \\  
\vspace{0.1cm}
~~             & ~~     & $0.12^{+0.09}_{-0.08}$\tnm{(i)} $\,$, $24.0^{+10.0}_{-8.0}$\tnm{(j)} &  $2.0^{+0.4}_{-0.3}$    & ~~  & ~~  & ~~ \\
\vspace{0.1cm}
~~             & ~~     & $0.11\pm0.05$\tnm{(i,k)}    & $2.21^{+0.19}_{-0.14}$\tnm{(l)} & ~~  & ~~             & ~~ \\
\vspace{0.1cm}
~~             & ~~     & $0.19^{+0.13}_{-0.10}$\tnm{(i)} & $2.6\pm 0.5$\tnm{(i)}, 2\tnm{(j,c)}  & ~~  & ~~ & ~~ \\
\vspace{0.1cm}
~~             & ~~     & $0.09^{+0.08}_{-0.06}$\tnm{(i)} & $2.2\pm 0.4$  & ~~  & ~~         & ~~ \\
\vspace{0.1cm}
IERS B2128+048 & 5.2     & $0.3^{+0.81}_{-0.3}$    & $1.5^{+0.6}_{-0.7}$   & 2       & \nodata          & \nodata\\ 
\vspace{0.1cm}
~~             & 5.2     & \nodata                &  $1.28^{+0.42}_{-0.41}$ & 9      & ~~              & ~~  \\
\vspace{0.1cm}
IERS B2352+495 & 12.4   & $\mathbf{0.66\pm0.27}$    & 1.75\tnm{(c)}        & 3     & 0.28; $\mathbf{2.56}$\tnm{(g,n)}  & 6 \\
\enddata
\tablenotetext{(a)}{Equivalent total hydrogen column density, assumed to be located at the 
                    source redshift; the values in boldface are those used for the correlation analysis 
                    (see Section \ref{sec_NH_sample}, and Tables \ref{tab_correlation1} and \ref{tab_correlation2}).}
\tablenotetext{(b)}{Neutral hydrogen column density derived from low resolution measurements, unable to determine the location
                    of the absorbing gas; the values in boldface are those used for the correlation analysis 
                    (see Section \ref{sec_NH_sample}, and Tables \ref{tab_correlation1} and \ref{tab_correlation2}).}
\tablenotetext{(c)} {Fixed.} 
\tablenotetext{(d)}{Derived from the value of \NHImath$_{,2\sigma}$ given by (6), which was obtained from the relation: 
                  $1.82\times10^{18}\times T_{\mathrm{s}} \times \tau_{2\sigma}\times \Delta V$, under the assumption 
                  $T_{\mathrm{s}}$=100 K, $\Delta V = 100$ km \persec, and $c_f=1$; no detection of absorption lines.}
\tablenotetext{(e)} {$3\sigma$ upper limit.}
\tablenotetext{(f)} {\NHImath$\sim 10^{22}\times (T_{\mathrm{s}}/100$ K$)$ \percmq was obtained by \citet{morganti2004b} from high-resolution  
                     measurements that constrain the location of the absorbing gas.}
\tablenotetext{(g)}{Derived from the value of \NHImath$_{,2\sigma}$ given by (6), which was obtained from the relation 
                    $1.82\times10^{18}\times T_{\mathrm{s}} \times \tau_{2\sigma}\times \Delta V$, under the assumption 
                    of $T_{\mathrm{s}}$=100 K, and $c_f=1$.}
\tablenotetext{(h)} {For this source, we report the X-ray spectral parameters derived by \citet{guainazzi2004} by means of different 
                     models for the X-ray emission; the first three parameter sets refer to the bow-ties represented 
                     in Fig.\ \ref{fig_seds}; the third set is bes fitted by our model.}
\tablenotetext{(i)} {Soft X-rays ($E > E_{\mathrm{break}}$).}
\tablenotetext{(j)} {Hard X-rays ($E>E_{\mathrm{break}}$).}
\tablenotetext{(k)} {Associated with a Compton-reflection model for the hard X-rays, 
                     with \NHmath$^{\mathrm{hard}}>9\times10^{23}$ \percmq.}
\tablenotetext{(l)} {Intrinsic, for the Compton-reflection model.}
\tablenotetext{(m)} {$3\sigma$ errors.}
\tablenotetext{(n)} {After submission of this paper, \citet{araya2010} published high resolution
                     measurements that constrain the location of the absorbing gas, and derived 
                     the values \NHImath$=7.3\times 10^{21}$ \percmq 
                     and \NHImath$=5.2\times 10^{22}$ \percmq, by assuming $T_s=8000$ K.}
\tablerefs{(1) \citet{odea2000}; (2) \citet{guainazzi2006}; (3) \citet{vink2006}; (4) \citet{guainazzi2004};  
           (6) \citet{vermeulen2003}; (7) \citet{mirabel1989}; (8) \citet{carilli1998}; 
           (9) \citet{siemiginowska2008}; (9b) Siemiginowska 2008, priv.\ comm.; (10) \citet{pihlstroem2003}.}
\end{deluxetable}


\clearpage
\begin{deluxetable}{ccccccccccc}
\tabletypesize{\scriptsize}
\tablecaption{\NHmath-\NHI connection: Correlation and regression analysis for detections \label{tab_correlation1}}
\tablewidth{0pt}
\tablehead{
\colhead{Sample\tnm{a}} & 
\colhead{N} & 
\colhead{~~} & 
\multicolumn{2}{c}{Pearson}  & 
\multicolumn{2}{c}{Spearman} & 
\multicolumn{2}{c}{Kendall} & 
\multicolumn{2}{c}{Linear regression~~~~~~~}\\
\colhead{~~} & 
\colhead{~~} &  
\colhead{~~} &  
\colhead{$r$} & 
\colhead{Prob.\tnm{b}} & 
\colhead{$\rho$}     &  
\colhead{Prob.\tnm{b}} & 
\colhead{$\tau$}    &  
\colhead{Prob.\tnm{b}} & 
\colhead{Slope}  &  
\colhead{Intercept} 
}
\startdata
{\it D5}    & 6  &~~ & 0.9966  & 1.739e-05 & 0.3947 & 0.4387 & 0.2981 & 0.4008 & 0.837$\pm$0.109 & 5.151$\pm$2.269 \\
{\it D5+AD} & 7  &~~ & 0.9967  & 1.229e-06 & 0.6301 & 0.1294 & 0.5143 & 0.1048 & 0.981$\pm$0.100 & 2.122$\pm$2.067 \\
\enddata
\tablenotetext{a}{{\it D5} is our sub-sample of \NHmath-\NHI detections;
                  {\it AD} indicates \NHmath-\NHI detections for additional sources reported 
                  by \citet{tengstrand2009};
                  see Section \ref{sec_NH_sample} and Table \ref{tab_xrad} for details.}
\tablenotetext{b}{Probability of the null hypothesis of no correlation being true. It is a 
                  two-sided significance level: because we are looking {\it a priori} for a positive correlation,
                  this value should actually be divided by 2, improving the significance by a factor of 2.} 
\end{deluxetable}


\begin{deluxetable}{lccccccccc}
\tabletypesize{\scriptsize}
\tablecaption{\NHmath-\NHI connection: Correlation and regression analysis for detections 
              and upper limits \label{tab_correlation2}}
\tablewidth{0pt}
\tablehead{
\colhead{Sample\tnm{a}} & 
\colhead{N}  & 
\colhead{~~} &  
\multicolumn{2}{c}{Generalized Spearman} & 
\multicolumn{2}{c}{Generalized Kendall} & 
\multicolumn{2}{c}{Schmitt's linear regression}\\
\colhead{~~}     & 
\colhead{~~} &  
\colhead{~~ }& 
\colhead{$\rho$ } &   
\colhead{Prob.\tnm{b,c}} &  
\colhead{$S$\tnm{d}} & 
\colhead{Prob.\tnm{b}} & 
\colhead{Slope} & 
\colhead{Intercept} 
}
\startdata 
{\it D5+U2}     & 8 &~~ &  0.670 & 0.0764 &  26. ($z$=1.749) & 0.0804  & 0.9433$\pm$0.0621 & 2.8378$\pm$1.3403 \\  
{\it D5+U2+ADU} & 11 &~~ & 0.738 & 0.0197 &  54. ($z$=2.402)  & 0.0163 &  1.0711$\pm$0.0034 & 0.0775$\pm$0.0695 \\
\enddata
\tablenotetext{a}{{\it D5} is our sub-sample of \NHmath-\NHI detections;
                    {\it U2} is our sub-sample of \NH and/or \NHI upper limits; 
                    {\it ADU} is the set of detections of and upper limits to, 
                    \NH and/or \NHI for additional sources 
                    reported by \citet{tengstrand2009};
                    see Section \ref{sec_NH_sample} and Table \ref{tab_xrad} for details.}
\tablenotetext{b}{Probability of the null hypothesis of no correlation being true. It is a 
                    two-sided significance level: because we are looking {\it a priori} for a positive correlation,
                    this value should actually be divided by 2, improving the significance by a factor of 2.} 
\tablenotetext{c}{According to the ASURV Rev.\ $1.2$ software package, this value is accurate only if $N>30$.}
\tablenotetext{d}{$S$ is defined in \citet{isobe1986}.}
\end{deluxetable}

\end{document}